\documentclass[12pt,a4paper]{article}
\pdfoutput=1
\usepackage{jheppub}
\usepackage[utf8]{inputenc}
\usepackage[T1]{fontenc}

\usepackage[only,llbracket,rrbracket]{stmaryrd}
\usepackage{tikz, pgf}  
\usetikzlibrary{patterns}
\usetikzlibrary{positioning}
\usetikzlibrary{decorations.markings}
\usetikzlibrary{intersections}
\usetikzlibrary{arrows}
\usepackage{amsmath,physics,float}  
\usepackage{amssymb}  
\usepackage{latexsym}
\usepackage[normalem]{ulem}
\usepackage{booktabs}
\usepackage{multirow}
\usepackage{amsfonts}
\usepackage{braket}
\usepackage{mathrsfs}

\usepackage{subfig} 
\usepackage{empheq}

\usepackage[shortlabels]{enumitem}

\usepackage{pgfplots}
\pgfplotsset{compat=1.13}
\usepackage{natbib} 
\usepackage{comment}

\usepackage{bigints}
\definecolor{ashgrey}{rgb}{0.6, 0.5, 0.5}
\definecolor{ashgrey2}{rgb}{0.7, 0.73, 0.71}

\newcommand{\bg}{\begin{equation}}
\newcommand{\nd}{\end{equation}}
\newcommand{\beq}{\begin{equation}}
\newcommand{\eeq}{\end{equation}} 
\newcommand{\ef}{\rm eff}

\newcommand{\del}{\partial}
\newcommand{\lc}{\left(}
\newcommand{\rc}{\right)}

\newcommand{\ca}{\bf}

\begin{document}

\title{Wheeler-De Witt equation and the Canonical Construction of the Glauber-Sudarshan States in Quantum Gravity}

\author{{Keshav Dasgupta}, {Fang-Yi Guo}, and {Bohdan Kulinich}}

\emailAdd{keshav@hep.physics.mcgill.ca, bohdan.kulinich@mail.mcgill.ca, fguo@physics.mcgill.ca}  

\affiliation{Department of Physics, McGill University, Montr\'{e}al, Qu\'{e}bec, H3A 2T8, Canada}

\begin{abstract}
{Quantum gravity is fundamentally different from the non-gravitational quantum field theories in the sense that most of the techniques derived for the latter cannot be easily extended to the former. For example, correlation functions in quantum gravity become hard to define properly if the bulk Hamiltonian $-$ in the absence of a boundary Hamiltonian, and as a consequence of the Wheeler-De Witt equation $-$ itself annihilates the states thus rendering the evolution operators to identities. An even harder problem is the back reactions of the fluctuations on the background itself. We argue that the construction of the Glauber-Sudarshan states takes care of these two issues in rather interesting ways. For the former, the displacement operators are defined at every instant of time, without directly invoking the Hamiltonian, so that they naturally extend to the path-integral description as sum over histories. For the latter, the back reactions of the fluctuations are carefully accounted for by the Schwinger-Dyson equations. In fact these back reactions are in part responsible to convert the ambient supersymmetric Minkowski background to a transient de Sitter phase. Expectedly this transient de Sitter phase is defined well within the validity regime of the trans-Planckian bound.}
\end{abstract}

\maketitle

\hskip1.3in \textcolor{blue}{\it Hamiltonian is dead and we must bury it with all the honor it deserves.}

\vskip.1in

\hskip2in $-$ Lev Landau, Kiev Conference 1954

\section{Introduction and summary}

Glauber-Sudarshan states \cite{wdwpaper, desitter2, coherbeta, coherbeta2, borel, joydeep} provide, probably for the first time, a controlled laboratory to perform non-supersymmetry computations in a temporally varying background in string theory and M-theory. Due to the various constraints discussed in \cite{wdwpaper, desitter2, coherbeta, coherbeta2, borel, joydeep}, the aforementioned computations are to be performed in the far IR limit of string/M theory. Moreover the construction of these states are defined respecting the Hamiltonian constraints appearing from the underlying Wheeler-De Witt equation \cite{wdwpaper}. These states are constructed as excited states over a local supersymmetric minimum, and as such break the underlying supersymmetry spontaneously. 

A natural question then is the connection between the Glauber--Sudarshan
states and the states created by vertex operators in string theory.  The
latter create string excitations which, in suitable circumstances, may be
interpreted as coherent deformations of a background, but one important
difference is that the standard vertex-operator construction is most
naturally formulated around perturbative string backgrounds, whereas the
Glauber--Sudarshan states considered here are designed to describe
transient, non-supersymmetric and temporally varying backgrounds in the
far-IR limit of string/M theory.  In fact, the natural starting point for
the Glauber--Sudarshan construction is M-theory---where no comparably simple
world-sheet vertex-operator formalism is known---and the corresponding
string-theory descriptions are obtained through duality transformations.
In section \ref{sec2}, and especially in section \ref{sec2.3}, we will
elaborate on these differences in more detail.  For the moment, it suffices
to emphasize that although the displacement operators defining the
Glauber--Sudarshan states bear a structural resemblance to integrated
vertex operators, the two constructions are conceptually distinct.  The
allowed Glauber--Sudarshan configurations are constrained by the full
quantum Schwinger--Dyson equations associated with the exact
effective action, whereas conventional physical string vertex operators
are characterized, at leading order, by the corresponding world-sheet
physical-state or marginality conditions, which reproduce the spacetime
equations of motion in the appropriate semiclassical limit.  More
importantly, despite the formal resemblance, the manner in which the
Glauber--Sudarshan states act is different.  Their construction uses the
degrees of freedom of the underlying supersymmetric warped-Minkowski
minimum, so that a Wilsonian effective description can be maintained, and
the transient de Sitter configuration is obtained from the corresponding
on-shell deformation of the full semiclassical background.  The associated
backreaction is therefore determined by solving the Schwinger--Dyson and
bootstrap equations for the full fluctuating configuration, as we shall
elaborate in section \ref{sec2.5}.

An important set of ingredients in the construction of the Glauber-Sudarshan states are the ghosts that are more general than the standard Faddeev-Popov ghosts. In section \ref{sec2.4} we will provide a detailed study of them wherein we will show that, unlike usual gauge theories, both the gauge fixing terms as well as the ghosts themselves require {\it additional} ghosts. In fact the additional ghosts, coming from the Faddeev-Popov ghosts, have gauge redundancies which require further ghosts to be added to the system. This implies a more general Batalin-Vilkovisky \cite{batalin} like formalism to tackle the system which we shall discuss in section \ref{sec2.4}. Moreover, the non-abelian degrees of freedom appearing from the localized G-fluxes in the far IR limit of M-theory put in extra ghosts that interact with the on-shell degrees of freedom and the aforementioned set of ghosts in a non-trivial way. 
All in all, and not surprisingly, the story gets pretty involved here.

The ghosts, as well as the gauge-fixing terms, are not just there to remove the gauge redundancies in the system. They play important roles in the underlying Wheeler-De Witt equation at the level of the warped-supersymmetric Minkowski background itself. This will be the subject of discussion in sections \ref{sec2.3} and \ref{sec2.4}. In fact the Wheeler-De Witt equations appear at {\it two} levels here. One, that we mentioned above, is at the supersymmetric warped-Minkowski level, and two, at the emergent de Sitter level. The latter is non-supersymmetric and is a transient phase that lies well within the validity regime of the trans-Planckian bound \cite{tcc}. The {\it bulk} Hamiltonian, on the other hand, is annihilated at both the Minkowski and the emergent de Sitter levels. For the former, the Hamiltonian is described by \eqref{oma4mey} and by \eqref{leylajoh}, and so has contributions from the ghosts and the gauge-fixing terms. For the latter, the Hamiltonian is an emergent one and is described from the action that determines the EOM \eqref{bormey}. The action itself appears from a sequence of transformations given in \eqref{aspenb}. The Hamiltonian constraints at {\it both} the levels put strong restrictions on the dynamics of the system. Interestingly, the construction of the Glauber-Sudarshan states that we study in section \ref{sec2} is arranged to precisely take care of the aforementioned constraints.

The {\it boundary} Hamiltonian, along with its ghosts and the gauge-fixing terms, is now responsible for organizing the temporal dependence of the eigenstates in the gravitational and the matter sector. The fact that this could be done was already anticipated in \cite{wdwpaper}, and in sections \ref{sec2.2} and \ref{sec2.3} we elaborate the story in more details. Interestingly the temporal dependence of the eigenstates continues to persist even in the unlikely scenario with decoupled ghosts. The fact that the Faddeev-Popov ghosts should always be present comes from an observation by Weinberg that in a theory like General Relativity (GR), the Faddeev-Popov ghosts {\it cannot} decouple \cite{weinberg}. The theory that we consider is much more complicated than GR because of the presence of higher order corrections for both matter and the gravitational sector, plus it takes a trans-series form\footnote{For more details on trans-series the readers may refer to \cite{transseries}. For the appearance of the action as a trans-series, the readers may refer to \cite{joydeep, borel, csvsgs}.}, so decoupling of ghosts is an unlikely scenario. However the aforementioned situation does not prohibit us to find field redefinitions using which the path-integral may be expressed in an apparent ghost-free way. As explained in \cite{rahman} (see also \cite{faddeev}), such a scenario is defined by the presence of {\it instantaneous} ghost propagators so that they serve as constraints in the system\footnote{We thank Renata Kallosh and Adel Rahman for discussion on this.}. In this language, the dimension of the Hilbert space is reduced and the action, expressed in terms of the available degrees of freedom, are now constructed with {\it non-dynamical} ghosts (thus not violating the Weinberg condition \cite{weinberg}). However there do exist degrees of freedom that do not involve Faddeev-Popov ghosts. In section \ref{sec2.4} we argue that the emergent degrees of freedom may indeed be expressed in a ghost-free way (but most likely involving the BRST ghosts), using an action that is more complicated than the original action because of the renormalization effects discussed in \cite{wdwpaper}. The reason why (at least) the Faddeev-Popov ghosts do not appear there is because the construction of the emergent degrees of freedom itself involve the ghosts. They remove the gauge redundancies from the emergent degrees of freedom resulting in an almost ghost free action. The wave-functional controlling the emergent degrees of freedom is given by a Wheeler-De Witt equation \eqref{ryanfurn} which may be compared to the evolution equation \eqref{nestkathia} governing the degrees of freedom at the supersymmetric warped-Minkowski level. In the absence of the boundary Hamiltonian, the problem of time\footnote{There are many papers discussing this issue. A small selection of them that provide a balanced view on the subject are \cite{teitelboim, page, suvrat, witten}.} re-appears which, interestingly, can be shown to be absent for the warped-Minkowski case because it supports a boundary Hamiltonian from its conformal completion.

The problem of time at the emergent space-time level is not a serious issue for us because, as mentioned above, from the supersymmetric warped-Minkowski point of view we have a well-defined notion of temporal evolution. What is however more important is the backreaction problem that we alluded to earlier. The question is how do the Glauber-Sudarshan states back-react on the ambient supersymmetric vacuum to convert it to a non-supersymmetric transient de Sitter spacetime?

This is what we discuss in detail in section \ref{sec2.5}. Our analysis
reveals that, at least for a simple toy model that we take here, one may quantify both the back-reaction effects as well as the value of the four-dimensional cosmological constant precisely using the powerful technique of the Borel-\'Ecalle resummation procedure \cite{borelborel, ecalle}. Interestingly we find that both the results are interconnected and without this resummation technique we would not have been able to infer them correctly here. In fact the resummation procedure is absolutely necessary to quantify the value of the four-dimensional cosmological constant and in turn argue for it's positivity. This suggests that maybe the cosmological constant is an {\it emergent} phenomena that is completely invisible at order-by-order perturbative level. Once we sum up all the non-perturbative effects, the positive cosmological constant emerges out of it.  
We end with a short discussion of our results in section \ref{sec3} where we also point out future avenues for research.

\section{Canonical construction of the Glauber-Sudarshan states \label{sec2}}
To define properly the displacement operator and to see why the conventional ways of expressing correlation functions from non-gravitational field theories are insufficient, one needs to go back to the description of correlation functions in field theories. For the present case we will take simple scalar field theories to illustrate the 
point. A two-point function defined for scalar fields on two different Cauchy slices takes the following well-known form:
\bg\label{cochi}
{\cal Z}_{12} \equiv \langle\Omega\vert \mathbb{T}\{\varphi(x_1) \varphi(x_2)\}\vert\Omega\rangle, \nd
where $x_i = ({\bf x}_i, x_i^0)$; and $\mathbb{T}$ is the time-ordering. The latter
requires an appropriate temporal ordering, which may be defined whenever a suitable
global time function or foliation exists.  In a stationary spacetime such as Minkowski
space this structure is particularly natural because there is, in addition, a global
timelike Killing vector and hence a preferred notion of time translation and conserved
energy.

However there exist a different problem once we try to express \eqref{cochi} using path-integral formalism. Conventional wisdom tells us that:
\bg\label{sabsuzu}
\begin{split}
{\cal Z}_{12} & =  \langle \varphi_a\vert e^{-i{\bf H}_{\rm tot} {\rm T}}\mathbb{T}\{\varphi_{\rm H}(x_1) \varphi_{\rm H}(x_2)\} e^{-i{\bf H}_{\rm tot} {\rm T}} \vert\varphi_b\rangle {\cal A}_{\rm over}\\
& =  \langle \varphi_a\vert e^{-i{\bf H}_{\rm tot} ({\rm T} - x_2^0)}\varphi_{\rm S}({\bf x}_2)e^{-i{\bf H}_{\rm tot} (x_2^0 - x_1^0)}\varphi_{\rm S}({\bf x}_1) e^{-i{\bf H}_{\rm tot}(x_1^0 + {\rm T})} \vert\varphi_b\rangle {\cal A}_{\rm over}\\
& = \lim_{{\rm T}\to \infty(1 - i\epsilon)} {\int {\cal D}\varphi ~\varphi(x_1) \varphi(x_2) ~{\rm exp}\left(i{\bf S}_{\rm tot}[\varphi; \pm{\rm T}]\right)\over\int {\cal D}\varphi ~{\rm exp}\left(i{\bf S}_{\rm tot}[\varphi; \pm{\rm T}]\right)},
\end{split}
\nd
where we have taken $x_2^0 > x_1^0$, $({\rm H, S})$ denote Heisenberg and Schr\"oedinger representations, $({\bf H}_{\rm tot}, {\bf S}_{\rm tot})$ are the total Hamiltonian and action respectively\footnote{There is a minor subtlety here. While ${\bf S}_{\rm tot}[\varphi; \pm{\rm T}]$ is defined in the conventional way, the Hamiltonian ${\bf H}_{\rm tot} \equiv {\bf H}_{\rm tot}(\varphi)$ is the integrated Hamiltonian density over a Cauchy slice. Alternatively one could use a more conventional approach with explicit time dependence of the {\it free} scalar fields entering the Hamiltonian and express the evolution operator using space-time integrals. This is called the {\it interaction picture} which works well for non-gravitational quantum field theories. With gravity the story is more involved as we shall see.}; and ${\cal A}_{\rm over}$ is the overlap integral defined in the following suggestive way:
\bg\label{dadari}
\begin{split}
{\cal A}^{-1}_{\rm over} & = \lim_{{\rm T}\to \infty(1 - i\epsilon)} 
\langle\varphi_a\vert\Omega\rangle \langle \Omega\vert\varphi_b\rangle ~
{\rm exp}\left(-i{\rm E}_0 {\rm T}\right)\\
& = \lim_{{\rm T}\to \infty(1 - i\epsilon)} \langle \varphi_a\vert {\rm exp}\left(-i{\bf H}_{\rm tot} {\rm T}\right)\vert\varphi_b\rangle = \lim_{{\rm T}\to \infty(1 - i\epsilon)} \int {\cal D}\varphi ~{\rm exp}\left(i{\bf S}_{\rm tot}[\varphi; \pm{\rm T}]\right),
\end{split}
\nd
using the interacting vacuum $\vert\Omega\rangle$ and the local vacua $\vert\varphi_a\rangle$ and $\vert\varphi_b\rangle$. The above analysis also relies on the fact that the local vacua may be expanded in terms of the eigenstates of the total Hamiltonian ${\bf H}_{\rm tot}$ in the following way:
\bg\label{molshnnon}
\vert\varphi_a\rangle = c_a \vert\Omega\rangle + \sum_{{\rm N}= 1}^\infty c_{a{\rm N}} \vert {\rm N}\rangle, \nd
with constant $(c_a, c_{a{\rm N}})$. One can then express $\vert\varphi_a\rangle$ completely in terms of $\vert\Omega\rangle$ using the fact that ${1\over c_a} \lim\limits_{{\rm T}\to \infty(1-i\epsilon)} {\rm exp}\left[-i({\bf H}_{\rm tot} - {\rm E}_0){\rm T}\right]\vert\varphi_a\rangle = \vert\Omega\rangle$, with ${\rm E}_0$ being the energy of the lowest eigenstate $\vert\Omega\rangle$ of the Hamiltonian ${\bf H}_{\rm tot}$.

What we said above is familiar and rather well-defined for non-gravitational quantum field theories with irrelevant or marginal couplings. (For relevant couplings, the story is a bit different but nevertheless follow somewhat similar line of analysis.) Unfortunately once we include gravitational degrees of freedom such a simple picture doesn't really work well. For example, if the Hamiltonian ${\bf H}_{\rm tot}$ annihilates the states, then all what we said above is not going to work. However there is a misconception that we can still define propagating states using some {\it boundary} Hamiltonian. Such an argument relies on the fact that the boundary Hamiltonian is {\it non-zero}. This is not always the case. In the following let us first illustrate that.

\subsection{Hamiltonian constraint from the WdW equation \label{sec2.1}}

In quantum gravity, providing an action {\it without} a boundary piece generically does not suffice. A boundary piece becomes essential and therefore ${\bf S}_{\rm tot}$ should allow for such a boundary piece. We should also specify the allowed degrees of freedom. Following \cite{wdwpaper}, we can express the on-shell degrees of freedom by ${\bf \Xi}$ and the set of ghosts\footnote{Ghosts didn't appear in \eqref{sabsuzu} because we took scalar fields. Once we go away from such a simple set-up and incorporate tensor and fermionic fields, ghosts would naturally appear. We will have more to say on this soon.} by ${\bf \Upsilon}$, and express ${\bf S}_{\rm tot}({\bf \Xi, \Upsilon})$ in the following suggestive way:
\bg\label{jojamezz}
{\bf S}_{\rm tot}({\bf \Xi, \Upsilon}) = \underbrace{\int_{\mathbb{M}} d^{n+1}x\sqrt{-{\bf g}}\left[{\cal L}_{\rm tot}({\bf \Xi, \Upsilon}) - \nabla_{\rm N} {\bf W}^{\rm N}({\bf \Xi, \Upsilon})\right]}_{{\bf S}_{\rm bulk}({\bf \Xi, \Upsilon})} + \underbrace{\int_{\del\mathbb{M}} d\Sigma^{(n)}_{\rm P} {\bf W}^{\rm P}({\bf \Xi, \Upsilon})}_{{\bf S}_{\rm bnd}({\bf \Xi, \Upsilon})},\nd
where note that have simply extracted the total derivative pieces from 
${\cal L}_{\rm tot}({\bf \Xi, \Upsilon})$ and re-written ${\bf S}_{\rm tot}({\bf \Xi, \Upsilon})$ in a more suggestive way. The fact that we could implement such a procedure can be easily seen directly from the {\it perturbative} part of the total action, namely\footnote{Meaning the zero instanton sector of the total action expressed as a trans-series \cite{joydeep, desitter2, coherbeta, coherbeta2, borel}. Note that for the simple case with Einstein gravity, this leads to the familiar Gibbons-Hawking-York boundary term \cite{GHY}. Moreover, a cleaner notation for \eqref{qotom} would be:
\begin{equation}
\prod_{k=1}^{N_R}
\left(
{\bf R}_{A_kB_kC_kD_k}
\right)^{l_k}
\prod_{p=1}^{N_G}
\left(
{\bf G}_{A_pB_pC_pD_p}
\right)^{m_p},
\end{equation}
with $N_R$ and $N_G$ understood as arbitrary bookkeeping integers. The numerical factors $61$ and $100$ have no
independent dynamical significance; they merely label two different classes
of insertions in the schematic trans-series expression that appeared in \cite{ desitter2, coherbeta, coherbeta2}.}:

{\footnotesize
\bg\label{qotom}
{\cal L}_{\rm tot}({\bf \Xi}) = \sum_{\{l_i\}, \{n_j\}} {{\cal C}_{\{l_i\}, \{n_j\}} \over {\rm M}_p^{\sigma_{nl}}} \left[{\bf g}^{-1}\right] \prod_{j = 0}^3 \left[\partial\right]^{n_j} \prod_{k = 1}^{60} \left({\bf R}_{\rm A_k B_k C_k D_k}\right)^{l_k} \prod_{p = 61}^{100} \left({\bf G}_{\rm A_p B_p C_p D_p}\right)^{l_p} + {\cal O}\left({\rm exp}(-{\rm M}_p^{\sigma'})\right), \nd}
where ${\cal C}_{\{l_i\}, \{n_j\}}$ are constant coefficients such that 
$(l_i, n_j) \in (+\mathbb{Z}, +\mathbb{Z})$ and $(n_0, n_1, n_2, n_3)$ refer to the number of derivatives along the temporal, ${\bf R}^2, {\cal M}_6$ and ${\mathbb{T}^2\over {\cal G}}$ directions respectively for M-theory on ${\bf R}^{2, 1} \times {\cal M}_6 \times {\mathbb{T}^2\over {\cal G}}$ (for details on the construction the readers may refer to \cite{joydeep}, or the earlier works \cite{desitter2, coherbeta, coherbeta2}). The other parameter is $\sigma_{nl}$ that determines the ${\rm M}_p$ suppressions of each terms in the above series and is given by
$\sigma_{nl} \equiv \sum\limits_{j = 0}^3 n_j + 2\sum\limits_{k = 1}^{60} l_k + \sum\limits_{p = 61}^{100} l_p$ (with $\sigma'$ being a similar factor but in the non-zero instanton sectors). Note three things: \textcolor{blue}{one}, the absence of ghosts in the expression \eqref{qotom}. This is intentional as ghosts will only appear once we insert this in the path-integral. \textcolor{blue}{Two}, it is easy to extract ${\bf W}^{\rm P}({\bf \Xi})$ terms from \eqref{qotom}. They are just the total derivative terms along the temporal, ${\bf R}^2, {\cal M}_6$ and ${\mathbb{T}^2\over {\cal G}}$ directions, implying that the {\it bulk} and the {\it boundary} terms are naturally contained in the form of ${\bf S}_{\rm tot}$. In fact even in the full trans-series form for ${\bf S}_{\rm tot}$, given by eq. (7.55) in \cite{joydeep}, it is not hard to extract the boundary contributions. These boundary pieces would signal the non-perturbative contributions to the boundary. And \textcolor{blue}{three}, we have only taken the {\it on-shell} degrees of freedom\footnote{By this we mean the degrees of freedom contributing to the on-shell equations that will eventually determine the de Sitter background in say the flat-slicing. The off-shell degrees of freedom are then the ``cross-terms'' that do not appear in the EOMs. This slightly unconventional way of expressing the action and the Hamiltonian turns out to be more useful. \label{karkapur}}. This is justified by the detailed analysis in \cite{joydeep} wherein the off-shell pieces are integrated over to give rise to the non-local contributions that are in-turn expressed using the on-shell degrees of freedom. That they contribute to the boundary action can also be easily inferred from the trans-series form given by eq. (7.55) in \cite{joydeep}. The Hamiltonian that appears from the aforementioned choice of the degrees of freedom, takes the following form:
\bg\label{sarbleck}
{\bf H}_{\rm tot}({\bf \Xi, \Upsilon}) \equiv {\bf H}_{\rm tot}(\hat{\bf \Xi}) = \int d^n{\bf x}\left(\Pi_{\hat{\bf \Xi}}(\hat{\bf \Xi}({\bf x}))\cdot {\dot{\hat{\bf \Xi}}}({\bf x}) - {\cal L}_{\rm tot}(\hat{\bf \Xi}({\bf x}))\right), \nd
which generically could have an {\it implicit} time-dependence and $\Pi_{\hat{\bf \Xi}}(\hat{\bf \Xi}({\bf x}))$ is the conjugate momenta for the degrees of freedom $\hat{\bf \Xi} = ({\bf \Xi, \Upsilon})$. By construction the Hamiltonian is fixed over a Cauchy slice at say $t = 0$. Note that we are not expressing the Hamiltonian using fields in the interaction picture. As mentioned in \cite{wdwpaper} there is no simple {\it free field} picture which can be used efficiently to express the Hamiltonian using temporally evolving {free fields}\footnote{In conventional non-gravitational field theories, the definition \eqref{sarbleck} suffices. For example if we identify 
${\bf H}_{\rm tot}(\hat{\bf \Xi}) = {\bf H}_{\rm tot}(\hat{\bf \Xi})\vert_{t = 0} = {\bf H}_{\rm tot}(\hat{\bf \Xi}; 0)$, then the temporal evolutions may be quantified as:
\bg\label{keriruss}
\hat{\bf \Xi}({\bf x}, t) = e^{i{\bf H}_{\rm tot}(\hat{\bf \Xi})t}~\hat{\bf\Xi}({\bf x}, 0)~e^{-i{\bf H}_{\rm tot}(\hat{\bf \Xi})t}, ~~~~ {\bf H}_{\rm tot}(\hat{\bf \Xi}; t) =e^{i{\bf H}_{\rm tot}(\hat{\bf \Xi})t}~{\bf H}_{\rm tot}(\hat{\bf \Xi})~ e^{-i{\bf H}_{\rm tot}(\hat{\bf \Xi})t}, \nd
implying that all temporal evolutions can be controlled by ${\bf H}_{\rm tot}(\hat{\bf \Xi})$ from \eqref{sarbleck}, and there is no need to go to the interaction picture. The problem with this simple yet elegant picture is that from the bulk point of view, and in the absence of any asymptotic boundary, ${\bf H}_{\rm tot}(\hat{\bf \Xi}) = 0$ so temporal evolutions within the framework of the Hamiltonian constraint becomes harder to interpret. This is of course one of the many problems in the canonical formalism that we shall discuss soon.}.

To see how any generic state behaves once both gravitational and matter degrees of freedom are incorporated in, we have to define the corresponding wave-functional associated with the state. This is basically the wave-functional of the universe, and may be defined using path-integral in the following way:
\bg\label{katnobi}
\Psi_{\rm tot}({\bf \Xi}_2, {\bf \Upsilon}_2; t) = \sum_{\mathbb W}\int_{{\bf \hat\Xi}_1, t_1}^{{\bf \hat\Xi}_2, t_2} {\cal D}{\bf \Xi} {\cal D} {\bf \Upsilon} ~e^{-i{\bf S}_{\rm tot}[{\bf \Xi, \Upsilon}]}\Psi_{\rm tot}({\bf \Xi}_1, {\bf \Upsilon}_1; t_1) {\cal D}{\bf \hat\Xi}_1, \nd
where ${\bf \hat\Xi} \equiv ({\bf \Xi, \Upsilon})$, $t = t_2 - t_1$;  and the path-integral over Faddeev-Popov ghosts becomes necessary to extract a finite answer from \eqref{katnobi}. Note two things: \textcolor{blue}{one}, the sum is over all histories that include sum over all configurations as well as sum over all topologies $\mathbb{W}$. And \textcolor{blue}{two}, the dependence on the final wave-functional over the space-time coordinates, and especially on time, at least formally\footnote{This is going to change soon.}. While the existence of a global time-like direction over a warped Minkowski background in M-theory guarantees the temporal dependence, this doesn't quite extend to the Hamiltonian. To see this we will perform a Weiss variation\footnote{Weiss variation is described in \cite{weiss}. To see how such a variation can be used to determine the Wheeler-De Witt equation, the readers may refer to the beautiful series of paper in \cite{feng}.} of the bulk and the boundary terms from \eqref{jojamezz}. This gives us \cite{feng, wdwpaper}:

{\footnotesize
\bg\label{nobil2meye}
\Delta_{\rm weiss}\Big[{\bf S}_{\rm bulk}({\bf \hat\Xi}) + 
{\bf S}_{\rm bnd}({\bf \hat\Xi})\Big] = 
\int_{\mathbb{M}} d^{n +1} x \sqrt{-{\bf g}} ~ {\delta{\bf S}_{\rm tot}({\bf \hat\Xi})\over \delta {\bf \hat\Xi}}\cdot \delta {\bf \hat\Xi} + 
\int_{\del \mathbb{M}} \left(\Pi^{\rm P} \cdot \Delta{\bf \hat\Xi} - \mathbb{H}^{\rm P}\cdot \Delta {\rm x}\right) d\Sigma^{(n)}_{\rm P}, \nd}
for a manifold $\mathbb{M}$ of dimension $n + 1 \equiv 11$. Note that, as expected from the Weiss variation \cite{weiss, feng}, all the currents appear at the boundary. In fact the boundary has the following contributions.

\vskip.1in

\noindent $\bullet$ The full Lagrangian (with the boundary piece) for
${\bf \hat\Xi} = ({\bf \Xi}, {\bf \Upsilon})$. The difference is that, instead of integration over the full $n+1$ dimensional space-time, the integral appears with a  measure given by $\delta x \cdot {\rm d}\Sigma^{(n)}$ captured by the variation $\delta x^{\rm M}$.

\vskip.1in

\noindent $\bullet$ The contribution coming from the deformation of the boundary action ${\bf S}_{\rm bnd}({\bf \hat\Xi})$ itself from \eqref{jojamezz}. Note that this is in addition to the Lagrangian piece that we discussed above that also includes the boundary term.

\vskip.1in

\noindent $\bullet$ The contribution coming from a total-derivative piece that appears once we separate the bulk variation from 
${\delta{\bf S}_{\rm bulk}({\bf \hat\Xi})\over \delta {\bf \hat\Xi}}$. This is of course the usual piece designed in conjunction with the second contribution discussed above. 

\vskip.1in

\noindent All the aforementioned three contributions then conspire to give the conjugate momenta $\Pi^{\rm P}_{{\rm M}...{\rm N}}$ and the energy-momentum tensor $\mathbb{H}^{\rm P}_{{\rm M}}$ at the boundary $\partial \mathbb{M}$. Clearly the latter is related to the Noether current associated with the transformation $\Delta {\rm x}^{\rm M}$, whereas the former is related to the Noether currents associated with the variations of the fields $\Delta{\bf \hat\Xi}^{\rm M...N}$. (The difference between $\delta\zeta$ and $\Delta\zeta$ is explained in \cite{feng}.) For the simpler diagonal ADM choice $n^\mu=(1,{\bf 0})$, or equivalently
$N=1$ and $N^i=0$, that we study here, with
$x=({\bf x},t)$, the total Hamiltonian may be written schematically as:
\bg\label{doboudi}
\int d\Sigma^{(n)}\cdot
\mathbb H_{00}(\hat{\bf\Xi})
=
{\bf H}_{\rm tot}(\hat{\bf\Xi})
=
\int d^n{\bf x}\,
{\cal H}_{\perp}(\hat{\bf\Xi})
+
{\bf H}_{\partial\mathbb M},
\nd
where $n = 10$ (not to be confused with the ADM choice $n^\mu$), and with our simpler ADM choice the momentum constraint is simply $\mathcal H_i = 0$. Moreover, 
in this gauge the Hamiltonian constraint is proportional, up to
convention-dependent numerical factors, to:
\begin{equation}
{\cal H}_{\perp}
=
\sqrt{-{\bf g}}\,
{\delta{\bf S}_{\rm tot}(\hat{\bf\Xi})
\over
\delta{\bf g}^{00}} \ \simeq\
{\delta{\bf S}_{\rm tot}(\hat{\bf \Xi})\over \delta{\bf g}^{00}}  ,
\end{equation}
which could also be expressed in the ADM language (see \cite{csvsgs} for more details), and where the second relation should be understood as the leading-order
background-field approximation\footnote{The cleanest way to formulate this is to write
${\bf g}_{\mu\nu}
=
\eta_{\mu\nu}
+
\kappa\,h_{\mu\nu}$,
where $\eta_{\mu\nu}$ is the Minkowski metric,
$h_{\mu\nu}$ denotes the metric fluctuation, and $\kappa$ is introduced so
that $\kappa h_{\mu\nu}$ is dimensionless. For example, in four dimensions
one conventionally takes
$\kappa^2
=
32\pi G_N$. The determinant of the full metric may then be expanded systematically as:
\begin{equation}
\sqrt{-{\bf g}}
=
1
+
{\kappa\over2}h
+
{\kappa^2\over8}
\left(
h^2
-
2h_{\mu\nu}h^{\mu\nu}
\right)
+
{\cal O}(\kappa^3h^3), \nonumber
\label{eq:sqrtgexpansion}
\end{equation}
where
$h
\equiv
\eta^{\mu\nu}h_{\mu\nu}$,
and indices on the fluctuations are raised and lowered with
$\eta_{\mu\nu}$. Note that we have used
${\rm det}~ {\bf g}
=
\det\eta\,
\det\left(
\delta^\mu{}_\nu
+
\kappa h^\mu{}_\nu
\right)$,
together with
$\log\det(1+X)
=
{\rm Tr}\log(1+X)$ to expand the determinant factor above. }
$\sqrt{-{\bf g}}
=
1+{\cal O}(\kappa h)$.
Over a Minkowski minimum, we can replace $\sqrt{-{\bf g}} \approx 1$
 The boundary Hamiltonian ${\bf H}_{\partial\mathbb M}$ is the additional
surface term required to make the canonical generator functionally
differentiable under the chosen asymptotic boundary conditions. Equivalently,
it compensates the surface contribution generated when the bulk Hamiltonian
is varied. On the bulk constraint surface,
${\cal H}_{\perp}={\cal H}_i=0$, so the nontrivial physical Hamiltonian
reduces to ${\bf H}_{\partial\mathbb M}$. In fact this is also the Hamiltonian that appears in the Wheeler-De Witt equation. To see this, we can insert the Weiss variation of the path-integral \eqref{nobil2meye} in the definition of the wave-functional from \eqref{katnobi} to get:

{\scriptsize
\bg\label{sedomink2}
\begin{split}
{\rm d}{\bf \Psi}_{\rm tot}({\bf \Xi}_2, {\bf \Upsilon}_2; t) & = 
{\partial{\bf \Psi}_{\rm tot}({\bf \hat\Xi}; t)\over \partial {\bf \hat\Xi}} \cdot \Delta {\bf \hat\Xi}\Big\vert_{{\bf \hat\Xi}_2; t_2} +{\partial{\bf \Psi}_{\rm tot}({\bf \hat\Xi}; t)\over \partial {\rm x}} \cdot \Delta {\rm x}\Big\vert_{{\bf \hat\Xi}_2; t_2}\\
& = \int {\cal D}{\bf \Xi} ~{\cal D}{\bf \Upsilon} \int_{\del \mathbb{M}} \left(\Pi^{\rm P}_{{\bf \hat\Xi}} \cdot \Delta{\bf \hat\Xi} - \mathbb{H}^{\rm P}\cdot \Delta {\rm x}\right) d\Sigma^{(n)}_{\rm P}\Big\vert_{{\bf \hat\Xi}_2; t_2}~{\rm exp}\left(-i{\bf S}_{\rm tot}[{\bf \Xi, \Upsilon}]\right) {\bf \Psi}_{\rm tot}({\bf \Xi_1, \Upsilon_1}; t_1) {\cal D}{\bf \Xi}_1 {\cal D}{\bf \Upsilon}_1, 
\end{split}
\nd}
where note that ${\rm d}{\bf \Psi}_{\rm tot}({\bf \Xi}_2, {\bf \Upsilon}_2; t)$ is only proportional to the contributions from the boundary that involve, among other currents, the Hamiltonian from \eqref{doboudi}. Comparing the two lines from \eqref{sedomink2}, then leads to the following two conditions:
\bg\label{nestkathia}
\begin{split}
& {\bf H}_{\rm tot}({\bf \hat\Xi}) \vert{\bf \Psi}_{\rm tot}({\bf \hat\Xi}; t)\rangle = \left(\int d^n{\bf x}\sqrt{-{\bf g}}~{\delta {\bf S}_{\rm tot}({\bf \hat\Xi})\over \delta{\bf g}^{00}({\bf x})}  + {\bf H}_{\partial\mathbb M}\right)\vert {\bf \Psi}_{\rm tot}({\bf \hat\Xi}; t)\rangle = i {\partial \over \partial t}\vert
{\bf \Psi}_{\rm tot}({\bf \hat\Xi}; t)\rangle\nonumber\\
& i {\partial \over \partial t}\vert
{\bf \Psi}_{\rm tot}({\bf \hat\Xi})\rangle = 0 = \int d^n{\bf x}\sqrt{-{\bf g}}~{\delta {\bf S}_{\rm tot}({\bf \hat\Xi})\over \delta{\bf g}^{00}({\bf x})}\vert {\bf \Psi}_{\rm tot}({\bf \hat\Xi})\rangle, ~~~{\rm iff}~~~ {\bf H}_{\partial\mathbb M} = 0,
\end{split}
\nd
where we have now treated ${\bf H}_{\rm tot}({\bf \hat\Xi}), {\bf H}_{\partial\mathbb M}$ and ${\delta {\bf S}_{\rm tot}({\bf \hat\Xi})\over \delta{\bf g}^{00}(x)}$ as {\it operators}\footnote{\textcolor{blue}{We will not use hats to distinguish functions from operators to avoid confusion with hats appearing elsewhere and with different meanings. Which is which should be clear from the context}.}. We have also removed the explicit time dependence in the wave-functional in the second line so that ${\bf \Psi}_{\rm tot}({\bf \hat\Xi})$ only depends on $t$ implicitly, if any. The two conditions in \eqref{nestkathia} are already implemented at the level of the background Minkowski spacetime, so they would effect the dynamics in the warped Minkowski spacetime itself. In particular $-$ and in the absence of the boundary Hamiltonian $-$ since the Hamiltonian annihilates the wave-functional at every instant of time, the evolution operator becomes identity. Alternatively:
\bg\label{lindmorgue}
{\rm exp}\left(-i{\bf H}_{\rm tot}(\hat{\bf \Xi}) t\right) \vert{\bf \Psi}_{\rm tot}({\bf\hat\Xi}; t)\rangle 
 = {\rm exp}\left(-i{\bf H}_{\partial\mathbb M}(\hat{\bf \Xi}) t\right) \vert{\bf \Psi}_{\rm tot}({\bf\hat\Xi}; t)\rangle
 \xrightarrow{\;\;{\bf H}_{\partial\mathbb M} = 0\;\;} 1 , \nd
implying that states cannot be evolved using the Hamiltonian ${\bf H}_{\rm tot}({\bf \Xi, \Upsilon})$ if ${\bf H}_{\partial\mathbb M}= 0$, at least not in the conventional way we are accustomed to do so in the non-gravitational quantum field theories. As discussed above, evolution with respect to a boundary time $t$ can be given a meaningful interpretation only if one introduces an appropriate boundary Hamiltonian. This, however, crucially requires a non-vanishing ${\bf H}_{\partial\mathbb M}$. It is therefore natural to ask whether, in an AdS background, the Hamiltonian of the dual gauge theory can be used to define the propagation of bulk states.

Irrespective of whether such a construction works in AdS, the situation is different for a warped Minkowski background. In the absence of a known gauge-theory dual, the advantages provided by gauge/gravity duality in AdS cannot be straightforwardly carried over to the warped Minkowski case. In particular, if there is no boundary Hamiltonian, then there is no obvious analogue of the boundary-time evolution that one might try to define using the dual gauge-theory Hamiltonian.

\subsection{Glauber-Sudarshan states and the WdW equation \label{sec2.2}}

Our discussion in the previous section raises some severe conundrums. If the evolution operators all become identities in the absence of a boundary Hamiltonian, how can we ever define  correlation functions? We have successfully used correlation functions in non-gravitational field theories to study properties of elementary particles, so how can, on one hand things seem to work well but on the other hand, once gravitational degrees of freedom are considered, everything seems to go awry. Since no system can be isolated from gravity, how do we then justify the successes we had so far dealing with the study of quantum field theories? Unfortunately there are no simple resolutions or universal agreements on the answers to the aforementioned questions with vanishing boundary Hamiltonian. So here, instead of resolving these deep conundrums completely, we will take a slightly different route wherein more emphasis will be given on dynamics that do not involve the Hamiltonian directly. In the process we will see that there may be a way to define temporally evolving states even within the Hamiltonian constraint.

Before moving ahead let us make one thing clear. In the absence of the boundary Hamiltonian, the bulk Hamiltonian ${\bf H}_{\rm tot}({\bf \hat\Xi})$ vanishes {\it on-shell}, but does not have to vanish {\it off-shell} implying that the path-integral is still controlled by ${\bf S}_{\rm tot}({\bf \hat\Xi})$. This however doesn't mean that the coast is clear: ${\bf H}_{\rm tot}({\bf \hat\Xi})$ as an {\it operator} always annihilates\footnote{This is of course a familiar consequence of the underlying Schwinger-Dyson equation as elaborated by eq. (2.19) in \cite{wdwpaper}. See also \cite{feng}.} the states so \eqref{lindmorgue} continues to be applicable. 
On the other hand, we can express the second set of equations from \eqref{nestkathia} in the following suggestive way:

{\footnotesize
\bg\label{leylajoh}
{\bf H}_{\rm tot}({\bf \hat\Xi}) \vert{\bf \Psi}_{\rm tot}({\bf \hat\Xi})\rangle 
= \bigg({\bf H}_{\rm grav}({\bf \Xi})+ {\bf H}_{\rm matter}({\bf \Xi}) + 
{\bf H}_{\rm mixed}({\bf \Xi}) + {\bf H}_{\rm ghost}({\bf \hat\Xi}) + {\bf H}_{\rm gf} ({\bf \Xi})\bigg)\vert {\bf \Psi}_{\rm tot}({\bf \hat\Xi})\rangle = 0, 
\nd}
where as before ${\bf \hat\Xi} = ({\bf \Xi, \Upsilon})$. The distinction between ${\bf H}_{\rm grav}({\bf \Xi}), {\bf H}_{\rm matter}({\bf \Xi})$ and ${\bf H}_{\rm mixed}({\bf \Xi})$ should be clear directly from the pertubative sector in \eqref{qotom}: ${\bf H}_{\rm grav}({\bf \Xi})$ denotes all terms with $l_p = 0$ for $p \in \mathbb{Z}$ and $61 \le p \le 100$; ${\bf H}_{\rm matter}({\bf \Xi})$ denotes everything with $l_k = 0$ for $k \in \mathbb{Z}$ and $1 \le k \le 60$. The mixed pieces in 
${\bf H}_{\rm mixed}({\bf \Xi})$ come from $l_k \ne 0, l_p \ne 0$ and can be arranged to contain part of the boundary contributions. Similar distribution extends to the full trans-series form of the Hamiltonian. If we now define ${\bf H}({\bf \Xi}) = {\bf H}_{\rm grav}({\bf \Xi})+ {\bf H}_{\rm matter}({\bf \Xi}) + 
{\bf H}_{\rm mixed}({\bf \Xi})$, then:
\bg\label{oma4mey}
{\bf H}_{\rm tot}({\bf \hat\Xi}) = {\bf H}({\bf \Xi}) + {\bf H}_{\rm ghost}({\bf \hat\Xi})+ {\bf H}_{\rm gf} ({\bf \Xi}),
\nd
which as an {\it operator} annihilates the states, but neither of the three pieces, namely ${\bf H}({\bf \Xi})$, ${\bf H}_{\rm ghost}({\bf \hat\Xi})$ or ${\bf H}_{\rm gf} ({\bf \Xi})$, are required to individually annihilate the states. In terms of the wave-functional of the universe, we expect:
\bg\label{portia}
{\partial{\bf \Psi}_{\rm tot}({\bf \hat\Xi})\over \partial t}  = 
{\partial \over \partial t} \bigg({\bf \Psi}_{\rm g + matter}({\bf \Xi}; t)  {\bf \Psi}_{\rm ghost}({\bf \hat\Xi}; t)\bigg) = -i{\bf H}_{\partial\mathbb M} {\bf \Psi}_{\rm tot}({\bf \hat\Xi}) , \nd
which means that the wave-functional associated with gravity + matter can allow for an evolution with respect to the temporal coordinate $t$ if and only if ${\bf H}_{\partial\mathbb M} \ne 0$. For ${\bf H}_{\partial\mathbb M} = 0$,  one cannot use an identity of the following form:
\bg\label{coolig}
{\partial\over \partial t}~{\rm log}~ {\bf \Psi}_{\rm g + matter}({\bf \Xi}; t) = - {\partial\over \partial t}~{\rm log}~ {\bf \Psi}_{\rm ghost}({\bf \hat\Xi}; t),
\nd
to provide a meaning of time in this set-up as the coordinate $t$ ceases to have a simple interpretation. Naive conclusion like trading the temporal dependence with the ghost sector to  get some leverage doesn't really work. In particular, a decomposition of the following form:
\bg\label{coolig}
{\bf \Psi}_{\rm tot}({\bf \hat\Xi};t)
=
{\bf \Psi}_{\rm g+matter}({\bf \Xi};t)\,
{\bf \Psi}_{\rm ghost}({\bf \hat\Xi};t)
\nd
may certainly lead to compensating $t$-dependences between the two factors,
but such a cancellation should not be interpreted as the origin of physical
time evolution. The Faddeev--Popov ghost sector implements gauge fixing and
does not, by itself, provide a physical clock. A genuine notion of temporal evolution may instead be defined relationally,
for example by identifying a dynamical degree of freedom $T({\bf \Xi})$
that can serve as an internal clock and expressing the remaining observables
relative to $T({\bf \Xi})=\tau$. If the Hamiltonian constraint can be
deparametrized with respect to this clock, it takes schematically the form
$P_T+{\bf H}_{\rm rel}=0$, where $T_T$ is the conjugate momentum, and ${\bf H}_{\rm rel}$ is the relational
Hamiltonian governing the remaining physical degrees of freedom. Upon
quantization one then obtains:
\begin{equation}
i{\partial\over\partial\tau}
{\bf \Psi}_{\rm phys}(\tau)
=
{\bf H}_{\rm rel}(\tau)
{\bf \Psi}_{\rm phys}(\tau),
\end{equation}
even though ${\bf H}_{\partial\mathbb M}=0$.
The essential relation is therefore
${\bf H}_{\rm tot}=0$ implies
$P_T=-{\bf H}_{\rm rel}$,
after a successful deparametrization. The {\it evolution} generated by
${\bf H}_{\rm rel}$ is therefore evolution of the remaining degrees of
freedom \emph{with respect to the chosen internal clock}, and not evolution
with respect to an asymptotic boundary time. Even when the Wheeler--DeWitt constraint is second order in an internal
clock in the following way:
\bg\label{sunss}
\left(P_T^2+{\bf H}_{\rm rest}\right){\bf \Psi}_{\rm phys}(T)=0,\nd 
one need not select a
positive- or negative-frequency branch in order to obtain a first-order
evolution equation. By enlarging the wave-functional to the two-component
object
${\bf \Psi}_{D}=({\bf \Psi}_{\rm phys},i\partial_T{\bf\Psi}_{\rm phys})^{\rm T}$, the Wheeler--DeWitt equation can
be rewritten exactly as
$i\partial_T{\bf \Psi}_{D}={\bf H}_{D}{\bf \Psi}_{D}$.
Both frequency branches are then retained. The resulting
${\bf H}_{D}$ is a relational first-order evolution operator and should
not be confused with a boundary Hamiltonian:
${\bf H}_{\partial\mathbb M}$ may remain identically zero throughout.

The above discussion still does not by itself settle the issue of
correlation functions. The important point, however, is that the absence of
a boundary Hamiltonian does not imply the absence of correlation functions.
What changes is their interpretation. Coordinate-time correlators may be
introduced after gauge fixing, whereas physical correlators should
ultimately be expressed in terms of diffeomorphism-invariant or relational
observables. Since the path integral sums over complete histories and is
controlled by ${\bf S}_{\rm tot}$ rather than by a nonvanishing
${\bf H}_{\partial\mathbb M}$, it remains the natural framework for this
purpose.

One might think that \eqref{sabsuzu} and \eqref{dadari} should be expressed
using the operator form for ${\bf H}({\bf \Xi})$ given in
\eqref{oma4mey}, but in a scalar field theory where no Faddeev--Popov ghosts
appear, ${\bf H}({\bf \Xi})$ may vanish on-shell without preventing the
path-integral representation of correlation functions. When gravitational
degrees of freedom, and consequently gauge fixing and ghosts, are present,
the canonical representation becomes more subtle (see for example \cite{gupta}). It is therefore safer to
express ${\cal Z}_{12}$ in \eqref{sabsuzu} directly through a path integral
involving ${\bf S}_{\rm tot}$ rather than to infer its properties from an
ordinary evolution operator generated by
${\bf H}_{\rm tot}({\bf \hat\Xi})$.

\begin{figure}[h]
\centering
\begin{tabular}{c}
\includegraphics[width=3in]{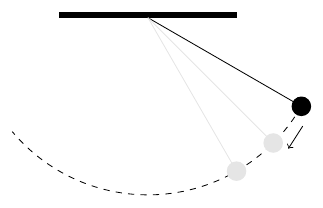}
\end{tabular}
\caption[]{Oscillation of a classical pendulum.}
\label{pendulum}
\end{figure}

Our discussion above then raises the following question: what if we want to
express the displacement operator for the Glauber--Sudarshan states using
the canonical approach? In
\cite{desitter2, coherbeta, coherbeta2, borel, joydeep}
we have always expressed the displacement operator
$\mathbb{D}(\sigma)$ using path-integral language, so the question is how to
proceed knowing that
${\bf H}_{\rm tot}({\bf \hat\Xi})$ annihilates physical states and that,
in the absence of a boundary Hamiltonian, there is no nontrivial physical
evolution operator of the form considered in \eqref{lindmorgue}.

A hint comes directly from the study of a classical pendulum using the
Hamiltonian equations of motion shown in {\bf figure \ref{pendulum}}. If we
displace the pendulum by a small angle $\theta$, then gravitational effects
will make it oscillate. The on-shell Hamiltonian equations governing the
dynamics are:
\bg\label{miavalent}
\dot{\theta}
=
{\partial{\bf H}\over\partial p_\theta},
\qquad
\dot{p}_\theta
=
-{\partial{\bf H}\over\partial\theta},
\nd
where $(\theta,p_\theta)$ are the standard conjugate variables. In quantum
mechanics, the equations \eqref{miavalent} appear from the path integral by
isolating the dominant path in the $\hbar\to0$ limit. If the canonical
Hamiltonian is constrained to vanish, one should not conclude that all
dynamics disappears; rather, the dynamics may instead be encoded in
relations among the dynamical variables. At the classical level these
relations continue to follow from the Lagrangian equation of motion
${\delta{\bf S}\over\delta\theta}=0$.
In the quantum theory the corresponding Schwinger--Dyson equation is:
\bg\label{reillyk}
\int{\cal D}\theta\,
e^{-i{\bf S}[\theta]}\,
{\delta{\bf S}\over\delta\theta}
\equiv
\left\langle
{\delta{\bf S}\over\delta\theta}
\right\rangle
=
0,
\nd
irrespective of whether we consider coherent states or not. What must be
specified separately, in a constrained theory, is which dynamical variable
is being used as the clock relative to which this motion is described. ({\bf Figure \ref{pendulum}} is intended only as a pedagogical illustration of this distinction and plays no
independent role in the subsequent derivation.)

In field theories, and especially those involving gravitational degrees of
freedom, the extension of this idea is more non-trivial. One may introduce
operators associated with a sequence of slices labelled by a parameter
$\lambda$:
\bg\label{mialotus}
\vert\varphi(\lambda_2)\rangle
=
\overleftarrow{\prod_{{\rm N}=0}^{{\lambda_2-\lambda_1\over\Delta\lambda}}}
{\cal O}(\lambda_1+{\rm N}\Delta\lambda)
\vert\varphi_a\rangle .
\nd
At this stage $\lambda$ is only an ordering or foliation parameter; it
should not yet be identified with physical time. In the limit
$\Delta\lambda\to0$, the ordered product becomes a path-ordered exponential.
No intermediate factor
$\exp(-i{\bf H}_{\partial\mathbb M}\Delta\lambda)$ appears because
${\bf H}_{\partial\mathbb M}=0$. Constraint-generated transformations may
still relate different embeddings, but these are gauge transformations and
should not be confused with physical evolution. If a relational clock
$T=\tau$ is subsequently introduced, then $\lambda$ may be replaced by the
physical relational parameter $\tau$, and the sequence of operators acquires
the interpretation of successive physical ``kicks.''

\begin{figure}[h]
\centering
\begin{tabular}{c}
\includegraphics[width=5in]{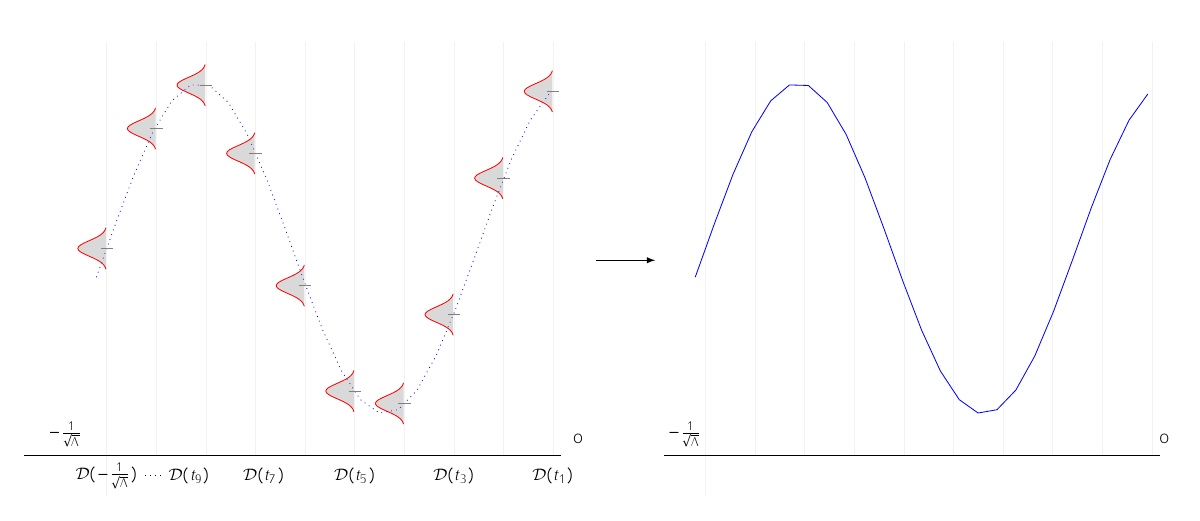}
\end{tabular}
\caption[]{On the left we have defined the displacement operators
${\cal D}(\sigma,\lambda)$ at successive values of a foliation parameter
$\lambda$ in configuration space. Their ordered product defines the
Glauber--Sudarshan state. Once an internal clock is chosen,
$\lambda$ may be replaced by a relational time $\tau$, in which case the
same construction admits the interpretation of a sequence of physical
displacements. On the right we show schematically the trajectory followed
by the dominant amplitudes of the resulting state.}
\label{gswdw}
\end{figure}

To proceed with our case, we will start by defining a displacement operator
on a given member $\Sigma_\lambda$ of the chosen foliation, integrated over
all spatial directions. In other words, to create a graviton
Glauber--Sudarshan state we define the following operator:

{\footnotesize
\bg\label{dwillem}
{\cal D}(\sigma,\lambda)\vert\Omega\rangle
=
{\rm exp}\left(
\int d^n{\bf x}\sqrt{-{\bf g}}\,
\sigma_{\rm MN}({\bf x},\lambda)
{\bf g}^{\rm MN}({\bf x},\lambda)
\right)\vert\Omega\rangle
=
{\rm exp}\left(
\int d^n{\bf k}\,
\widetilde{\sigma}_{\rm MN}({\bf k},\lambda)
\widetilde{\bf g}^{{\rm MN}}({\bf k},\lambda)
\right)\vert\Omega\rangle ,
\nd}
where $ \widetilde{\bf g}^{{\rm MN}}({\bf k},\lambda)$ and $\widetilde{\sigma}_{\rm MN}({\bf k}, \lambda)$ are related to the Fourier transforms of $ \sqrt{-{\bf g}({\bf x}, \lambda)}~{\bf g}^{\rm MN}({\bf x}, \lambda)$ and $\sigma_{\rm MN}({\bf x}, \lambda)$ respectively;  
and $\lambda$ should initially be regarded as the foliation parameter
rather than as an external physical time. The integral is well-defined
because the metric configuration operators commute at equal slice label.
Note that the definition \eqref{dwillem} has no $i$, while
$\sigma_{\rm MN}({\bf x},\lambda)$ or
$\widetilde\sigma_{\rm MN}({\bf k},\lambda)$ may be taken to be complex functions.
Thus ${\cal D}(\sigma,\lambda)$ is not unitary. We have taken $n=10$
because we want to study the Glauber--Sudarshan states in the low-energy EFT
of M-theory, which means
$({\rm M,N})\in{\bf R}^{2,1}\times{\cal M}_6\times{\mathbb T^2\over{\cal G}}$.
We have also absorbed the Jacobian factor in the definition of the metric
Fourier modes
$\widetilde{\bf g}^{\rm MN}({\bf k},k_0)$.
The Glauber--Sudarshan state is now defined using the ordered product of
${\cal D}(\sigma,\lambda)$:
\bg\label{deebrian}
\vert\sigma\rangle
=
\mathbb D(\sigma)\vert\Omega\rangle
=
\overleftarrow{\prod_{{\rm N}=0}^{|{\rm T}|/\Delta\lambda}}
{\cal D}(\sigma,{\rm N}\Delta\lambda)
\vert\Omega\rangle ,
\nd
where acting with $\mathbb D(\sigma)$ on a physical vacuum does not, by itself, guarantee that
the resulting state remains in the physical Hilbert space.  The allowed displacement
data are therefore restricted so that $\mathbb D(\sigma)$ preserves the gravitational
constraint surface.  Thus, if $|\Omega\rangle$ satisfy the Hamiltonian and the momentum constraints, then the resulting GS state must also satisfy the Hamiltonian and momentum
constraints\footnote{In a BRST formulation the corresponding condition is
${\bf Q}_{\rm BRST}|\sigma\rangle=0$ modulo BRST-exact states, with the appropriate Batalin-Vilkovisky
extension understood for the reducible gauge symmetries of supergravity. This will be discussed in section \ref{sec2.3}.}.
At this stage ${\rm T}$ denotes the range of the foliation parameter in \eqref{deebrian}. If the
chosen gauge is subsequently tied to a relational clock $\tau$, one may set
$\lambda=\tau$ and, for the flat-slicing configuration of interest, choose
the corresponding relational interval to be the following trans-Planckian domain: 
\begin{equation}
-{1\over\sqrt{\Lambda}}\leq\tau\leq0 ,
\end{equation}
motivated from \cite{desitter2,coherbeta,coherbeta2,borel,joydeep,wdwpaper}.
The infinite product for
$\Delta\lambda\to0$ and ${\rm N}\to\infty$ then gives the
Glauber--Sudarshan state. Before a clock is chosen this product should be
viewed as a single foliation-ordered state-preparation operator; after
deparametrization it may equivalently be interpreted as a continuous
sequence of kicks in relational time. Equivalently:
\begin{equation}
\mathbb D(\sigma)
=
{\cal P}
\exp\left[
\int d\lambda
\int_{\Sigma_\lambda}d^n{\bf x}\,
\sqrt{-{\bf g}}\,
\sigma_{\rm MN}({\bf x},\lambda)
{\bf g}^{\rm MN}({\bf x},\lambda)
\right],
\label{GSpathorder}
\end{equation}
where ${\cal P}$ denotes ordering with respect to the foliation parameter. This is almost what we had in \cite{desitter2,coherbeta,coherbeta2,borel,joydeep,wdwpaper}, except for one last step where we identify the clock.
After choosing a relational clock this becomes:
\begin{equation}
\mathbb D(\sigma)
=
{\cal T}_{\tau}
\exp\left[
\int d\tau
\int_{\Sigma_\tau}d^n{\bf x}\,
\sqrt{-{\bf g}}\,
\sigma_{\rm MN}({\bf x},\tau)
{\bf g}^{\rm MN}({\bf x},\tau)
\right],
\label{GSreltime}
\end{equation}
where ${\cal T}_{\tau}$ now has the physical meaning of relational-time
ordering. Using the Glauber--Sudarshan state one can construct the expectation value
of the metric operator by sandwiching it within the state
$\vert\sigma\rangle$. In the canonical formalism this may be expressed as:
\bg\label{mersirn}
\begin{split}
\langle{\bf g}_{\rm PQ}\rangle_\sigma
&=
{\langle\sigma\vert{\bf g}_{\rm PQ}\vert\sigma\rangle
\over
\langle\sigma\vert\sigma\rangle}
=
{\langle\Omega\vert
\mathbb D^\dagger(\sigma)
{\bf g}_{\rm PQ}
\mathbb D(\sigma)
\vert\Omega\rangle
\over
\langle\Omega\vert
\mathbb D^\dagger(\sigma)
\mathbb D(\sigma)
\vert\Omega\rangle}
\\
&=
{\langle\Omega\vert
{\mathbb P}\Big\{
\prod\limits_{{\rm N}=0}^{|{\rm T}|/\Delta\lambda}
{\cal D}^\dagger(\sigma,{\rm N}\Delta\lambda)\,
{\bf g}^{\rm PQ}({\bf x},\lambda_i)\,
\prod\limits_{{\rm N}'=0}^{|{\rm T}|/\Delta\lambda}
{\cal D}(\sigma,{\rm N}'\Delta\lambda)
\Big\}
\vert\Omega\rangle
\over
\langle\Omega\vert
{\mathbb P}\Big\{
\prod\limits_{{\rm N}=0}^{|{\rm T}|/\Delta\lambda}
{\cal D}^\dagger(\sigma,{\rm N}\Delta\lambda)
\prod\limits_{{\rm N}'=0}^{|{\rm T}|/\Delta\lambda}
{\cal D}(\sigma,{\rm N}'\Delta\lambda)
\Big\}
\vert\Omega\rangle}
\\
&=
{\langle\Omega\vert
{\mathbb P}\Big\{
{\bf g}^{\rm PQ}({\bf x},\lambda_i)
\prod\limits_{{\rm N}=0}^{|{\rm T}|/\Delta\lambda}
{\cal D}^\dagger(\sigma,{\rm N}\Delta\lambda)
{\cal D}(\sigma,{\rm N}\Delta\lambda)
\Big\}
\vert\Omega\rangle
\over
\langle\Omega\vert
{\mathbb P}\Big\{
\prod\limits_{{\rm N}=0}^{|{\rm T}|/\Delta\lambda}
{\cal D}^\dagger(\sigma,{\rm N}\Delta\lambda)
{\cal D}(\sigma,{\rm N}\Delta\lambda)
\Big\}
\vert\Omega\rangle},
\end{split}
\nd
where the denominator is defined without involving the source term, and 
$\mathbb P$ denotes ordering with respect to $\lambda$. Once a
relational clock is chosen, one may allow for the following replacements:
\begin{equation}
\lambda\longrightarrow\tau,
\qquad
\mathbb P\longrightarrow\mathbb T_\tau,
\end{equation}
and $\lambda_i$ becomes the relational time at which the metric operator is
inserted. In that representation the above expression takes exactly the
usual temporally ordered form, but the temporal ordering is now with respect
to an internal physical clock rather than an asymptotic boundary time. It is then useful to summarize the situation explicitly:

{\footnotesize
\begin{equation}
\begin{array}{|c|c|c|}
\hline
\textcolor{blue}{\bf data} &
\textcolor{blue}{{\bf H}_{\partial\mathbb M}\neq 0}
&
\textcolor{blue}{{\bf H}_{\partial\mathbb M}=0}
\\
\hline
\text{coordinate }t
&
\text{may be tied to boundary time}
&
\text{generally gauge/foliation label}
\\
\hline
\text{physical Hamiltonian}
&
{\bf H}_{\partial}
&
0\ \text{unless deparametrized}
\\
\hline
\hat {\bf G}(t)
&
{\rm U}_{\partial}^{\dagger}\hat {\bf G}(t_0){\rm U}_{\partial}
&
\text{not physical Heisenberg evolution}
\\
\hline
\mathcal D_{t_n}
&
\text{physical-time kick}
&
\text{foliation-labelled insertion}
\\
\hline
\prod\limits_n \mathcal D_{t_n}
&
\text{successive driven evolution}
&
\text{single ordered state-preparation operator}
\\
\hline
{\cal T}
&
\text{physical-time ordering}
&
\text{foliation/path ordering}
\\
\hline
e^{-i{\bf H}_{\partial}\Delta t}
&
\text{physical propagation}
&
\text{absent}
\\
\hline
\text{relational clock}
&
\text{optional}
&
\text{needed for physical temporal interpretation}
\\
\hline
\end{array} \nonumber
\end{equation}}

\vskip.1in

\noindent From the above table the important conceptual point is the following. When ${\bf H}_{\partial\mathbb M}=0$, the Glauber--Sudarshan displacement
operator can still be represented as an ordered product of slice-supported
operators. What must be changed is its interpretation. The slice label is
initially only a foliation parameter, and the complete product defines a
single spacetime-supported state-preparation operator. It cannot yet be
interpreted as a sequence of physical-time kicks. If a dynamical variable
can be used as an internal clock, the theory may be deparametrized and the
same ordered product can then be written in terms of relational time
$\tau$. In that representation the individual displacement operators do
acquire the meaning of successive kicks, with their temporal dependence
generated by the relational Hamiltonian ${\bf H}_{\rm rel}$ rather than by
a boundary Hamiltonian.

One might nevertheless worry that we have defined operators at the same
points on individual slices, so that there could be short-distance
singularities. As we will show in the next subsection by lifting
\eqref{mersirn} to the path integral, this can be treated in the usual
renormalized composite-operator sense. Note also that, since the
configuration operators commute on a given slice,
{\it i.e.}
$[{\bf g}_{\rm PQ}({\bf x},\lambda),
{\bf g}_{\rm PQ}({\bf y},\lambda)]=0$\footnote{Or more generically
$[{\bf \Xi}({\bf x},\lambda),
{\bf \Xi}({\bf y},\lambda)]_{\pm}=0$, where
${\bf \Xi}=
\{{\bf g}_{\rm PQ},{\bf C}_{\rm PQR},
{\bf \Psi}_{\rm P},\overline{\bf \Psi}_{\rm P}\}$
covers the on-shell degrees of freedom, and $\pm$ denotes a commutator or
anti-commutator depending on whether
${\bf \Xi}({\bf x},\lambda)$ is bosonic or fermionic.},
\bg\label{sirendee}
{\cal D}^\dagger(\sigma,{\rm N}\Delta\lambda)
{\cal D}(\sigma,{\rm N}\Delta\lambda)
=
{\cal D}(\sigma,{\rm N}\Delta\lambda)
{\cal D}^\dagger(\sigma,{\rm N}\Delta\lambda)
\neq1 ,
\nd
$\forall\,{\rm N}$, implying that the displacement operator
$\mathbb D(\sigma)$ is not unitary. The ordering between operators belonging
to distinct slices, however, remains essential. Before deparametrization
this is foliation ordering; after the introduction of an internal clock it
becomes relational-time ordering. The lift to the path integral, to which we
turn next, provides the natural covariant implementation of this structure.

\subsection{Glauber-Sudarshan states and the vertex operators \label{sec2.3}}

The difficulty we faced in interpreting the temporal evolution of the states,
as well as in giving meaning to the notion of time itself, suggests that a
boundary Hamiltonian provides the cleanest way of defining an external
physical time. This is not, of course, the only logical possibility, since
one may instead introduce a relational clock as discussed above. Fortunately,
our choice of a Minkowski, or warped-Minkowski, minimum naturally admits an
asymptotic boundary Hamiltonian. There is, however, a subtlety here.
Minkowski spacetime itself has no ordinary manifold boundary:
$\partial{\cal M}_{1,d-1}=\varnothing$. Nevertheless, it is noncompact and
has well-defined asymptotic regions. Conformal compactification maps these
regions at infinite physical distance to finite coordinate locations and
adjoins them to the spacetime as
$\partial\widetilde{\cal M}_{1,d-1}$. Thus the ``asymptotic boundary'' is
not an ordinary boundary of Minkowski spacetime; it is the boundary of its
conformal completion. More precisely, it is the asymptotic fall-off
structure represented by this completion that allows one to define
quantities such as the ADM Hamiltonian, or more generally the boundary
Hamiltonian, and the corresponding asymptotic charges\footnote{Unfortunately
this idea cannot be carried over to a de Sitter, or more general accelerating,
spacetime in the same Hamiltonian sense. There is of course no obstruction
to conformally compactifying de Sitter or a more general accelerating
spacetime, and such spacetimes may indeed possess conformal boundaries.
What fails, in general, is the identification of that conformal boundary
with a spatial boundary carrying a globally defined Hamiltonian analogous
to the ADM Hamiltonian of asymptotically flat Minkowski space. In global de
Sitter the spatial slices are compact and have no boundary, while
${\cal I}^{\pm}$ are spacelike boundaries at the past and future ends of
the complete spacetime. Moreover, de Sitter possesses no globally timelike
Killing vector. Hence the existence of a conformal boundary does not by
itself provide the nonzero boundary Hamiltonian required to generate
ordinary global time evolution. This is precisely the advantage, for the
present construction, of starting with a Minkowski, or warped-Minkowski,
spacetime and then constructing excited states over it.}.

The lift of \eqref{mersirn} to a path integral is slightly more
non-trivial. Comparing the third line of \eqref{mersirn} with \eqref{cochi},
we see an immediate similarity, but this is not enough to lift it to a path
integral. The problem is related to the first line of \eqref{cochi}: we
need to express $\vert\Omega\rangle$ in terms of local Minkowski minima.
For this, however, we would need an evolution operator. If only the bulk
constraint Hamiltonian is retained, its action on physical states is
trivial, and the above line of thinking leads to a dead end. Once the
boundary Hamiltonian
${\bf H}_{\partial\mathbb M}({\bf \Xi,\Upsilon})$ is included, the situation
changes. The canonical Hamiltonian may be written schematically as
\begin{equation}
{\bf H}_{\rm tot}({\bf \Xi, \Upsilon)}
=
{\bf H}_{\rm bulk}({\bf \Xi, \Upsilon})
+
{\bf H}_{\partial\mathbb M}({\bf \Xi, \Upsilon}),
\end{equation}
where ${\bf H}_{\rm bulk}({\bf \Xi, \Upsilon})$ is the bulk part that is essentially given by \eqref{oma4mey} and contains gravity, matter, ghosts as well as the gauge-fixing term. Moreover, 
the bulk part consists of the Hamiltonian and momentum constraints
and therefore annihilates physical states:
\bg\label{tagchurch}
{\bf H}_{\rm bulk}
\vert\Psi_{\rm phys}\rangle
=
0 ,
\nd
thus leading to all the issues that we discussed in the previous section. In the absence of the boundary term this leads to the Wheeler-DeWitt (WdW) equation and, from the discussion above, we see that ghosts and gauge-fixing terms do contribute to the WdW equation.
Consequently, on the physical Hilbert space the nontrivial generator reduces
to the boundary Hamiltonian:
\bg
{\bf H}_{\rm tot}
\vert\Psi_{\rm phys}\rangle
=
{\bf H}_{\partial\mathbb M}
\vert\Psi_{\rm phys}\rangle,
\nd
and there is no requirement that
${\bf H}_{\partial\mathbb M}$ itself annihilate physical states.
In fact, the gauge-fixed representative
${\bf H}_{\partial\mathbb M}({\bf \Xi,\Upsilon})$ may depend on both the
physical fields ${\bf \Xi}$ and the corresponding Faddeev--Popov sector
${\bf \Upsilon}$\footnote{The physical boundary charge itself is represented
by a BRST cohomology class and is therefore independent of the choice of
gauge, even though a particular gauge-fixed representative may depend
explicitly on ghosts. See \cite{csvsgs} for more details.}. The projection
onto the interacting ground state may therefore be written, more cleanly,
using the physical boundary Hamiltonian:
\bg\label{streeshrad}
\vert\Omega\rangle
\propto
{1\over c_1}
\lim\limits_{{\rm T}\to\infty(1-i\epsilon)}
{\rm exp}\left[
-i\left(
{\bf H}^{\rm phys}_{\partial\mathbb M}({\bf \Xi})
-
{\rm E}_0
\right){\rm T}
\right]
\vert0\rangle_{\rm min1},
\nd
where the {\it physical} boundary Hamiltonian only depends on the on-shell field ${\bf \Xi}$ that we took above and not on the Faddeev-Popov (FP) ghosts. This will be elaborated below. Additionally, 
${\rm E}_0$ is the lowest physical eigenvalue of the boundary
Hamiltonian, {\it i.e.}:
\begin{equation}
{\bf H}^{\rm phys}_{\partial\mathbb M}
\vert\Psi^0_{\rm phys}\rangle
=
{\rm E}_0
\vert\Psi^0_{\rm phys}\rangle,
\end{equation}
$\vert0\rangle_{\rm min1}$ is {\it one} of the local Minkowski minima and
$(c_1,c_{1{\rm N}})$ are the coefficients replacing $(c_a,c_{a{\rm N}})$
in \eqref{molshnnon}\footnote{The full identification of the interacting
vacuum with the free harmonic vacuum is a bit more subtle, as shown in
\cite{csvsgs}. For the present treatment, \eqref{streeshrad} suffices. We
will also study the case with ${\rm T}=-\vert{\rm T}\vert$ soon.}. One may
equivalently use a gauge-fixed representative of
${\bf H}_{\partial\mathbb M}$, provided all statements are understood in
BRST cohomology.

Interestingly, the decomposition of
${\bf H}_{\partial\mathbb M}({\bf \hat\Xi})$ in terms of gravitational,
matter, mixed, ghost and gauge-fixing contributions does not require the
ghosts to remain explicitly coupled in every gauge. In fact, what we expect
schematically is:
\bg\label{graciejool}
\begin{split}
{\bf H}_{\partial\mathbb M}({\bf \Xi,\Upsilon})
&=
{\bf H}_{\partial\mathbb M}^{\rm grav}({\bf \Xi})
+
{\bf H}_{\partial\mathbb M}^{\rm matter}({\bf \Xi})
+
{\bf H}_{\partial\mathbb M}^{\rm mixed}({\bf \Xi})
+
{\bf H}_{\partial\mathbb M}^{\rm ghost}({\bf \Xi,\Upsilon})
+
{\bf H}_{\partial\mathbb M}^{\rm gf}({\bf \Xi})
\\
& = {\bf H}_{\partial\mathbb M}^{\rm grav,phys} + {\bf H}_{\partial\mathbb M}^{\rm matter,phys} + ... + 
\delta_{\rm BRST}\left({\bf K}_{\rm grav} + {\bf K}_{\rm matter} + ...\right)
\\
&=
{\bf H}_{\partial\mathbb M}^{\rm phys}
+
\delta_{\rm BRST}{\bf K}_{\partial} = 
{\bf H}_{\partial\mathbb M}^{\rm phys}
\left(
{\bf \Xi}
\right)
+
\Big[
{\bf Q}_{\rm BRST},
{\bf K}_{\partial}
\left(
{\bf \Xi},
{\bf \Upsilon}
\right)
\Big]_+  ,
\end{split}
\nd
where the first line is the decomposition of the gauge-fixed boundary
Hamiltonian according to its gravitational, matter, mixed, ghost and
gauge-fixing contributions, while the second and the third lines are their corresponding
BRST decompositions. ${\bf Q}_{\rm BRST}$ is the BRST charge and
${\bf K}_{\partial}$ is a Grassmann-odd boundary operator whose precise
form depends on the gauge fixing and on the boundary conditions. In the
third line, the first term represents the nontrivial BRST cohomology class
of the physical boundary charge, while the second collects gauge-dependent
BRST-exact contributions. The latter need not coincide solely with the
explicit ghost and gauge-fixing terms; pieces from the gravitational,
matter and mixed sectors may participate in the BRST-exact combination as
well, as clear from the second line. As usual, this statement assumes that the relevant BRST
transformations are proper gauge transformations rather than asymptotic
symmetries carrying independent boundary charges.

Even if the ghosts decouple in a particular gauge, the first decomposition
in \eqref{graciejool} may still be expressed in terms of the remaining
gravitational, matter, mixed and gauge-fixing contributions, while its
physical content continues to be captured by
${\bf H}^{\rm phys}_{\partial\mathbb M}$. Thus
${\bf H}_{\partial\mathbb M}({\bf \Xi,\Upsilon})$ is not required to
annihilate physical states and may be nonzero on-shell. What does that buy
us in terms of the constraint on the wave-functional of the universe?

The crucial point is that the Wheeler--DeWitt constraint and the boundary
Schr\"odinger equation now play different roles. The physical state must
continue to satisfy the bulk gravitational constraints, {\it i.e.} \eqref{tagchurch}
while its evolution with respect to the asymptotic boundary time is
generated by the physical boundary Hamiltonian:
\bg\label{dixlinema}
{\bf H}^{\rm phys}_{\partial\mathbb M}({\bf \Xi})
\left\vert
\Psi_{\rm g+mat+gf}({\bf \Xi},t)
\right\rangle
=
i{\partial\over\partial t}
\left\vert
\Psi_{\rm g+mat+gf}({\bf \Xi},t)
\right\rangle .
\nd
Thus the Wheeler--DeWitt equation does not imply the absence of temporal
evolution once a nontrivial boundary Hamiltonian is present: the bulk
constraint annihilates the physical state, while the boundary charge
generates its physical time dependence.

If instead one works with a particular gauge-fixed representative of the
boundary Hamiltonian, the representative of the state may differ from the
one obtained using
${\bf H}^{\rm phys}_{\partial\mathbb M}$ by a BRST-exact contribution.
Schematically one may write
\bg\label{cartsteve}
\Delta{\bf \Psi}({\bf \Xi,\Upsilon},t)
=
{\bf \Psi}_{\rm gf}({\bf \Xi,\Upsilon},t)
-
{\bf \Psi}_{\rm phys}({\bf \Xi},t)
=
{\bf Q}_{\rm BRST}
{\bf \Phi}({\bf \Xi,\Upsilon},t),
\nd
provided the two states are representatives of the same BRST cohomology
class. Equation \eqref{cartsteve} should therefore be understood as an
equivalence between representatives rather than as a physical deformation
of the state. The physical wave-functional continues to satisfy the
Wheeler--DeWitt and momentum constraints together with the boundary
Schr\"odinger equation \eqref{dixlinema}. Coupling the ghost sector can
change the gauge-fixed representative, but not the corresponding physical
BRST cohomology class. The question is whether these observations provide
some leverage on the path-integral formalism from \eqref{mersirn}, namely,
whether they provide a way to uplift \eqref{mersirn} to a path integral.

To discuss the uplift we will study the numerator and the denominator in
the last line of \eqref{mersirn} separately. Since
${\bf H}^{\rm phys}_{\partial\mathbb M}$ does not annihilate physical
states, we can use it to express the interacting vacuum
$\vert\Omega\rangle$ in terms of the local Minkowski vacuum
$\vert0\rangle_{\rm min1}$ as in \eqref{streeshrad}. Plugging this in
\eqref{mersirn} gives us:

{\scriptsize
\bg\label{gshalom}
\langle{\bf g}_{\rm PQ}\rangle_\sigma
=
\lim_{\vert{\rm T}\vert\to\infty(1+i\epsilon)}
{{}_{\rm min1}\langle0\vert
{\rm exp}\left[
i
\widetilde{\bf H}^{\rm phys}_{\partial\mathbb M}({\bf \Xi})
\vert{\rm T}\vert
\right]
\mathbb{T}\Bigg\{
{\bf g}^{\rm PQ}({\bf x},t_i)
\prod\limits_{{\rm N}=0}^{\vert{\rm T}\vert/\Delta t}
{\cal D}^{\dagger}(\sigma,{\rm N}\Delta t)
{\cal D}(\sigma,{\rm N}\Delta t)
\Bigg\}
{\rm exp}\left[
i
\widetilde{\bf H}^{\rm phys}_{\partial\mathbb M}({\bf \Xi})
\vert{\rm T}'\vert
\right]
\vert0\rangle_{\rm min1}
\over
{}_{\rm min1}\langle0\vert
{\rm exp}\left[
i
\widetilde{\bf H}^{\rm phys}_{\partial\mathbb M}({\bf \Xi})
\vert{\rm T}\vert
\right]
\mathbb{T}\Bigg\{
\prod\limits_{{\rm N}=0}^{\vert{\rm T}\vert/\Delta t}
{\cal D}^{\dagger}(\sigma,{\rm N}\Delta t)
{\cal D}(\sigma,{\rm N}\Delta t)
\Bigg\}
{\rm exp}\left[
i
\widetilde{\bf H}^{\rm phys}_{\partial\mathbb M}({\bf \Xi})
\vert{\rm T}'\vert
\right]
\vert0\rangle_{\rm min1}},
\nd}
where $ \widetilde{\bf H}^{\rm phys}_{\partial\mathbb M}({\bf \Xi}) \equiv {\bf H}^{\rm phys}_{\partial\mathbb M}({\bf \Xi}) -{\rm E}_0$; ${\rm T}=-\vert{\rm T}\vert$ and
$\vert{\rm T}'\vert=\vert{\rm T}\vert+{1\over\sqrt{\Lambda}}$ for the
temporal domain
$-{1\over\sqrt{\Lambda}}\leq t\leq0$. The precise $i\epsilon$
prescription on the two asymptotic legs is understood to be the one
appropriate to the vacuum projection. In this form, \eqref{gshalom}
resembles \eqref{sabsuzu} but without the overlap factor
${\cal A}_{\rm over}$ and with
${\bf H}^{\rm phys}_{\partial\mathbb M}({\bf \Xi})$ replacing the bulk
constraint Hamiltonian. The former cancels between the numerator and the
denominator, while the latter provides the physical generator needed to
define the correlation function over a local Minkowski minimum without
violating the underlying Wheeler--DeWitt constraints.

The above form for the correlation function in \eqref{gshalom} is still
incomplete because it hides the ordering issue that appears once we go to
the Lorentzian setting. This problem has recently been analyzed in
\cite{csvsgs} by resorting to Schwinger--Keldysh contours. As shown there,
the issue of the reality of the metric expectation value can be treated
cleanly using the in-in formalism. Interested readers may look up
\cite{csvsgs} for more details. Fortunately, many of these ordering
subtleties are less explicit in the Euclidean formalism, and therefore here
we shall work primarily in that setting.

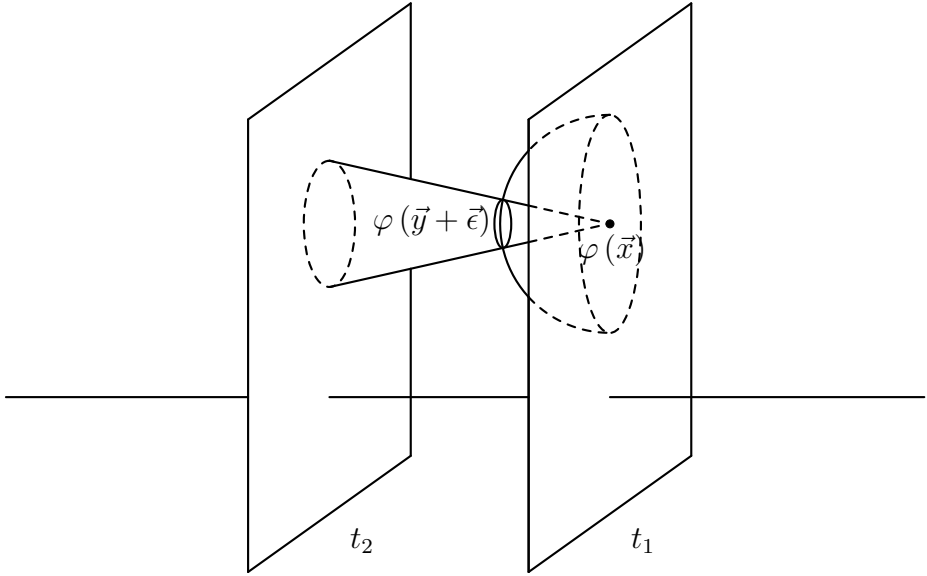
\begin{figure}
    \centering
    \begin{tikzpicture}[line cap=round,line join=round,>=triangle 45,x=1cm,y=1cm]
        \clip(-6.525716612895435,-2.7370231052405583) rectangle (6.6790182064531765,6.325962290594211);
        \draw [thick] (-2.897209302325581,3.6813953488372095)-- (-2.897209302325581,-2.3186046511627905);
        \draw [thick] (-2.897209302325581,3.6813953488372095)-- (-0.75721,5.2214);
        \draw [thick] (-2.897209302325581,-2.3186046511627905)-- (-0.75721,-0.7786);
        \draw [thick] (0.7939708581714067,3.6813953488372095)-- (2.9339701604969877,5.2214);
        \draw [thick] (2.9339701604969877,5.2214)-- (2.9339701604969872,-0.7786);
        \draw [thick] (0.7939708581714067,3.6813953488372095)-- (0.7939708581714067,-2.3186046511627905);
        \draw [thick] (0.7939708581714067,-2.3186046511627905)-- (2.9339701604969872,-0.7786);
        \draw [rotate around={90:(-1.8272096511627909,2.2971878931220746)},thick,dashed] (-1.8272096511627909,2.2971878931220746) ellipse (0.836604365070054cm and 0.3372701988144682cm);
        \draw [rotate around={90:(1.8639705093341972,2.2971878931220844)},thick,dashed] (1.8639705093341972,2.2971878931220844) ellipse (1.4451761618270786cm and 0.40763005876631575cm);
        \draw [thick] (-1.8272096511627904,0)-- (0.7939708581714067,0);
        \draw [thick] (-2.8972093023255816,0)-- (-6.083660247060193,0);
        \draw [thick] (1.8639705093341974,0)-- (6,0);
        \draw [rotate around={90:(0.4545421879804282,2.2971878931220724)},thick] (0.4545421879804282,2.2971878931220724) ellipse (0.31944631110585825cm and 0.11070816063653462cm);
        \draw [thick] (-0.75721,5.2214)-- (-0.75721,2.891277306370949);
        \draw [thick] (-0.75721,1.7030984798732018)-- (-0.75721,-0.7786);
        \draw [thick] (0.7939708581714067,3.6813953488372095)-- (0.7939708581714067,3.2685966048519886);
        \draw [thick] (0.7939708581714067,1.3257791813921798)-- (0.7939708581714067,-2.3186046511627905);
        \draw [shift={(1.8639705093341972,2.2971878931220777)},thick,dashed] plot[domain=1.5707963267948966:2.4044525433942505,variable=\t]({1*1.4451761618270846*cos(\t r)},{1*1.4451761618270846*sin(\t r)});
        \draw [shift={(1.8639705093341972,2.2971878931220777)},thick,dashed] plot[domain=3.878732763785329:4.71238898038469,variable=\t]({1*1.4451761618270742*cos(\t r)},{1*1.4451761618270742*sin(\t r)});
        \draw [shift={(1.8639705093341972,2.2971878931220777)},thick] plot[domain=2.4044525433942505:3.878732763785329,variable=\t]({1*1.4451761618270826*cos(\t r)},{1*1.4451761618270826*sin(\t r)});
        \draw [thick] (-1.8266610435225243,3.133791151420077)-- (0.7939708581714068,2.539702844943258);
        \draw [thick] (-1.8189777123374735,1.4608327596814656)-- (0.7939708581714068,2.0546729413008955);
        \draw [thick,dashed] (0.7939708581714068,2.539702844943258)-- (1.8639705093341972,2.2971878931220777);
        \draw [thick,dashed] (0.7939708581714068,2.0546729413008955)-- (1.8527864650603076,2.2997227518533867);
        \begin{scriptsize}
            \draw [fill=black] (1.8639705093341972,2.2971878931220777) circle (1.5pt);
        \end{scriptsize}
        \draw (1.3,2.3) node[anchor=north west] {$\varphi\left(\vec{x}\right)$};
        \draw (-1.4,2.6843618974650907) node[anchor=north west] {$\varphi\left(\vec{y}+\vec{\epsilon}\right)$};
        \draw (2,-1.6) node[anchor=north west] {$t_1$};
        \draw (-1.7,-1.6) node[anchor=north west] {$t_2$};
    \end{tikzpicture}
    \caption{Fields on two Cauchy slices within and outside the light cone of the first field.}
    \label{causality}
\end{figure}

To perform the path-integral uplift of \eqref{gshalom}, we will first have
to understand how the product of operators
$\prod\limits_{{\rm N}=0}^{{\rm T}/\Delta t}
{\cal D}^{\dagger}(\sigma,{\rm N}\Delta t)
{\cal D}(\sigma,{\rm N}\Delta t)$
appears in the path integral. There are two points to consider.
\textcolor{blue}{One}, the fundamental configuration operators commute on a
given Cauchy slice, subject to the usual qualifications associated with
constraints and gauge fixing; and \textcolor{blue}{two}, local observables
satisfy microcausality and therefore commute at spacelike separation. For
the gauge-fixed metric itself this statement should be understood within
the chosen perturbative gauge, since strictly local metric components are
not by themselves diffeomorphism-invariant observables.

The latter point nevertheless leads to nontrivial issues because the
operator \eqref{dwillem} switches on an infinite set of field operators on a
given Cauchy slice whose configuration-space picture is shown at every
instant in {\bf figure \ref{gswdw}}. Thus, for any given field operator on
a Cauchy slice, we have to examine its local neighborhood and its relation
to neighboring slices. In a sufficiently small normal neighborhood one may,
for example, introduce Riemann normal coordinates about $x^\rho$ and write
the local expansion

{\scriptsize
\bg\label{lucytrue}
{\bf g}_{\rm PQ}(x^\rho+\epsilon^\rho)
{\bf g}_{\rm RS}(y^\rho)
=
{\bf g}_{\rm PQ}(x^\rho)
{\bf g}_{\rm RS}(y^\rho)
-
{1\over3}
{\bf R}_{\rm PMQN}(x^\rho)
{\bf g}_{\rm RS}(y^\rho)
\epsilon^{\rm M}\epsilon^{\rm N}
-
{1\over6}
{\bf D}_{\rm U}{\bf R}_{\rm PMQN}(x^\rho)
{\bf g}_{\rm RS}(y^\rho)
\epsilon^{\rm U}\epsilon^{\rm M}\epsilon^{\rm N}
+
{\cal O}((\epsilon^\sigma)^4),
\nd}
where ${\bf D}_{\rm U}$ is the covariant derivative with respect to the
metric connection, $x^\rho$ is the full eleven-dimensional vector, and
$\epsilon^0$ measures the separation between neighboring slices. Equation
\eqref{lucytrue} is to be understood as a local normal-coordinate expansion;
it does not by itself replace the causal or operator-ordering information
contained in the unequal-slice products.

For free or Gaussian fundamental fields, unequal-time commutators are
$c$-number distributions. In a generic interacting theory, and especially
for nonlinear composite operators such as the renormalized metric
combinations entering the displacement operator, the corresponding
commutators need not be $c$-numbers. Thus, only in the sectors for which
the commutator closes to a $c$-number may one write schematically
\bg\label{asacrific}
\Big[
{\bf \Xi}_1({\bf x}_1,t_1),
\sum_j{\bf \Xi}_j({\bf y}_j,t_j)
\Big]_\pm
=
c(x_1,\{y_j\}),
\nd
where $c(x_1,\{y_j\})$ is a $c$-number distribution with tensor and spinor
indices properly contracted. In the interacting theory the right-hand side
may instead be operator-valued. In that case one must retain the complete
Baker--Campbell--Hausdorff, or equivalently the time-ordered Dyson,
structure when computing, for example,
${\cal D}(\sigma,t_1){\cal D}(\sigma,t_2)$. This is a complicated process,
so as a first step we will suppress these operator-ordering subtleties and
express \eqref{gshalom} as:
\bg\label{gilsalaam}
\langle{\bf g}_{\rm PQ}(y)\rangle_\sigma
=
{\int{\cal D}{\bf \Xi}\,
{\rm exp}\Big(-{\bf S}[{\bf \Xi}]\Big)
{\rm exp}\Big(
2\int d^{n+1}x\sqrt{-{\bf g}}\,
{\bf Re}\left(
\sigma_{\rm MN}{\bf g}^{\rm MN}
\right)
\Big)
{\bf g}_{\rm PQ}(y)
\over
\int{\cal D}{\bf \Xi}\,
{\rm exp}\Big(-{\bf S}[{\bf \Xi}]\Big)
{\rm exp}\Big(
2\int d^{n+1}x\sqrt{-{\bf g}}\,
{\bf Re}\left(
\sigma_{\rm MN}{\bf g}^{\rm MN}
\right)
\Big)},
\nd
where we have used Euclidean space to represent the path integral and
$d^{n+1}x\equiv dt\,d^n{\bf x}$ with $n=10$, implying integration over the
full eleven-dimensional spacetime. A more explicit treatment, in which the
operator ordering is handled carefully, has recently been carried out in
\cite{csvsgs} for both the Lorentzian and Euclidean formulations. In the
Lorentzian case, the natural description is provided by the in-in
Schwinger--Keldysh contour, with doubled fields on the forward and backward
branches and the GS displacement operators represented as contour
insertions. The Euclidean formulation admits a corresponding description in
terms of the appropriate thimble, or Picard--Lefschetz, decomposition.
Together these constructions provide a systematic path-integral realization
of the operator expressions discussed above. The readers may refer to
section 6 of \cite{csvsgs} for the detailed derivations.

As expected from the canonical expression \eqref{gshalom}, the
Faddeev--Popov sector is not displayed explicitly in
\eqref{gilsalaam}, and the path integral as it stands is therefore not yet
the complete gauge-fixed gravitational path integral. The appropriate
completion is not to insert ghost fields independently, but rather to
restore the full gauge-fixed Faddeev--Popov, or equivalently BRST,
measure. This converts \eqref{gilsalaam} to:

{\scriptsize
\bg\label{gilsalaam2}
\langle{\bf g}_{\rm PQ}(y)\rangle_\sigma
=
{\int{\cal D}{\bf \hat\Xi}\,
{\rm exp}\Big(-{\bf S}_{\rm tot}[{\bf \hat\Xi}]\Big)
{\rm exp}\Big(
2\int d^{n+1}k\,
\left(
{\bf Re}\,\widetilde\sigma_{\rm MN}(k)\,
{\bf Re}\,\widetilde{\bf g}^{\rm MN}(k)
+
{\bf Im}\,\widetilde\sigma_{\rm MN}(k)\,
{\bf Im}\,\widetilde{\bf g}^{\rm MN}(k)
\right)
\Big)
{\bf g}_{\rm PQ}(y)
\over
\int{\cal D}{\bf \hat\Xi}\,
{\rm exp}\Big(-{\bf S}_{\rm tot}[{\bf \hat\Xi}]\Big)
{\rm exp}\Big(
2\int d^{n+1}k\,
\left(
{\bf Re}\,\widetilde\sigma_{\rm MN}(k)\,
{\bf Re}\,\widetilde{\bf g}^{\rm MN}(k)
+
{\bf Im}\,\widetilde\sigma_{\rm MN}(k)\,
{\bf Im}\,\widetilde{\bf g}^{\rm MN}(k)
\right)
\Big)},
\nd}
where the integral is over the full $n+1$ dimensions and
${\bf \hat\Xi}=({\bf \Xi,\Upsilon})$, with ${\bf \Upsilon}$ denoting the
Faddeev--Popov sector. For $n=10$ we are studying the eleven-dimensional
low-energy limit of M-theory, while for $n<10$ the corresponding expression
describes its dimensional reduction to $n+1$ dimensions. We have also
defined the total gauge-fixed action
${\bf S}_{\rm tot}({\bf \hat\Xi})$ as
\bg\label{gilartg}
{\bf S}_{\rm tot}({\bf \hat\Xi})
\equiv
{\bf S}_{\rm tot}({\bf \Xi,\Upsilon})
=
{\bf S}({\bf \Xi})
+
{\bf S}_{\rm gf}({\bf \Xi})
+
{\bf S}_{\rm ghost}({\bf \hat\Xi}),
\nd
as given in \cite{wdwpaper} and also in our earlier works
\cite{coherbeta, coherbeta2, borel, joydeep}. This form of the expectation value matches what we had in
\cite{coherbeta, coherbeta2, borel, joydeep}, where the study was performed
directly from the path-integral point of view without invoking the canonical
analysis presented here. As we saw above, the canonical analysis requires
special care because the bulk Hamiltonian is a constraint, whereas the
nontrivial temporal evolution is generated by the boundary Hamiltonian. The
path-integral representation provides a particularly natural description
because it involves the full action
${\bf S}_{\rm tot}({\bf \hat\Xi})$, which does not vanish, while the
canonical bulk Hamiltonian continues to annihilate physical states. In other words, the path-integral approach\footnote{In \eqref{gilsalaam2} we have used the
Euclidean formalism. Since
$\mathbb{D}^{\dagger}(\sigma)\mathbb{D}(\sigma)$ is real for the source
choice under consideration, the reality properties of
$\langle{\bf g}_{\rm PQ}\rangle_\sigma$ are considerably more transparent
in the Euclidean case. In a Lorentzian formulation one instead has to use
the appropriate in-in prescription, or equivalently the
Schwinger--Keldysh contour, in order to establish the corresponding reality
condition. See \cite{csvsgs} for details.} involves the total action
${\bf S}_{\rm tot}({\bf \hat\Xi})$, whereas the canonical description
separates the bulk constraints from the nonvanishing physical boundary
generator.

One might nevertheless worry that the integral representation of the
displacement operators contains an integral over the full $n+1$
dimensional spacetime, and hence it might appear that the system is
``pushed'' continuously by an externally prescribed force. This is not the
interpretation intended here, as mentioned earlier and depicted in
{\bf figure \ref{gswdw}}. The source entering the GS operator is ultimately
constrained by the self-consistency conditions discussed below rather than
being an arbitrary external drive. This is beautifully captured by a {\it bootstrapping} procedure as shown recently in section 7 of \cite{csvsgs}.

Moreover, the form of
$\mathbb{D}^{\dagger}(\sigma)\mathbb{D}(\sigma)$ appearing in
\eqref{gilsalaam2} should be familiar. For example, putting $n=1$, we see
that the product of the displacement operators takes the following form:
\bg\label{solocke}
\mathbb{D}^{\dagger}(\sigma)\mathbb{D}(\sigma)
=
{\rm exp}\Big(
2\int d^2k\,
\left(
{\bf Re}\,\widetilde\sigma_{\rm MN}(k)\,
{\bf Re}\,\widetilde{\bf g}^{\rm MN}(k)
+
{\bf Im}\,\widetilde\sigma_{\rm MN}(k)\,
{\bf Im}\,\widetilde{\bf g}^{\rm MN}(k)
\right)
\Big),
\nd
which has a structure reminiscent of exponentiated integrated vertex
operators in string theory. Integrated graviton vertex operators deform the
world-sheet sigma-model background and generate closed-string excitations;
the analogy with \eqref{solocke} is therefore suggestive, although the GS
operator should not be identified literally with an ordinary perturbative
string vertex operator\footnote{Let us recall how an integrated graviton
vertex operator acts in string theory. It deforms the world-sheet sigma
model through a change of its target-space metric coupling. Exponentiating
an appropriate integrated vertex operator may therefore be viewed as
generating a finite background deformation or coherent closed-string
excitation. The difference between the two-dimensional integration of the
world-sheet vertex operator and the ten- or eleven-dimensional
spacetime-supported GS insertion reflects, among other things, the
first-quantized nature of the perturbative string description. The analogy
is consequently structural rather than an exact identification. This is shown in more detail in section 8.2 of \cite{csvsgs}.}. The
integration over the full world-sheet is essential in the usual string
construction; it should not be interpreted as evidence that an intermediate
Hamiltonian is absent, but rather as the covariant world-sheet
representation of the corresponding insertion.

Also, since in string theory strings are not necessarily the most useful
fundamental variables at all points in the moduli space of the theory
\cite{bbs}, there may be corners of the theory where higher-dimensional
branes provide a more natural description. For such a scenario one might
conjecture a generalized ``vertex'' operator $\mathbb V$ whose exponentiation
takes the schematic form:
\bg\label{aviloftus}
{\rm exp}(\mathbb V)
=
{\rm exp}\Big(
2\int d^{n+1}k\,
\left(
{\bf Re}\,\widetilde\sigma_{\rm MN}(k)\,
{\bf Re}\,\widetilde{\bf g}^{\rm MN}(k)
+
{\bf Im}\,\widetilde\sigma_{\rm MN}(k)\,
{\bf Im}\,\widetilde{\bf g}^{\rm MN}(k)
+\ldots
\right)
\Big),
\nd
possibly creating coherent excitations once inserted in the path integral;
the dotted terms denote other bosonic and fermionic degrees of freedom from
the set ${\bf \Xi}$. Since a complete first-quantized formulation of
interacting M2- and M5-branes analogous to the perturbative string
world-sheet is not presently available, the operator form
\eqref{aviloftus} should be regarded as conjectural at this stage, motivated
by the corresponding membrane and five-brane vertex-operator ideas in
\cite{bbs}.

This nevertheless raises the question whether the displacement operator
$\mathbb D(\sigma)$, or more accurately
$\mathbb D^{\dagger}(\sigma)\mathbb D(\sigma)$, could be viewed as a
generalized vertex-like operator responsible for creating the
Glauber--Sudarshan state. This is an interesting possibility, but for
$n=10$ it cannot simply be identified with the usual perturbative string
vertex operators. The origin of the Glauber--Sudarshan states in our
construction is the low-energy description of M-theory, and the
Glauber--Sudarshan states in lower dimensions, for example in type-II
theories, are obtained by suitable dimensional reductions and duality
transformations\footnote{Another example is the construction of
Glauber--Sudarshan states in heterotic theories \cite{gshet}.}. It is
conceivable that a more fundamental version of such an operator could be
formulated within string field theory or a future microscopic formulation
of M-theory. String field theory naturally incorporates the usual string
vertex structure; see, for example, the recent review \cite{sensft}. A
concrete eleven-dimensional construction of the type needed here,
however, remains to be established.

What we have achieved here is to show that the low-energy limits of
M-theory, and hence the corresponding string theories after dimensional
reduction and duality transformations, admit the class of
Glauber--Sudarshan states considered in this work. These states are
constructed through spacetime-supported displacement operators whose
structure has a suggestive resemblance to integrated vertex operators.
Their role, however, is different. Whereas conventional string vertex
operators describe perturbative string states or background deformations,
the GS operators considered here are used to construct the transient
non-supersymmetric configurations of interest\footnote{It is also possible
to create stable supersymmetric brane configurations using the
Glauber--Sudarshan states in M-theory, as shown in \cite{dileep}.}. These
operators are integrated over the full eleven-dimensional spacetime rather
than being localized on individual Cauchy slices. This covariant
spacetime-supported description is compatible with the underlying bulk
Hamiltonian constraints, while the physical notion of time, whenever
needed in the canonical description, is supplied by the asymptotic boundary
Hamiltonian.

Moreover, the displacement and the vertex operators appear in the far IR of M-theory or string theory,
where the massive stringy modes have been integrated out. Another key
difference between the Glauber--Sudarshan states and ordinary string vertex
operators is that the behavior of $\widetilde\sigma_{\rm MN}(k)$ for the
Glauber--Sudarshan states is controlled by a remnant Schwinger--Dyson, and
therefore intrinsically quantum, equation of the form
\cite{wdwpaper, coherbeta, coherbeta2, borel, joydeep}:
\bg\label{bormey}
{\delta\check{\bf S}
\left(
\langle{\bf \Xi}\rangle_\sigma
\right)
\over
\delta\langle{\bf \Xi}\rangle_\sigma}
=
0,
\nd
where
$\check{\bf S}(\langle{\bf \Xi}\rangle_\sigma)$ differs from
${\bf S}(\langle{\bf \Xi}\rangle_\sigma)$ by the renormalization effects
described in \cite{wdwpaper}, compared with the standard classical
supergravity equations of motion associated with conventional background
deformations. In fact the chain of transformations from the action in
\eqref{gilartg} is:
\bg\label{aspenb}
{\bf S}_{\rm tot}({\bf \hat\Xi})
~\to~
{\bf S}({\bf \Xi})
~\to~
{\bf S}(\langle{\bf \Xi}\rangle_\sigma)
~\to~
\check{\bf S}(\langle{\bf \Xi}\rangle_\sigma),
\nd
which is described in detail in \cite{wdwpaper}\footnote{There are quite a
few steps that we have not shown here. For example, to see how one goes from
the full set of Schwinger--Dyson equations:
\bg\label{lenamun}
\left\langle
{\delta{\bf S}_{\rm tot}({\bf \hat\Xi})
\over\delta{\bf \Xi}}
\right\rangle_\sigma
=
\left\langle
{\delta\over\delta{\bf \Xi}}
\log\left(
\mathbb D^\dagger({\bf \Xi})
\mathbb D({\bf \Xi})
\right)
\right\rangle_\sigma, \nonumber
\nd
to \eqref{bormey}, the reader may refer to
\cite{wdwpaper, joydeep, coherbeta2, csvsgs}. There are also ghost
Schwinger--Dyson equations that we do not elaborate on here. Interested
readers may find the necessary details in
\cite{wdwpaper, coherbeta, coherbeta2, borel, joydeep, csvsgs}. The above equation
also illustrates the difference between the backgrounds generated by the
Glauber--Sudarshan construction and conventional background deformations
described by classical supergravity equations. The latter satisfy the
classical equations
${\delta{\bf S}({\bf \Xi})/\delta{\bf \Xi}}=0$, whereas the GS expectation
values are generically constrained by the above equation and, after the appropriate
renormalized reduction, further by the ``on-shell" condition \eqref{bormey}. In this sense the backgrounds
considered here are intrinsically quantum rather than solutions obtained by
simply taking a classical limit of the GS state.}.
Solving \eqref{bormey} determines the allowed behavior of
$\sigma_{\rm MN}({\bf x}, t)$, which in turn governs the temporal profile represented
schematically in {\bf figure \ref{gswdw}}. In the covariant path-integral
description this determination does not require inserting an independent
bulk Hamiltonian between successive displacement operators. When the same
construction is interpreted canonically in terms of physical boundary time,
the corresponding temporal evolution is generated by
${\bf H}^{\rm phys}_{\partial\mathbb M}$, so there is no contradiction
between the Schwinger--Dyson description and the boundary-Hamiltonian
description \cite{csvsgs}. Finally, despite similarities with vertex operators, the
Glauber--Sudarshan states act in a rather different way. Their construction
uses ingredients of the underlying supersymmetric warped-Minkowski minima,
so that a Wilsonian effective description can be maintained, while the
backreaction of the GS states is used to construct the transient de Sitter
configurations. The backreaction is computed through the chain
\eqref{aspenb} originating from the Schwinger--Dyson equations. This will be
elaborated in section \ref{sec2.5}.

We finally note that spacetime integration of the displacement operators
does not, by itself, eliminate all short-distance singularities. Local
coincident-point singularities and contact terms may still arise for
composite operators and must be treated by the same renormalization
prescription used in defining the corresponding operators in
${\bf S}_{\rm tot}$. The integrated form does, however, smear such
distributions against the source and can therefore soften their appearance
in the final expressions. Moreover, the displacement insertion considered
here couples directly to the physical fields ${\bf \Xi}$ and not directly
to the ghost fields ${\bf \Upsilon}$. The ghosts nevertheless remain
present indirectly through the gauge-fixed measure and the full action
${\bf S}_{\rm tot}({\bf \Xi,\Upsilon})$. Thus the source insertion modifies
the weighting of the physical-field configurations in the path integral,
while BRST invariance and the ghost sector continue to ensure the consistent
gauge-fixed description.

\subsection{Glauber-Sudarshan states and the Faddeev-Popov ghosts \label{sec2.4}}

As discussed above, the Faddeev-Popov ghosts play a very important role in our construction. In fact the non-decoupling of the Fadeev-Popov ghosts at the level of the supersymmetric warped Minkowski background can actually make the temporal evolution of the wave-functionals $-$ in the sector with on-shell degrees of freedom (represented by the set ${\bf \Xi}$) $-$ much more manifest. Of course, as we saw earlier, even with the decoupled ghosts scenario, we can still describe the temporal evolutions of the wave-functionals as long as we allow for non-zero boundary Hamiltonian. Despite that, it would be useful to figure out how the ghosts appear in our set-up. In the likely scenario where the ghosts do not decouple, our analysis would form the basis to quantify the discussion
in section \ref{sec2.3}.

Before moving forward, let us clarify one other detail that we kept under the rug so far. This has to do with the off-shell degrees of freedom and should in-principle appear at the level of \eqref{gilsalaam}, namely:

{\footnotesize
\bg\label{gilsalaam3}
\langle{\bf g}_{\rm PQ}(y)\rangle_\sigma = {\int {\cal D}{\bf \Xi} {\cal D}({\bf \Sigma})~ {\rm exp}\Big(-{\bf S}[{\bf \Xi, \Sigma}]\Big)~{\rm exp}\Big(2\int d^{n+1}{x} \sqrt{-{\bf g}} ~{\bf Re}\left(\sigma_{\rm MN} ~{\bf g}^{\rm MN}\right)\Big) {\bf g}_{\rm PQ}(y) \over\int {\cal D}{\bf \Xi} {\cal D}({\bf \Sigma})~{\rm exp}\Big(-{\bf S}[{\bf \Xi, \Sigma}]\Big)~{\rm exp}\Big(2\int d^{n+1}{x} \sqrt{-{\bf g}} ~{\bf Re}\left(\sigma_{\rm MN}~ {\bf g}^{\rm MN}\right)\Big)}, \nd}
where ${\bf \Sigma}$ is the sector containing all the off-shell degrees of freedom. In string/M theory the off-shell degrees of freedom are massless, so once we integrate them out, they will appear as non-local contribution to ${\bf S}({\bf \Xi})$ expressed in terms of the on-shell degrees of freedom. This has been carefully described in \cite{joydeep} which the readers can refer for more details. Thus \eqref{gilsalaam} as it stands is complete once we assume that ${\bf S}({\bf \Xi})$ contains both the local and non-local interactions. This would also mean that the Faddeev-Popov ghosts will have local and non-local contributions. This fits perfectly with what we discussed in \cite{joydeep}.

There is another subtlety that we kept under the rug here, and it corresponds to the steps that lead from \eqref{gilsalaam} to \eqref{gilsalaam2}. Since $\mathbb{D}^\dagger({\bf \Xi})\mathbb{D}({\bf \Xi})$ is not gauge invariant, and neither are the on-shell fields ${\bf \Xi}$, the question of how the Faddeev-Popov procedure is being imposed here now appears. Recall that the standard procedure of imposing the Faddeev-Popov technique is on {\it gauge-invariant} correlation functions, but here the story is a bit different because we are looking at one-point functions that are not gauge invariant. The answer however lies on our definition of the displacement operators ${\cal D}(t)$ from \eqref{deebrian} and the way we defined the correlation function in say \eqref{mersirn}. The displacement operators ${\cal D}(t)$ are defined at
every instant of time between the temporal domain $-{1\over \sqrt{\Lambda}} \le t \le 0$ in a way that they propagate gauge {\it inequivalent} configurations. In fact this is exactly what ${\rm exp}(-i{\bf H}_{\rm tot}(\hat{\bf \Xi}){\rm T})$, or even ${\rm exp}(-i{\bf H}({\bf \Xi}){\rm T})$, could not do in the absence of the boundary Hamiltonian. In the path-integral form, this is only possible if the displacement operator $\mathbb{D}(\sigma)$, or more appropriately $\mathbb{D}^\dagger(\sigma) \mathbb{D}(\sigma)$, is defined for a {\it gauge-fixed} theory. To see how this works, let us define a quantity $\mathbb{Z}[\sigma]$ in a gauge-fixed theory in the following way:
\bg\label{daiseyT}
\mathbb{Z}[\sigma] \equiv \int {\cal D}{\bf \Xi}~{\cal D}{\bf \Upsilon}~ {\rm exp}\Big(-{\bf S}_{\rm tot}[{\bf \Xi, \Upsilon}]\Big)~{\rm exp}\Big(2\int d^{n+1}{x} \sqrt{-{\bf g}} ~{\bf Re}\left(\sigma_{\rm MN} ~{\bf g}^{\rm MN}\right)\Big), \nd
where ${\bf S}_{\rm tot}({\bf \Xi, \Upsilon})$ is the total action containing the Faddev-Popov ghosts and the gauge fixing term as in \eqref{gilartg} with ${\bf \Upsilon}$ denoting the ghost sector, as well as the non-local terms, and $n = 10$. The form of $\mathbb{Z}(\sigma)$ should be reminiscent of how we add a source term to the path-integral in usual field theories, and here the source is the de Sitter excited state. We can now extract, for example, the expectation value of the metric in the following way:

{\footnotesize
\bg\label{habibi}
\begin{split}
\langle {\bf g}_{\rm PQ}(z)\rangle_\sigma & = {1\over 2\sqrt{-{\bf g}(z)}~\mathbb{Z}[\sigma]}\cdot{\delta \mathbb{Z}[\sigma]\over \delta {\bf Re}~\sigma_{\rm PQ}(z)}\\
& = {\int {\cal D}{\bf \Xi} {\cal D}({\bf \Upsilon})~ {\rm exp}\Big(-{\bf S}_{\rm tot}[{\bf \Xi, \Upsilon}]\Big)~{\rm exp}\Big(2\int d^{11}{x} \sqrt{-{\bf g}} ~{\bf Re}\left(\sigma_{\rm MN} ~{\bf g}^{\rm MN}\right)\Big) {\bf g}_{\rm PQ}(z) \over\int {\cal D}{\bf \Xi} {\cal D}({\bf \Upsilon})~{\rm exp}\Big(-{\bf S}_{\rm tot}[{\bf \Xi, \Upsilon}]\Big)~{\rm exp}\Big(2\int d^{11}{x} \sqrt{-{\bf g}} ~{\bf Re}\left(\sigma_{\rm MN}~ {\bf g}^{\rm MN}\right)\Big)}, 
\end{split} \nd}
which will provide the emergent metric configuration in a gauge-fixed theory. In a similar vein, other one-point functions can also be extracted and together they would solve the Schwinger-Dyson equations \eqref{lenamun} and \eqref{bormey}. Remarkably, one of the immediate consequence of the above procedure is visible when we determine the Wheeler-de Witt equation governing the wave-functional over the emergent metric configurations: both the Hamiltonian constraint and the wave-functional of the universe have no Faddev-Popov ghost dependence anymore! 
This is precisely the prediction we had in eq. (2.34) of \cite{wdwpaper}.

The only thing now remains is to determine the precise form of the Faddeev-Popov ghosts that appear in $\mathbb{Z}[0]$ from \eqref{daiseyT}. For this we have to define our class of gauge fixing conditions. Since there are multiple on-shell degrees of freedom, we can define:
\bg\label{daisiefull}
\mathbb{G}^{(i)}({\bf \Xi}(x)) = \omega^{(i)}(x), \nd
with $i = 1, .., {\rm N}$ forming the set of equations over the on-shell fields and $\omega^{(i)}(x)$ are arbitrary functions. The gauge non-equivalent configurations of the on-shell fields are the ones that satisfy the equations from \eqref{daisiefull}. These enter the path-integral in the following way:

{\footnotesize
\bg\label{katanran}
\begin{split}
\int{\cal D}({\bf \Xi})~{\rm exp}\left[-{\bf S}({\bf \Xi})\right] & = 
\int {\cal D}{\bf \Xi}~\prod_{i = 1}^{\rm N} {\cal D}\{a\}~\delta\left( \mathbb{G}^{(i)}({\bf \Xi}^{(a)}(x)) - \omega^{(i)}(x)\right) ~ {\rm exp}\left[-{\bf S}({\bf \Xi})\right]\\
& = \int {\cal D}{\bf \Xi}^{(a)}~\prod_{i = 1}^{\rm N} {\cal D}\{a\}~\delta\big( \mathbb{G}^{(i)}({\bf \Xi}^{(a)}(x)) - \omega^{(i)}(x)\big) ~ {\rm exp}\big[-{\bf S}({\bf \Xi}^{(a)})\big]
\end{split}
\nd}
where ${\bf \Xi} \to {\bf \Xi}^{(a)} := {\bf \Xi} + \sum\limits_{k=1}^{\rm N} c_k {\cal O}_k\left(\{a\}\right)$ is the gauge transformations acting on the on-shell field configurations using function ${\cal O}_k(\{a\})$, and $\{a\}$ collectively denotes the corresponding bosonic
and fermionic gauge parameters.  Depending on the sector, these include the
diffeomorphism parameters $\xi^{\rm M}(x)$, the $p$-form gauge parameters, the non-Abelian
gauge parameters and the fermionic local-supersymmetry parameters. Since both the measure ${\cal D}{\bf \Xi}$ and the action ${\bf S}({\bf \Xi})$ are invariant under the transformations, the path-integral naturally takes the form as in \eqref{katanran} above. It is also easy to show that the path-integral \eqref{katanran} reduces to:

{\footnotesize
\bg\label{gility}
\int {\cal D}{\bf \Xi}^{(a)}~\prod_{i = 1}^{\rm N} {\cal D} \{a\}~{\rm det}\bigg[{\delta\over \delta \{a\}}\left(\mathbb{G}^{(i)}\left({\bf \Xi}^{(a)}(x)\right) - \omega^{(i)}(x)\right)\bigg]~\delta\big( \mathbb{G}^{(i)}({\bf \Xi}^{(a)}(x)) - \omega^{(i)}(x)\big) ~ {\rm exp}\big[-{\bf S}({\bf \Xi}^{(a)})\big], \nd}
when expressed in terms of the transformation parameters $\{a\}$ for all set of $i = 1, 2, ..., {\rm N}$ transformations.
Since the Faddeev--Popov operator is obtained by linearizing the gauge orbit in an
infinitesimal neighborhood of the identity, we restrict the transformation parameters
so that
$\sum\limits_k c_k\,{\cal O}_k(\{a\})\ll 1, \forall k\in\mathbb{Z}$.
The path integral in \eqref{gility} may therefore be expanded to leading order in the
infinitesimal gauge transformation, giving:

{\footnotesize
\bg\label{gility2}
\int {\cal D}{\bf \Xi}^{(a)}~\prod_{i = 1}^{\rm N} {\cal D} \{a\}~{\rm det}\bigg\llbracket\sum_k c_k {\delta {\cal O}_k(\{a\})\over \delta \{a\}} \cdot {\delta \mathbb{G}^{(i)}\left({\bf \Xi}^{(a)}(x)\right) \over \delta {\bf \Xi}(x)}\bigg\rrbracket~\delta\big( \mathbb{G}^{(i)}({\bf \Xi}^{(a)}(x)) - \omega^{(i)}(x)\big) ~ {\rm exp}\big[-{\bf S}({\bf \Xi}^{(a)})\big], \nd}
where no $\omega^{(i)}(x)$ appears inside the determinant, and the double-bracket denotes ordering of the terms. This is required because both ${\delta {\cal O}_k(\{a\})\over \delta \{a\}}$ and ${\delta \mathbb{G}^{(i)}\left({\bf \Xi}^{(a)}(x)\right) \over \delta {\bf \Xi}(x)}$ are {\it operators} and therefore would require certain orderings. It should also be clear that the double bracket satisfy the following properties:
\bg\label{skinecare}
\Big\llbracket \sum\limits_i {\rm A}_i\Big\rrbracket = \sum\limits_i\Big\llbracket{\rm A}_i\Big\rrbracket, ~~~
\Big\llbracket \prod_{i = 1}^{\rm N} {\rm A}_i\Big\rrbracket = \sum_{i = 1}^{{\rm N}!} b_i~\underbrace{\rm A_1 A_2 A_3 A_4......A_{\rm N-1} A_{\rm N}}_{i^{\rm th}-{\rm permutation}},
\nd
where ${\rm A}_i$ are matrices and $b_i \in (0, 1)$. 
The form \eqref{gility2} can be motivated by first Taylor expanding over ${\bf \Xi}(x)$ and then taking the functional derivative with respect to the set of parameters $\{a\}$. For simple gauge transformations, we expect:
\bg\label{sercrem}
{\delta^2 {\cal O}_k(\{a\})\over \delta^2\{a\}} = 0, ~~~ {\delta \mathbb{G}^{(i)}\left({\bf \Xi}^{(a)}(x)\right) \over \delta {\bf \Xi}(x)} = 
{\delta \mathbb{G}^{(i)}\left({\bf \Xi}^{(a)}(x)\right) \over \delta {\bf \Xi}^{(a)}(x)}, \nd
where for more general gauge transformations the first equality may not hold, but the second one continues to hold. If we ignore the generic picture, which we shall discuss soon, the path-integral in \eqref{gility2} takes the following form:

{\scriptsize
\bg\label{desyfoolai}
\int {\cal D}\{a\}\int {\cal D}{\bf \Xi}~{\cal D}\{\overline{\mathbb{C}}\} {\cal D}\{\mathbb{C}\}\prod_{i, k}{\rm exp}\left[ -{\bf S}({\bf \Xi}) - \int d^{n+1} x ~ \overline{\mathbb{C}}^{ik\alpha}(x) \cdot  
\Bigg\llbracket{\delta {\cal O}_k(\{a\})\over \delta \{a\}}  {\delta \mathbb{G}^{(i)}\left({\bf \Xi}(x)\right) \over \delta {\bf \Xi}(x)}\Bigg\rrbracket_{\alpha\beta} \cdot \mathbb{C}^{ik\beta}(x)\right] \delta\big( \mathbb{G}^{(i)}({\bf \Xi}(x)) - \omega^{(i)}(x)\big), \nd}
where we have expressed the ghost action using an Euclidean formalism, and absorbed $c_k$ in the definition of the ghost fields $\mathbb{C}^{ik\alpha}(x)$. We have also defined ${\cal D}\{\overline{\mathbb{C}}\} {\cal D}\{\mathbb{C}\} = \prod\limits_{j, l, \sigma}{\cal D}\overline{\mathbb{C}}^{jl\sigma} {\cal D}\mathbb{C}^{jl\sigma}$.  The dots take care of all the tensor and the fermionic indices of the ghosts. There is an awkward integral over ${\cal D}\{a\}$, but this simply provides a multiplicative factor that goes away once we divide the path-integral by the denominator term. We can now soak in the delta function using the following identity:
\bg\label{trottagra}
\int {\cal D}\{\omega\} ~{\rm exp}\left[-{1\over 2}\int d^{n+1} x ~\omega^{(i)}(x) \Delta_{ij}({\bf \Xi}(x)) \omega^{(j)}(x)\right] = {1\over \sqrt{{\rm det}~\Delta_{ij}({\bf \Xi})}}, \nd
and then express ${\rm det}~\Delta_{ij}({\bf \Xi})$ in terms of new set of ghosts (sometime called the {\it third ghosts}). Note that the new set of third ghosts are coming from the gauge fixing term itself. In cases like QED or QCD, the third ghosts decouple, so they are not important\footnote{In fact for QED or QCD, $\Delta_{ij}({\bf \Xi}(x)) = {\delta_{ij}\over \zeta}$ where $\zeta$ is a free parameter.}. Here however they matter because they would non-trivially couple to the on-shell degrees of freedom. This would imply the following identity derived from \eqref{trottagra}:

{\scriptsize
\bg\label{blink2}
\int {\cal D}\{b\} {\cal D}\{{\overline{b}}\} ~{\rm exp}\left[-{1\over 2}\int d^{n+1} x ~\overline{b}^{\sigma}(x) \cdot\Delta_{\sigma\rho}({\bf \Xi}(x))\cdot b^{\rho}(x)\right]  \int {\cal D}\{\omega\} ~{\rm exp}\left[-{1\over 2}\int d^{n+1} x ~\omega^{(i)}(x) \Delta_{ij}({\bf \Xi}(x)) \omega^{(j)}(x)\right] = 1, \nd}
where $b^\sigma(x), \overline{b}^\rho(x)$ are the fermionic ghosts with tensor and fermionic indices; and $\omega^i(x)$ are the bosonic fields. Plugging \eqref{blink2} to \eqref{desyfoolai} would replace 
$\omega^{(i)}(x)$ by $\mathbb{G}^{(i)}({\bf \Xi}(x))$, and will get rid of the measure factor ${\cal D}\{\omega\}$, although the $b$-ghost integral would continue to remain as it is in \eqref{blink2}. Thus the ghost structure would become more non-trivial because of both $\mathbb{C}$ and $b$ ghosts. Unfortunately this is not the only complication and the story is more involved than what we wrote above. Our analysis relied heavily on the first relation in \eqref{sercrem}. Question is: what if this is not true? For example, what if we allow for a class of gauge transformation parameters $\{a\}^{(k)}$, instead of just one set $\{a\}$ used above? For such a case, clearly:
\bg\label{aspengarage}
{\delta^2{\cal O}_k(\{a\}^{(l)})\over \delta\{a\}^{(i)} \delta\{a\}^{(j)}} \ne 0, ~~~~~~ i \ne j, \nd
and therefore the aforementioned analysis cannot be the full story, thus resulting in an even more complicated ghost structure. Such cases arise, for example, when we have $p$-form gauge fields. For the present case, gauge transformation on a $p$-form gauge field leads to the following sequence of transformations:

{\scriptsize
\bg\label{timecut}
\begin{split}
{\bf\Xi}_p &\to {\bf \Xi}_p + d\Lambda^{(1)}_{p-1} \to {\bf \Xi}_p + d\left(\Lambda^{(1)}_{p-1} + d\Lambda^{(2)}_{p-2}\right) \to 
{\bf \Xi}_p + d\left(\Lambda^{(1)}_{p-1} + d\left(\Lambda^{(2)}_{p-2} + d\Lambda^{(3)}_{p-3}\right)\right) \\
&\to 
{\bf \Xi}_p + d\left(\Lambda^{(1)}_{p-1} + d\left(\Lambda^{(2)}_{p-2} + d\left(\Lambda^{(3)}_{p-3} + d\Lambda^{(4)}_{p-4}\right)\right)\right) 
\to ... \to  {\bf \Xi}_p + d\left(\Lambda^{(1)}_{p-1} + d\left(\Lambda^{(2)}_{p-2} + ... + d\left(\Lambda^{(p-1)}_{1} + d\Lambda^{(p)}_{0}\right)\right)..\right),
\end{split}
\nd}
culminating in the zero form $\Lambda_0$. Since the original gauge transformation by a $p-1$ form $\Lambda_{p-1}$ is compensated by a set of ghosts, the second transformation would imply finding {\it ghosts of ghosts}. Similarly, the third transformation would be to find {\it ghosts of ghosts of ghosts}. This will continue till we hit the zero form. In other words, \eqref{sercrem} should be replaced by:
\bg\label{luclidia}
{\delta^p\over \delta\{a\}^{(p)} \delta\{a\}^{(p-1)} \delta\{a\}^{(p-2)} .... \delta\{a\}^{(1)}}~{\cal O}_k(\{a\}^{(l)})= 0, \nd
for a $p$-form gauge field\footnote{Although \eqref{luclidia} does not prohibit us to find a sub-sector with rank $r < p$ that annihilates ${\cal O}_k(\{a\}^{(l)})$. We will discuss a case with $r = 2$ soon.}. Since in M-theory we have a four-form G-flux with the corresponding three-form potential ${\bf C}_3$ in the far IR, the aforementioned issue becomes relevant for us. This generalized structure of ghosts is studied using the Batalin-Vilcovisky formalism \cite{batalin} and a full treatment in the presence of gravity is clearly beyond the scope of this paper. All in all, as expected, the structure of ghosts is pretty complicated here as both the gauge fixing terms and the ghosts themselves, contribute to additional ghosts in the system. These give us:

{\scriptsize
\bg\label{lioness}
\begin{split}
& {\bf S}_{\rm gf} = {1\over 2} \int d^{n+1}x \mathbb{G}^{(i)}({\bf \Xi}(x)) \Delta_{ij}({\bf \Xi}(x))\mathbb{G}^{(j)}({\bf \Xi}(x)) + ...\\
& {\bf S}_{\rm ghost} = \sum_{i, j, k} \int d^{n+1} x ~ \overline{\mathbb{C}}^{ijk\alpha}(x) \cdot  
\left\llbracket{\delta {\cal O}_k(\{a\}^{(l)})\over \delta \{a\}^{(j)}}  {\delta \mathbb{G}^{(i)}\left({\bf \Xi}(x)\right) \over \delta {\bf \Xi}(x)}\right\rrbracket_{\alpha\beta} \cdot \mathbb{C}^{ijk\beta}(x) + 
{1\over 2}\int d^{n+1} x ~\overline{b}^{\sigma}(x) \cdot\Delta_{\sigma\rho}({\bf \Xi}(x))\cdot b^{\rho}(x) + ..., 
\end{split}
\nd}
where the dotted terms come from additional ghosts and gauge fixing terms.
One contribution to the additional ghosts and gauge fixing terms is of course from the Batalin-Vilkovisky formalism that takes us beyond the sub-sector:
\bg\label{aaliyacruz}
{\delta^2\over \delta\{a\}^{(i)} \delta\{a\}^{(i)}}~{\cal O}_k(\{a\}^{(l)})={\delta^2\over \delta\{a\}^{(i)} \delta\{a\}^{(j)}}~{\cal O}_k(\{a\}^{(l)}) = 0, \nd
that we used in \eqref{lioness} to express the $\mathbb{C}$-ghost contributions. Other contributions come from the non-abelian gauge fluxes that would appear from the localized G-fluxes in the far IR limit of M-theory\footnote{A mathematical treatment of how such localized G-fluxes lead to 2-form gauge fluxes on branes has appeared in \cite{DRS}. The fact that the six-dimensional base of the internal eight manifold in the far IR limit of M-theory may not be K\"ahler or may not even allow for an integrable complex structure doesn't matter for the constructions in \cite{DRS}. In this paper we will not develop this story to avoid over-complicating the ghost structure of the system.}. In fact in the presence of wrapped membranes on vanishing 2-cycles of the internal manifold such non-abelian degrees of freedom can be easily constructed. Interestingly, as shown in \cite{dileep}, there exist a class of Glauber-Sudarshan states that partake in such a construction. A full analysis of this is beyond the scope of this paper and will be dealt elsewhere.

However there is yet another, more subtle, contribution that we briefly alluded to when we wrote the path-integral as \eqref{gilsalaam3}. This has to do with the off-shell degrees of freedom being integrated away and then expressed as non-local terms using the on-shell degrees of freedom. Such replacement also introduce {\it non-local ghosts} as well as {\it non-local gauge fixing terms} in the system\footnote{One way to infer this would be to first express the action in terms of all on- and off-shell degrees of freedom, then add in all the ghosts using the Faddeev-Popov and the Batalin-Vilkovisky formalisms. After which we can integrate out the off-shell degrees of freedom to determine the non-local ghost action ${\bf S}^{\rm nloc}_{\rm ghost}$.}. In \cite{coherbeta, joydeep, wdwpaper} we expressed the non-local ghosts' contributions using ${\bf S}^{\rm nloc}_{\rm ghost}$, such that:
\bg\label{aaliya}
{\bf S}_{\rm ghost}(\hat{\bf \Xi}) = {\bf S}^{\rm loc}_{\rm ghost}(\hat{\bf \Xi}) + 
{\bf S}^{\rm nloc}_{\rm ghost}(\hat{\bf \Xi}), \nd
with similar distribution for the gauge fixing terms. The advantage of expressing everything in terms of the on-shell degrees of freedom is that the {\it number} of such degrees of freedom is precisely known.  
We can use the aforementioned analysis to fix the ghost and the gauge-fixing terms for ${\bf S}_{\rm tot}(\hat{\bf \Xi})$ to express $\mathbb{Z}[\sigma]$ as in \eqref{daiseyT}, with ${\bf S}_{\rm tot}(\hat{\bf \Xi}) = {\bf S}({\bf \Xi}) + {\bf S}_{\rm ghost}(\hat{\bf \Xi}) + {\bf S}_{\rm gf}({\bf \Xi})$. Using $\mathbb{Z}[\sigma]$, one can show that \eqref{bormey} naturally arises from \eqref{lenamun}, with $\check{\bf S}(\langle {\bf \Xi}\rangle_\sigma)$ differing from ${\bf S}(\langle {\bf \Xi}\rangle_\sigma)$ by the renormalization procedure demonstrated in section 2.2 of \cite{wdwpaper}. Moreover, the ghost action should satisfy the following set of quantum 
equations:

{\scriptsize
\bg\label{seldena}
\left\langle {\delta{\bf S}_{\rm ghost}(\hat{\bf \Xi})\over \delta {\bf \Xi}}\right\rangle_\sigma = \left\langle{\delta \over \delta{\bf \Xi}}~{\rm log}\left(\mathbb{D}^\dagger({\bf \Xi})\mathbb{D}({\bf \Xi})\right)\right\rangle_\sigma - \sum_{\{\sigma_i\} \ne \sigma} 
\left\langle {\delta\check{\bf S}({\bf \Xi})\over \delta {\bf \Xi}}\right\rangle_{(\{\sigma'\}|\sigma)} -\left\langle {\delta\check{\bf S}_{\rm gf}({\bf \Xi})\over \delta {\bf \Xi}}\right\rangle_\sigma,~~~~~ \left\langle
{\delta{\bf S}_{\rm tot}({\bf \Xi},{\bf \Upsilon})
\over
\delta{\bf \Upsilon}(x)}
\right\rangle_\sigma
=
0 , \nd}
where ${\vert \sigma' -\sigma\vert \over \sigma} \ge 1$ and $\check{\bf S}_{\rm gf}$ differs from ${\bf S}_{\rm gf}$ because of the renormalization procedure \cite{wdwpaper}. This renormalization procedure does two things: \textcolor{blue}{one}, it allows to sum over the off-shell Glauber-Sudarshan states $\vert \sigma'\rangle$ with ${\vert \sigma' -\sigma\vert \over \sigma} \ge 1$ as mentioned above. And 
\textcolor{blue}{two}, all corrections from the off-shell states $\vert\sigma'\rangle$ with ${\vert \sigma' -\sigma\vert \over \sigma} < 1$
convert:
\bg\label{shystyl}
{\bf S}({\bf \Xi}) \to \check{\bf S}({\bf \Xi}), ~~~~~~ 
{\bf S}_{\rm gf}({\bf \Xi}) \to \check{\bf S}_{\rm gf}({\bf \Xi}), \nd
which is what enters \eqref{bormey} and \eqref{seldena}. Both these equations are quantum equations, and there are no classical analogues of them. This, as alluded to earlier, distinguishes the Glauber-Sudarshan states from the states created by the vertex operators. (In fact combining \eqref{bormey} with \eqref{seldena} forms the set of {\it bootstrap} equations recently developed in \cite{csvsgs}. These details are not necessary for the present treatment.)

The issue of decoupling or non-decoupling of the ghosts is important. For this first we need to quantify the behavior of the propagators. The two set of propagators in the ghost sector \eqref{lioness} may be expressed as:
\bg\label{demmore}
[{\rm D}]_{\alpha\beta} = \left[\Delta^{-1}(0)\right]_{\alpha\beta}, ~~~~~
[{\rm D}^{(i)}_{(jk)}]_{\alpha\beta} = \left[\left({\delta {\cal O}_k(\{a\}^{(l)})\over \delta \{a\}^{(j)}}  {\delta \mathbb{G}^{(i)}\left({\bf \Xi}(x)\right) \over \delta {\bf \Xi}(x)}\right)^{-1}_{{\bf \Xi} = 0}\right]_{\alpha\beta}, \nd
where the operators are described in \eqref{lioness}. The aformentioned list is not exhaustive because there are other propagators associated with the ghosts appearing from the Batalin-Vilkovisky method, non-abelian degrees of freedom and non-local counterterms that we alluded to earlier.
One may easily check that \eqref{demmore} gives the right propagator in the case of QED and QCD. For example in the QCD Lorentz gauge:
\bg\label{thepog}
{\delta{\cal O}_k(\{a\}^{(l)})\over \delta\{a\}^{(j)}} = 
\delta_{k1}\delta_{l1}\delta_{j1} {\rm D}_\mu, ~~~~~~ {\delta \mathbb{G}^{(i)}({\bf\Xi}(x))\over \delta{\bf \Xi}(x)} = \delta_{i1} \partial_\mu,~~~~~~~~ \Delta({\bf \Xi}(x))_{\alpha\beta} = {\delta_{\alpha\beta}\over \zeta}, \nd
and therefore the ghosts from the gauge-fixing terms decouple but the other set of ghosts have propagators proportional to $(\partial^\mu \partial_\mu)^{-1}$. Moreover in the QCD axial gauge, all ghosts decouple.
For the present case a subsector of the ghosts would decouple if the following conditions are met:

{\scriptsize
\bg\label{maikamon}
{\delta\Delta_{\alpha\beta}({\bf \Xi}(x))\over \delta{\bf \Xi}(x)} = 
\sum_\sigma \left\llbracket\left({\delta {\cal O}_k(\{a\}^{(l)})\over \delta\{a\}^{(j)}}\right)_{\alpha\sigma}\left({\delta^2\mathbb{G}^{(i)}\left({\bf \Xi}(x)\right) \over \delta {\bf \Xi}^2(x)}\right)^\sigma_\beta +  
\left({\delta^2 {\cal O}_k(\{a\}^{(l)})\over\delta {\bf\Xi}(x)\delta\{a\}^{(j)}}\right)_{\alpha\sigma}\left({\delta\mathbb{G}^{(i)}\left({\bf \Xi}(x)\right) \over \delta {\bf \Xi}(x)}\right)^\sigma_\beta\right\rrbracket = 0, \nd}
where note that we have taken ${\delta^2 {\cal O}_k(\{a\}^{(l)})\over\delta {\bf\Xi}(x)\delta\{a\}^{(j)}} \ne 0$, which is true even for QCD because of it's non-abelian nature. For the subsector of ghosts with abelian interactions this may still be true because the on-shell metric couples with all other on-shell degrees of freedom.  Since the behavior of ${\delta {\cal O}_k(\{a\}^{(l)})\over \delta\{a\}^{(j)}}$ is fixed from the underlying symmetries of the theory itself, the decoupling would solely depend on the behavior of ${\delta\mathbb{G}^{(i)}\left({\bf \Xi}(x)\right) \over \delta {\bf \Xi}(x)}$ satisfying \eqref{maikamon}. This then leads to the following three scenarios.

\vskip.1in

\noindent $\bullet$ If 
${\delta^2 {\cal O}_k(\{a\}^{(l)})\over\delta {\bf\Xi}(x)\delta\{a\}^{(j)}} = 0$, then in any gauge $\mathbb{G}^{(i)}({\bf \Xi}(x))$ where the dependence is {\it linearly} on ${\bf \Xi}(x)$, the Faddeev-Popov ghosts decouple because:
\bg\label{maikamaa}
\left({\delta^2\mathbb{G}^{(i)}\left({\bf \Xi}(x)\right) \over \delta {\bf \Xi}^2(x)}\right)_{\alpha\beta} = 0. \nd

\noindent $\bullet$ If $\mathbb{G}^{(i)}({\bf \Xi}(x))$ depends {\it linearly} on ${\bf \Xi}(x)$, but ${\delta^2 {\cal O}_k(\{a\}^{(l)})\over\delta {\bf\Xi}(x)\delta\{a\}^{(j)}} \ne 0$, then the Faddeev-Popov ghosts decouple only if:
\bg\label{longmaika}
\sum_\sigma \left\llbracket \left({\delta^2 {\cal O}_k(\{a\}^{(l)})\over\delta {\bf\Xi}(x)\delta\{a\}^{(j)}}\right)_{\alpha\sigma}\left({\delta\mathbb{G}^{(i)}\left({\bf \Xi}(x)\right) \over \delta {\bf \Xi}(x)}\right)^\sigma_\beta\right\rrbracket = 0. \nd

\noindent $\bullet$ If the operator $\Delta_{\alpha\beta}({\bf \Xi}(x))$ is linear in ${\bf \Xi}(x)$, then the ghosts from the gauge-fixing terms would decouple because:
\bg\label{alicwitt}
{\delta^2\Delta_{\alpha\beta}({\bf \Xi}(x))\over \delta{\bf \Xi}^2(x)} = 0.\nd

\vskip.1in

\noindent Thus, while the first and the third conditions seem possible under the aforementioned criteria, the second condition is rather non-trivial and is unclear whether it could be satisfied generically. However even if \eqref{longmaika} is satisfied, this does not imply decoupling of other ghosts related to Batalin-Vilcovisky, non-local counterterms and non-abelian degrees of freedom. Such a conclusion seems consistent with Weinberg's comment \cite{weinberg} that in a theory like General Relativity there is no way of choosing a coordinate system in which the ghosts decouple.

In the {non-decoupled} ghosts scenario one may however ask whether we can find  
field redefinitions using which we can express $\mathbb{Z}[0]$ from \eqref{daiseyT} in a ghost-free way. As explained in \cite{rahman}, such a choice doesn't exactly negate Weinberg's comment \cite{weinberg} (see also \cite{faddeev}) because it involves instantaneous ghost 
propagators\footnote{Meaning that the ghosts do not have any dynamics.} thus acting as constraints in the system. To implement this for our case, first note that we haven't lost any degrees of freedom when we expressed $\mathbb{Z}[0]$ in \eqref{daiseyT} using on-shell fields ${\bf \Xi}$. The reason is because the other off-shell degrees of freedom are integrated away to provide the non-local interactions (see footnote \ref{karkapur}). In the presence of both local and the non-local terms (but in the absence of their respective ghosts), the effects of all degrees of freedom are present albeit with the expected gauge redundancies. Therefore in this set-up a possible redefinition could be ${\bf \Xi}^\ast = {\bf \Xi}^\ast({\bf \Xi, \Upsilon})$ such that:
\bg\label{lizbank}
\mathbb{Z}[0] =  \int {\cal D}{\bf \Xi}~{\cal D}{\bf \Upsilon}~ {\rm exp}\Big(-{\bf S}_{\rm tot}[{\bf \Xi, \Upsilon}]\Big) = 
\int {\cal D}{\bf \Xi}^\ast~ {\rm exp}\Big(-{\bf S}^\ast[{\bf \Xi}^\ast]\Big), \nd
where despite the equality of the path-integral, the dimension of the Hilbert space for ${\bf \Xi}^\ast$ could be smaller than the dimension of the Hilbert space entering $\mathbb{Z}[0]$ in \eqref{daiseyT}. Moreover, once we compute $\mathbb{Z}[\sigma]$ as in \eqref{daiseyT} with the second path-integral, the ghosts would reappear because 
${\bf \Xi} = {\bf \Xi}({\bf \Xi}^\ast, {\bf \Upsilon})$. The Wheeler-De Witt equations can be derived by performing a Weiss variation \cite{weiss, feng} over the second path-integral (by including the wave-functional) to give us:
\bg\label{nestkathia3}
{\partial\widetilde{\bf \Psi}({\bf \Xi}^\ast)\over \partial t} = 0, ~~~~
{\bf H}^\ast({\bf\Xi}^\ast) \vert\widetilde{\bf \Psi}({\bf\Xi}^\ast)\rangle = \int d^n{\bf x}~{\delta {\bf S}^\ast({\bf\Xi}^\ast)\over \delta{\bf g}^{\ast 00}({\bf x})}\vert \widetilde{\bf \Psi}({\bf\Xi}^\ast)\rangle = 0, \nd
in addition to the Wheeler-De Witt equation \eqref{nestkathia} in the absence of the boundary Hamiltonian. Not surprisingly, the problem of time reappears in the ${\bf\Xi}^\ast$ degrees of freedom because we can only talk about {\it relational time}
as in sections \ref{sec2.2} and \ref{sec2.3} but not an absolute {\it boundary time}. The Schwinger-Dyson equations at the supersymmetric warped-Minkowski level takes the form:

{\footnotesize
\bg\label{clodugark5}
\left\langle {\delta\over \delta {\bf \Xi}(x)}\left({\bf S}_{\rm tot}({\bf \hat\Xi}) - {\rm log}~\mathbb{D}^\dagger({\bf \Xi}, \sigma) \mathbb{D}({\bf \Xi}, \sigma)\right)\right\rangle_\sigma = 0, ~~~ \left\langle
{\delta{\bf S}_{\rm tot}({\bf \Xi},{\bf \Upsilon})
\over
\delta{\bf \Upsilon}(x)}
\right\rangle_\sigma
=
0, ~~~
\left\langle {\delta{\bf S}^\ast({\bf \Xi}^\ast)\over \delta {\bf\Xi}^\ast(x)}\right\rangle_0 = 0, \nd}
valid for $\sigma > 0$ as well as $\sigma = 0$, with possible $\sigma$ extension with ${\bf \Xi}^\ast$ degrees of freedom that we do not show here. There is of course no classical analogue to \eqref{clodugark5}, but one can ask what happens if we go to the $\hbar \to 0$ limit. Naively, from the path-integral to classical correspondence in the $\hbar \to 0$ limit, we expect the following set of ``classical'' equation of 
motion\footnote{The stationarity condition in \eqref{clodugark55} is covariant at the level of the full functional
variation.  Any subsequent expression in terms of particular components of the metric,
however, is tied to the chosen slicing and gauge and should not be assigned an
independent coordinate-invariant meaning.}:
\bg\label{clodugark55}
{\delta\over \delta \hat{\bf \Xi}(x)}\left({\bf S}_{\rm tot}({\bf \hat\Xi}) - {\rm log}~\mathbb{D}^\dagger({\bf \Xi}, \sigma) \mathbb{D}({\bf \Xi}, \sigma)\right) = 
 {\delta{\bf S}^\ast({\bf \Xi}^\ast)\over \delta {\bf\Xi}^\ast(x)} = 0, \nd
where $\hat{\bf \Xi} = {\bf \Xi, \Upsilon}$ with ${\bf \Upsilon}$ being the ghost degrees of freedom. At the classical level, the ghost degrees of freedom should not exist, so ${\bf \Upsilon} = 0$, implying ${\bf S}_{\rm ghost} = {\bf S}_{\rm gf} = 0$ and ${\bf S}_{\rm tot}(\hat{\bf \Xi}) = {\bf S}({\bf \Xi})$ from \eqref{gilartg}. This will also make ${\bf \Xi}^\ast = {\bf \Xi}$, and ${\bf S}^\ast({\bf \Xi}^\ast) = {\bf S}({\bf \Xi})$. Together they leave us with:
\bg\label{clodugark6}
{\delta\over \delta {\bf \Xi}(x)}\left({\bf S}({\bf \Xi} - {\rm log}~\mathbb{D}^\dagger({\bf \Xi}, \sigma) \mathbb{D}({\bf \Xi}, \sigma)\right) = 0. \nd
For $\sigma = 0$, \eqref{clodugark6} clearly produces the supersymmetric warped Minkowski background that we have used so far. When $\sigma > 0$, it would appear that we have added an effective positive cosmological constant term to the action, and therefore we are essentially asking for a {\it classical} de Sitter vacuum solution. Such a solution will bring back all the problems that we listed, for example in \cite{joydeep}, related to the existence of an EFT satisfying Wilsonian (or an Exact) Renormalization Group procedure, et cetera. Moreover from the construction of the Glauber-Sudarshan states using the displacement operator as in \eqref{deebrian}, the arrangement of $\widetilde\sigma_{\rm MN}(k)$ is controlled by the quantum equation of motion \eqref{bormey}. Putting everything together would imply that a consistent picture emerges in the limit $\hbar \to 0$ {\it only} when we have $\widetilde\sigma_{\rm MN}(k) = 0$ and the EOM is ${\delta{\bf S}({\bf \Xi})\over \delta{\bf \Xi}} = 0$, leading to the supersymmetric vacuum configuration\footnote{A way to see this would be the following. Consider $\mathbb{Z}[\sigma]$ from \eqref{daiseyT}, and let us switch off the ghosts and the gauge fixing terms as mentioned above. The ``classical'' EOM coming from the action ${\bf S}({\bf \Xi})$ in the $\hbar \to 0$ limit, extended by the source term, takes the following form:
\bg\label{500casilos}
{\delta {\bf S}({\bf \Xi})\over \delta {\bf g}^{\rm AB}(y)} = \sqrt{-{\bf g}(y)}\Big(2 \sigma_{\rm AB}(y) - \sigma_{\rm MN}(y) {\bf g}_{\rm AB}(y) {\bf g}^{\rm MN}(y)\Big), \nd
which would appear as though we have added a cosmological constant term on the RHS, except that the term is a function of the coordinates $y^{\rm A}$ and therefore is not actually a {\it constant}. The question is whether we can find a functional form for $\sigma_{\rm MN}(y)$ which would lead to a four-dimensional de Sitter {vacuum} solution. The answer is {\it no} because such a solution is a sharply peaked state in the configuration space
and such a configuration will generically spread or deform under quantum
evolution, with the detailed behavior determined by its fluctuation spectrum and
interactions, thereby progressively losing any residual classical character. This helps with the conundrum because
existence of any such solution would have created all the issues mentioned in say \cite{joydeep} that we barely managed to avoid by considering four-dimensional de Sitter solution as an {\it excited state} in the theory. Saying differently, demanding (1) a Wilsonian effective action at the far IR and (2) a trans-Planckian bound over the temporal domain, would imply that we cannot take a time-dependent background at the classical vacuum level. (One may consider resorting to a static patch but that unfortunately, as we showed in \cite{joydeep}, does not help.) Since $\sigma_{\rm MN}(y)$ or ${\bf g}_{\rm MN}(y)$ or both are necessarily time-dependent (see sections \ref{sec2.2} and \ref{sec2.3}) the only consistent solution at the vacuum level should be $\sigma_{\rm MN}(y) = 0$, implying that the EOM can only be ${\delta {\bf S}({\bf \Xi})\over \delta {\bf g}^{\rm AB}(y)} = 0$ with time-independent solutions. Viewing ${\bf Re}\left(\sigma_{\rm MN} {\bf g}^{\rm MN}\right) \equiv {\Lambda\over 2}$ with a constant $\Lambda$ in \eqref{daiseyT} does not help either. In fact it leads to the same problem as one gets by viewing de Sitter as a vacuum solution, or as a very sharply peaked state. However we do know that we can come close to getting a four-dimensional positive cosmological constant solution without involving a constant term, as shown in \cite{desitter2}, but the problem with such an analysis is that we cannot justify it in a Wilsonian framework (see discussion in \cite{desitter2, joydeep}). Thus the question is whether we can get a four-dimensional de Sitter solution within a Wilsonian framework, {\it without} involving any extra factors on the RHS of an equation of the form \eqref{500casilos}. This is exactly what \eqref{bormey} does: it determines the dynamics of $\langle{\bf g}_{\rm MN}\rangle_\sigma$ in the standard way, but in a Wilsonian framework, and any extra factors go to the ghost EOMs as in \eqref{seldena}. The only difference is that the action governing the dynamics of $\langle{\bf g}_{\rm MN}\rangle_\sigma$, or more generically $\langle{\bf \Xi}\rangle_\sigma$, is $\check{\bf S}(\langle{\bf \Xi}\rangle_\sigma)$ coming from the renormalization effect studied in \cite{wdwpaper}. Additionally the very fact that $\hbar > 0$ (and not exactly by $\hbar \to 0$), the system is controlled by \eqref{bormey} and \eqref{seldena} and {\it not} by any classical EOM. This is the advantage we get by viewing de Sitter as a Glauber-Sudarshan state. \label{elecneel}}. Thus the de Sitter transient state comes exclusively from the quantum EOM
given by the first equation in \eqref{clodugark5} leading to \eqref{bormey} which has no classical analogue. We can tabulate the two scenarios as:
\vskip.1in
\begin{table}[H]  
 \begin{center}
\renewcommand{\arraystretch}{1.5}
\begin{tabular}{|c||c||c|}\hline {\bf EOM}  & {\bf Backgrounds} \\ \hline\hline
${\delta{\bf S}({\bf \Xi})\over \delta{\bf \Xi}} = 0$ & { Supersymmetric warped-Minkowski background}\\ \hline
${\delta \check{\bf S}(\langle {\bf \Xi}\rangle_\sigma) \over \delta \langle{\bf \Xi}\rangle_\sigma} = 0$ & {Non-supersymmetric transient de Sitter background}\\ \hline
\end{tabular} 
\renewcommand{\arraystretch}{1}
\end{center}
\label{firzacutoo}
 \end{table} 
\noindent thus clearly distinguishing the two backgrounds. The actions are also slightly different: the second one appears from the renormalization effects discussed in \cite{wdwpaper}. Interestingly, existence of the EOM \eqref{bormey} suggests that the dynamics of the {\it emergent} degrees of freedom $\langle{\bf \Xi}\rangle_\sigma$ may be controlled by the following partition function:
\bg\label{lidipoet}
\widetilde{\mathbb{Z}}[0] = \int {\cal D}\langle{\bf \Xi}\rangle_\sigma ~{\rm exp}\Big(-\check{\bf S}(\langle{\bf \Xi}\rangle_\sigma)\Big), \nd
without involving any additional Faddeev-Popov ghosts as they have already been taken care of in the construction of $\langle{\bf \Xi}\rangle_\sigma$ (see discussions in \cite{wdwpaper})\footnote{A simple way to see this is the following. Consider the usual partition function for a non-abelian gauge theory, namely $\mathbb{Z}_1[0]= \int {\cal D}{\bf A}~{\rm exp}\left(-{1\over 4}\int {\rm tr}~{\bf F}({\bf A}) \wedge \ast {\bf F}({\bf A})\right)$. This is incomplete because of gauge redundancies and would therefore require inserting Faddeev-Popov ghosts to extract finite answer from it. Now consider a Glauber-Sudarshan state $\vert\sigma\rangle$, from which we compute $\langle {\bf A}\rangle_\sigma$. Clearly in the computation of 
$\langle {\bf A}\rangle_\sigma$ all the ghosts have been taken care of, and therefore in the partition function $\mathbb{Z}_2[0]= \int {\cal D}\langle{\bf A}\rangle_\sigma~{\rm exp}\left(-{a\over 4}\int {\rm tr}~{\bf F}(\langle{\bf A}\rangle_\sigma) \wedge \ast {\bf F}(\langle{\bf A}\rangle_\sigma)\right)$, where $a$ is the renormalization parameter coming from the neighboring off-shell Glauber-Sudarshan states, no additional Faddeev-Popov ghosts are required because we are summing over gauge {\it inequivalent} configurations (although BRST ghosts could still appear). \label{kimmet}}. The classical and the quantum EOMs appearing from \eqref{lidipoet} respectively turn out to be:
\bg\label{korina}
{\delta\check{\bf S}(\langle{\bf \Xi}\rangle_\sigma)\over \delta \langle{\bf \Xi}(x)\rangle_\sigma} = 0, ~~~~~~~~ \left\langle {\delta\check{\bf S}(\langle{\bf \Xi}\rangle_\sigma)\over \delta \langle{\bf \Xi}(x)\rangle_\sigma}\right\rangle = 0, \nd
where $x = ({\bf x}, t)$. The first one from \eqref{korina} is a {\it quantum} EOM from the point of view of the warped-Minkowski background as evident from \eqref{clodugark5} but is a classical EOM from the point of view of \eqref{lidipoet}, whereas the latter is a genuine quantum EOM from \eqref{lidipoet}. The latter leads to a {\it second} Wheeler-de Witt equation of the following form (see also \cite{wdwpaper}, \cite{csvsgs}):
\bg\label{ryanfurn}
i{\partial \Psi(\langle{\bf \Xi}\rangle_\sigma\over \partial t} = 
\underbrace{\int d^n{\bf x}\sqrt{-\langle {\bf g}\rangle_\sigma} \left\langle {\delta\check{\bf S}(\langle{\bf \Xi}\rangle_\sigma)\over \delta \langle{\bf g}^{00}({\bf x})\rangle_\sigma}\right\rangle_{\rm oper}}_{{\mathbb{H}(\langle{\bf \Xi}\rangle_\sigma)}} \Psi(\langle{\bf \Xi}\rangle_\sigma) = 0, \nd
where the subscript ``oper'' denotes the operator formalism for the bracket and $n = 10$ from M-theory point of view. The effective Hamiltonian for the emergent degrees of freedom may be denoted by 
${\mathbb{H}(\langle{\bf \Xi}\rangle_\sigma)}$, and in this language we reproduce exactly the Wheeler-de Witt equation \cite{HH, FPI} for the accelerating transient de Sitter background {\it without} involving any extra Faddeev-Popov ghosts! The Wheeler-De Witt equation \eqref{ryanfurn} should be compared to the one we got for the warped-Minkowski background in \eqref{nestkathia} with vanishing boundary Hamiltonian. We can also tabulate them in the following way:

\vskip.1in

\vskip.1in
\begin{table}[H]  
 \begin{center}
\renewcommand{\arraystretch}{1.5}
\begin{tabular}{|c||c||c|}\hline {\bf Backgrounds}  & {\bf Wheeler-De Witt Equations} \\ \hline\hline
Warped-Minkowski & $i{\partial{\bf \Psi}_{\rm tot}({\bf \hat\Xi})\over \partial t} = \int d^n{\bf x}\sqrt{-{\bf g}}~{\delta {\bf S}_{\rm tot}({\bf \hat\Xi})\over \delta{\bf g}^{00}({\bf x})}\Big\vert_{\rm oper} {\bf \Psi}_{\rm tot}({\bf \hat\Xi}) = {\bf H}_{\rm tot}({\bf \hat\Xi}) {\bf \Psi}_{\rm tot}({\bf \hat\Xi}) =  0$ \\ \hline
Transient de Sitter & $i{\partial \Psi(\langle{\bf \Xi}\rangle_\sigma\over \partial t} =\int d^n{\bf x} \sqrt{-\langle {\bf g}\rangle_\sigma}\left\langle {\delta\check{\bf S}(\langle{\bf \Xi}\rangle_\sigma)\over \delta \langle{\bf g}^{00}({\bf x})\rangle_\sigma}\right\rangle_{\rm oper}\Psi(\langle{\bf \Xi}\rangle_\sigma) =
{{\mathbb{H}(\langle{\bf \Xi}\rangle_\sigma)}} \Psi(\langle{\bf \Xi}\rangle_\sigma) = 0$ \\ \hline
\end{tabular} 
\renewcommand{\arraystretch}{1}
\end{center}
\label{firzacutoo2}
 \end{table} 
\noindent where $\hat{\bf \Xi} = ({\bf \Xi, \Upsilon})$. Comparing the two Wheeler-De Witt equations, we can clearly see how the dynamics of the wave-functionals appear. In the emergent de Sitter case, the problem of time may not appear because 
the dynamics is governed by the Hilbert space constructed over a Minkowski/warped-Minkowski minimum that allows for a boundary Hamiltonian. Therefore from the warped-Minkowski point of view, one {\it can} provide a definite meaning to the temporal evolutions of the corresponding eigenstates in the gravitational and the matter sector. The pertinent question now is: where do the backreaction effects play a role here? This is what we turn to next.

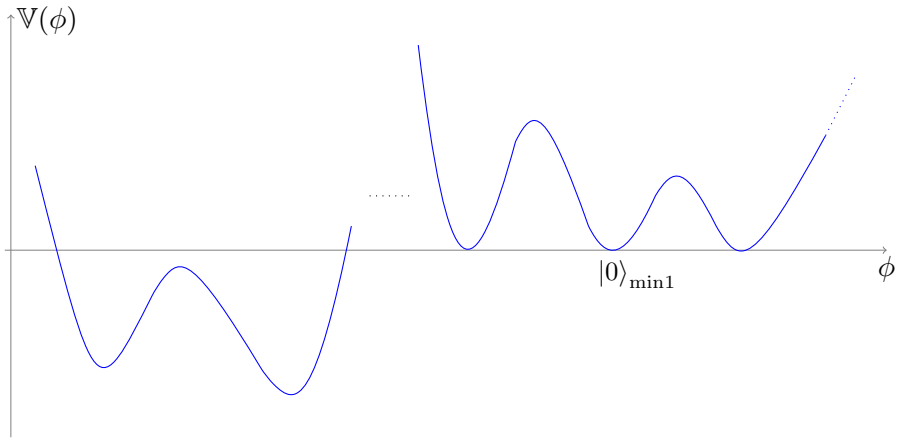
\begin{figure}
    \centering

  \begin{tikzpicture}[scale=0.8]
       \draw[ultra thin, gray, ->] (-0.4, -3.5) -- (-0.4, 3.5);
       \draw[ultra thin, gray, ->] (-0.5,-.4) -- (14,-.4);
       \draw [rounded corners, blue] (0,1) .. controls (1,-3) .. (2,-1) .. controls (2.4, -0.5) and (2.6, -0.5) .. (3.8, -2.5) .. controls (4.3, -3) and (4.6, -3) .. (5.2,0);
       \draw[dotted] (5.5,0.5) -- (6.2,0.5);
       \draw[blue] (6.3, 3) .. controls (6.8, -1.2) and (7.2,-1.2) .. (7.9,1.4) .. controls (8.2, 2) and (8.4, 2) ..  (9.1,0) .. controls (9.4,-0.6) and (9.7, -0.6) .. (10.2, 0.5) .. controls (10.5, 1) and (10.7, 1) .. (11.2, 0) .. controls (11.6, -0.7) and (11.8, -0.7) .. (13, 1.5);      
       \draw[dotted, blue] (13, 1.5) -- (13.5, 2.5);

       \draw[] (9.9, -0.8) node {\footnotesize $\ket{0}_{{\rm min}1}$};
       \draw[] (0.2, 3.4) node {\small $\mathbb{V}(\phi)$}; 
       \draw[] (14, -0.7) node {\small $\phi$};
  \end{tikzpicture}
    
    \caption{Potential for any on-shell degrees of freedom, here we show only a scalar modulus, consists of only supersymmetric Minkowski and AdS minima and {\it no} de Sitter minima. We also expect the AdS minima to be separated from the Minkowski minima by large field distance or by a potential wall to reduce the tunneling effects. A specific Minkowski minimum, labeled by $\vert 0\rangle_{\rm min1}$ and described in \eqref{streeshrad}, is chosen for illustrative purpose.}
    \label{bogoliubov}
\end{figure}

\subsection{Glauber-Sudarshan states and the backreaction effects  \label{sec2.5}}

Backreaction effects in quantum gravity \cite{backreaction} cannot be ignored and they play a crucial role here. In fact the question becomes more acute once we revisit the discussion in footnote \ref{elecneel}. The $\hbar \to 0$ limit seems to give us \eqref{500casilos} via \eqref{daiseyT} which, from various considerations\footnote{For example, it being sharply peaked in the configuration space and therefore spreading rapidly under quantum evolution thereby progressively losing any residual classical character.}, lead us to the only allowed supersymmetric time-independent warped-Minkowski solution as 
${\delta{\bf S}({\bf \Xi})\over \delta {\bf \Xi}}= 0$.

This similarity is not a coincidence. Taking one of the supersymmetric Minkowski minimum from {\bf figure \ref{bogoliubov}} labeled by $\vert0\rangle_{\rm min1}$, the effect of the Glauber-Sudarshan state should be to {\it replace} the Minkowski minimum by a transient de Sitter background. This {\it replacement} should be viewed as an additive correction coming from the Glauber-Sudarshan state itself. This additive correction is what we will quantify as the backreaction effects. In the following let us demonstrate this. 

We will take a scalar field model to show this, and from our experience so far, we can take one on-shell field $\varphi$. We will ignore the non-local part for simplicity here. Moreover, due to zero gauge redundancies with scalar fields, the complications of Faddeev-Popov (or BRST) ghosts do not enter here. This means the problems of time and temporal evolutions would exist even at the Minkowski level, but we will not worry about them here. The action and the displacement operator now will be:
\begin{equation}\label{chukamuk}
{\bf S}_{\rm tot}[\varphi] = \int d^{11} x \left[(\partial \varphi)^2 + {g\over 4!}\varphi^4\right], ~~~~ \mathbb{D}(\sigma, \varphi) = {\rm exp}\left(\int d^{11} x ~ \sigma \varphi\right), 
\end{equation}
which is the well-known $\varphi^4$ theory and therefore the coupling would be running. However if we consider only IR physics, this running is not important and we can take $g$ to be the renormalized coupling at that scale. Moreover, 
{if we ignore the tensor indices, $\varphi$ is like one of the on-shell metric component, say ${\bf g}_{00}$.  This means we can at least expect:}
\bg\label{wownerpas1}
\langle \varphi\rangle_\sigma \equiv \langle {\bf g}_{00}\rangle_\sigma = \left({1\over \Lambda t^2}\right)^{4\over 3} = \Lambda^{-4/3} t^{-8/3}
 \nd
where $\Lambda$ is the four-dimensional cosmological constant.
This way we might also be able to quantify $\sigma$.
 In terms of path-integral, which we can explicitly perform here, the emergent quantity $\langle\varphi\rangle_\sigma$ becomes (we will only show the numerator):

{\footnotesize 
\bg\label{wownerpasa2}
\langle\varphi\rangle_\sigma = \varphi_o({\bf x}) + {1\over {\cal D}}\int \prod_k d\varphi_k \prod_{k'} e^{-a_{k'}(\varphi_{k'} - \sigma_{k'}/a_{k'})^2} \Bigg[1 - {g\over 4!}\Bigg({1\over {\rm V}}\sum_{\tilde{k}}\varphi_{\tilde{k}}\Bigg)^4 + {\cal O}(g^2)\Bigg] \sum_{\hat{k}}{\varphi_{\hat{k}}\over {\rm V}} e^{-i\hat{k}.x},
\nd}
where ${\cal D}$ is the denominator; $\varphi_o({\bf x})$ is the classical background\footnote{For the simple $\varphi^4$ theory, the classical solitonic background $\varphi({\bf x}) = 0$. But we keep this for illustrative purpose only as \eqref{chukamuk} could be thought of as coming from a more complicated action that allows a non-trivial solitonic solution.}; $x = ({\bf x}, t)$; $\varphi_k$ are the Fourier components; $a_k = {k^2\over {\rm V}}$ is the propagator; ${\rm V}$ is the volume of the space; and $k_{\rm IR} \le k \le \mu$. We have ignored many subtleties as they are not very important for the present discussion. These have been carefully explained in \cite{joydeep, desitter2, coherbeta, coherbeta2, borel, gshet} which the interested readers may refer to. At ${\rm N}^{th}$ order \eqref{wownerpasa2} becomes:

{\footnotesize
\bg \label{wownerfasali}
\langle\varphi\rangle_\sigma = \varphi_o({\bf x}) + {1\over {\cal D}}\int \prod_k d\varphi_k \prod_{k'} e^{-a_{k'}(\varphi_{k'} - \sigma_{k'}/a_{k'})^2} \sum_{{\rm N} = 0}^\infty{(-g)^{\rm N}\over {\rm N}! (4!)^{\rm N} {\rm V}^{4{\rm N} + 1}}\left(\varphi_{k_1} + \varphi_{k_2} + ...\right)^{4{\rm N}} \sum_{\hat{k}}\varphi_{\hat{k}} e^{-i\hat{k}.x}, 
\nd}
where we have shown the explicit series representation of the scalar field in terms of it's momentum components. The expansion is done with respect to $-g$ to reflect the Euclidean nature. In fact the series can be studied using the so-called {\it nodal diagrams} that we initiated in \cite{borel}. They are like the Feynman diagrams but are less restricted because of the shifted vacua \cite{borel}. However as in Feynman diagrams, the nodal diagrams are controlled by a few rules. For the specific case in \eqref{wownerfasali}, they may be tabulated as:

\vskip.1in

\noindent $\bullet$ Because of the volume factor, the most dominant terms would be the ones where all the $4{\rm N} - 1$ momenta are different and only one of the interaction momentum match with the source momentum.

\vskip.1in

\noindent $\bullet$ If all of the $4{\rm N}$ momenta are different from the source momentum, such diagrams will be eliminated by the denominator of the path integral. This is much like the vacuum bubble story that one would expect with $\sigma = 0$.

\vskip.1in

\noindent $\bullet$ However any {\it two} of the interaction vertices could have the same momenta and be different from the source momentum. But then no other vertices could have the same momenta as the source, otherwise they will be suppressed by the volume factor.

 \vskip.1in

\noindent $\bullet$ If more than two vertices have the same momenta, then the amplitude will be suppressed by the volume factor and consequently become sub-dominant contributions.

\vskip.1in

\noindent Needless to say, this is new kind of quantum field theory computation because we are using shifted vacua. As mentioned earlier, this also keeps the one-point functions non-zero. For the present case we can work out the behavior of the one-point function by looking at the growth at the ${\rm N}^{\rm th}$ order.
At ${\rm N}^{\rm th}$ order the interaction and the source terms, for the most dominant amplitude, naively scale in the following way:
\bg\label{hkpgkm}
{(4{\rm N})! (-g)^{\rm N}\over {\rm N}! (4!)^{\rm N} {\rm V}^{4{\rm N} + 1}}, \nd
where we can clearly see some factorial growth, but is not the usual Borel one \cite{borelborel}. This is typically called the {Gevrey} growth \cite{gevrey} where the growth is some integer power of ${\rm N}!$, much like $({\rm N}!)^\alpha$. This is sometime referred to as Gevrey-$\alpha$. Interestingly one can show that any Gevrey-$\alpha$ growth eventually boils down to Gevrey-1, {\it i.e.} to $(\alpha{\rm N})!$ \cite{gevrey, borel, joydeep}. However \eqref{hkpgkm} cannot quite be the right answer because it ignores the momentum conservation in the nodal diagrams. How should we impose the momentum conservation here? And how would this influence the growth of the nodal diagrams?

One of the simplest way to conserve momentum is to identify a momentum $k$ line in the nodal diagram with a momentum $-k$ line (much like how we do in the standard Feynman diagrams), where $k$ is the eleven-momentum. Since $\varphi_{-k} = \varphi^\ast_k$, we can actually rewrite \eqref{wownerfasali} in the following way:

{\scriptsize
\bg \label{wownerfasali2}
\langle\varphi\rangle_\sigma = \varphi_o({\bf x}) + {1\over {\cal D}}\int \prod_k d\varphi_k \prod_{k'} e^{-a_{k'}(\varphi_{k'} - \sigma_{k'}/a_{k'})^2} \sum_{{\rm N} = 0}^\infty{(-g)^{\rm N}\over {\rm N}! (4!)^{\rm N} {\rm V}^{p{\rm N} + 1}}\left(\vert\varphi_{k_1}\vert^2 + \vert\varphi_{k_2}\vert^2 + ...\right)^{d} \sum_{\hat{k}}\varphi_{\hat{k}} e^{-i\hat{k}.x}, 
\nd}
with the last bit to be identified with the source momentum $\hat{k}$ to form $\vert\varphi_{\hat{k}}\vert^2$, implying that $d \approx 2{\rm N}$ and $p \le 3$. Unfortunately \eqref{wownerfasali2} still cannot be the full answer because momentum conservation can happen in multiple different ways, and the arrangement in \eqref{wownerfasali2} is only a special case. Thus, if we express the generic form for the expectation value as:
\bg\label{tagchin2}
\langle\varphi\rangle_\sigma = \varphi_o({\bf x}) + {1\over {\cal D}}\int \prod_k d\varphi_k \prod_{k'} e^{-a_{k'}(\varphi_{k'} - \sigma_{k'}/a_{k'})^2} \sum_{{\rm N} = 0}^\infty {\bf A}^{\rm N} \sum_{\hat{k}}\varphi_{\hat{k}} e^{-i\hat{k}.x}, 
\nd
then the simplest form for ${\bf A}^{\rm N}$ appears in \eqref{wownerfasali2}. To facilitate a more general construction we can use the technique developed in section 5.2 of \cite{joydeep} with distributions of 1's and 2's in a square bracket. For example:
\bg\label{kungkee}
\big[\underbrace{1,~ 1,~ 1,~ 1, ....,~ 1}_{p-1~{\rm fields}}, ~2\big]_{\varphi} ~ \equiv ~
\underbrace{\varphi_{k_1}~\varphi_{k_2} ~\varphi_{k_3}~... ~\varphi_{k_{p-1}}~\overbracket[1pt][7pt]{\varphi_{k}~\varphi_k}}_{p~{\rm fields} ~+ ~1~{\rm source}}, \nd
 where ``2'' denotes two scalar field components $-$ here $\varphi_{k_1}$ and $\varphi_{k_2}$ $-$ with equal (or opposite) momenta, {\it i.e.} $k_1 = \pm k_2$. We will also take $p = 4$ to be consistent with the action in \eqref{chukamuk}. Using this notation, ${\bf A}^{\rm N}$ takes the following form:

 {\footnotesize
\bg\label{tagchinke}
{\bf A}^{\rm N} \equiv\sum_{j = 0}^{\lfloor {4{\rm N} -1\over 2}\rfloor}\left({\bf A}^{\rm N}\right)_{1j} = {(-g)^{\rm N}\over {\rm V}^{p{\rm N} +1} {\rm N}! (4!)^{\rm N}}\sum_{j = 0}^{\lfloor {4{\rm N} -1\over 2}\rfloor} \big[\underbrace{1, ~1, ~ 1,...,~1}_{4{\rm N} - 2j -1}, ~\underbrace{2,~ 2,~ 2, ....,~ 2}_j, ~2\big]_{\varphi} ~ + ~ {\rm permutations}\, ,\nd}
from where we see that while some specific choices of $j$ reproduce the series in \eqref{wownerfasali2}, there are many more choices that go beyond \eqref{wownerfasali2}. In fact using \eqref{tagchinke} we can distinguish between the numerator ${\cal N}$ and the denominator ${\cal D}$ of the path-integral even for ${\rm N} = 1$. For example they take the form:

{\footnotesize
\bg\label{machelegoppa}
\begin{split}
&{\cal D} = {\bf I} - {g\over 4!}\left(\left[1, 1, 1, 1\right]_\varphi + \left[1, 1, 2\right]_\varphi + \left[2, 2\right]_\varphi\right) + {\cal O}(g^2)\\
& {\cal N} = \left[1\right]_\varphi - {g\over 4!}\left(\left[1, 1, 1, 1\right]_\varphi\left[1\right]_\varphi + \left[1, 1, 2\right]_\varphi\left[1\right]_\varphi+ \left[2, 2\right]_\varphi\left[1\right]_\varphi+ \left[1, 1, 1, 2\right]_\varphi + \left[1, 2, 2\right]_\varphi\right) + {\cal O}(g^2),
\end{split}
\nd}
where expectedly the bubble diagrams isolate out; and  the ${\rm V}^{-1}$ factors are arranged according to how the lines are contracted in the nodal diagram. One may also easily check the momentum conservation works well. For example in $[1, 1, 1, 1]_\varphi$ we allow field components with momenta $k_1, k_2, k_3$ and the fourth one with momentum $-k_1 - k_2-k_3$. This leads to a total momentum of zero. In the $[1, 1, 1, 2]_\varphi$, we identify $k_1 = k$ where $k$ is the source momentum. Then the momentum distribution is $(k_2, k_3, -k - k_2 - k_3, k, k)$ which leads to a total free momentum of $k$. Similarly in the $[1, 2, 2]_\varphi$ case, the momentum distribution becomes $(-k - 2k_2, k_2, k_2, k, k)$ again leading to a total free momentum of $k$, with $k$ being the source momentum. At ${\cal O}(g^2)$ momentum conservation demands that the distribution of 1's and 2's and the corresponding momenta for the field components can {\it atmost} be:
\bg\label{tranilond}
\left[1, 1, 1, 2, 2, 2\right]_\varphi = \varphi_{-k -2k_2}~ \varphi_{k_3'}~\varphi_{-2k_1'-k_3'}~\overbracket[1pt][7pt]{\varphi_{k_1'}\varphi_{k_1'}}~\overbracket[1pt][7pt]
{\varphi_{k_2}\varphi_{k_2}}~\overbracket[1pt][7pt]{\varphi_k\varphi_k} \nd
with the total momentum sum leading to $k$, the momentum of the source. A distribution like $[1, 2, 2, 2, 2]_\varphi$ is sub-dominant as one may easily verify, and $[2, 2, 2, 2, 2]_\varphi$ cannot happen because the number of nodes will not permit this distribution. Moreover, since the momentum conservation happens as every nodes with four lines (because we are taking $\varphi^4$ theory), the distribution of 1's and 2's in the nodal diagrams becomes more involved. Thus in the generic distribution given in \eqref{tagchinke} we expect $j < \left\lfloor {4{\rm N} -1\over 2}\right\rfloor$. (As an example, for ${\rm N} = 2$ discussed above, the maximum value for $j$ is $j = 2$ and not $j = 3$.) As ${\rm N} >> 1$, the maximum value of $j$ approaches 
$\left\lfloor {4{\rm N} -1\over 2}\right\rfloor$. However there are a couple of reasons that tell us that the aforementioned complications {\it may not} pose any issues in our computations. The reasons are as follows. 

\vskip.1in

\noindent $\bullet$ We have taken one scalar field, but in M-theory there are atmost 256 field components at the supersymmetric Minkowski level excluding the ghosts\footnote{Despite this, the growth of the stringy diagrams continues to be $({\rm N}!)^2$. This should be compared to the ${\rm N}!$ growth of the Feynman diagrams. This is explained more clearly in section 1.3 of the first reference in \cite{gshet}.}. Even if we, for simplicity, label these degrees of freedom as scalar fields, consistency require us to take more than one such scalar fields. We can then relocate the momentum conservation to one of the scalar fields and allow the remaining fields to have Fourier components with arbitrary momenta respectively. This is of course the approach we took in \cite{borel}, and \cite{joydeep} with three and four scalar fields respectively.

\vskip.1in

\noindent $\bullet$ In string/M theory, all the bosonic fields appear with spatial and temporal {\it derivatives}. This allows us to provide precise ${\rm M}_p$ scalings for the interaction terms since the fields are taken to be dimensionless. (The fermionic terms do not pose any problems because they are taken to be dimensionful.) Thus in the simpler set-up where we use only scalar fields, we should also insert derivatives so as to match up with the string/M theory interactions. In the nodal diagrams these derivatives bring in extra powers of the momentum factors accompanying $({\bf A}^{\rm N})_{1j}$ in \eqref{tagchinke}. Therefore the aforementioned computations should have additional contributions from the momentum factors (see \cite{borel, joydeep} for more details).

\vskip.1in

\noindent The two points discussed above suggest that we could relax the momentum conservation at the $\varphi^4$ vertex by relocating it to another scalar interaction to incorporate a bigger picture. This means we can relax any constraint on $j$ and take $j_{\rm max} = \left\lfloor {4{\rm N} -1\over 2}\right\rfloor$ in \eqref{tagchinke}. But since 
$({\bf A}^{\rm N})_{1j} \ne ({\bf A}_{1j})^{\rm N}$ for $j \ge 0$, we are not out of the water yet. However for $j = 0$, it is easy to see that by defining:
\bg\label{penelopecar}
\begin{split}
& \big[\underbrace{1,~ 1,~ 1,~ 1, ....,~ 1}_{p~{\rm fields}}, ~1\big]_{\varphi} ~ \equiv ~
\underbrace{\varphi_{k_1}~\varphi_{k_2} ~\varphi_{k_3}~... ~\varphi_{k_{p-1}}~{\varphi_{k_p}~\varphi_k}}_{p~{\rm fields} ~+ ~1~{\rm source}}\\
& ~~~~~~{\bf A}_{0j} \equiv -{g\over 4!{\rm V}^p} \big[\underbrace{1, ~1, ~ 1,...,~1}_{p - 2j -1}, ~\underbrace{2,~ 2,~ 2, ....,~ 2}_j, ~1\big]_{\varphi}
\end{split}
\nd
for an interaction of the form $\varphi^p$, where $p = 4$ here, $\left({\bf A}^{\rm N}\right)_{10} = {\bf A}^{\rm N -1}_{00} {\bf A}_{10} = {\bf A}_{00}^{\rm N} \left({\bf A}_{00}^{-1} {\bf A}_{10}\right)$. Comparing \eqref{tagchinke} with \eqref{penelopecar}, we see that the numerator of the path-integral representation of $\langle\varphi\rangle_\sigma$ is controlled by $({\bf A}^{\rm N})_{1j}$ whereas the denominator is controlled by $({\bf A}^{\rm N})_{0j}$. For $j = 0$, $({\bf A}^{\rm N})_{00} = ({\bf A}_{00})^{\rm N}$, and in the following we will first take the simplest case of $j = 0$. 

Even for this simple case there are a large number of subtleties pointed out in \cite{borel, joydeep}. For example, the prominent ones are: (1) the no-go theorem \cite{nogo} puts a constraint on the tree-level result, (2) the ratio 
${\bf A}_{00}^{-1} {\bf A}_{10}$ introduces some additive constant that is generically suppressed by the powers of the interaction term, (3) there are other interaction terms that contribute to ${\bf A}_{00}^{\rm N}$ at the ${\rm N}^{\rm th}$ level, and (4) the factorial growth of the nodal diagrams. 
 After we integrate out the fields in the path-integral, and take care of the aforementioned (and a few other) subtleties, the amplitude now grow as:
\begin{equation}
\sum_{{\rm N} = 0}^\infty (-g)^{\rm N} (\alpha{\rm N})! {\cal A}^{\rm N} \int_{k_{\rm IR}}^\mu d^{11} k~{\sigma(k)\over a(k)} ~e^{-ik.x} \nonumber
\end{equation}
where $(\sigma(k), a(k))$ are the parameter appearing in the displacement operator \eqref{chukamuk} and the path integral \eqref{wownerpasa2} respectively. We have also expressed the factorial growth as $(\alpha{\rm N})!$. This may be motivated in the following way. The naive growth can be inferred from \eqref{hkpgkm} as ${(4{\rm N})!\over {\rm N}!}$. On the other hand if we consider \eqref{wownerfasali2}, the growth is roughly ${(2{\rm N})!\over {\rm N}!}$. The $(4{\rm N})!$ factorial growth appears from a binomial expansion ignoring the momentum constraint at the $\varphi^4$ vertex. As mentioned earlier, this cannot quite be right in the absence of other fields (to absorb the momenta). On the other hand the $(2{\rm N})!$ growth is a bit too restrictive. Thus we express the factorial growth as $(\alpha {\rm N})!$ where $1 \le \alpha \le 2$ \cite{gshet}. 
More precisely, what is required for the resummation is the asymptotic Gevrey class
rather than the exact finite-order factorial.  Thus the large-order behavior should be
understood schematically as:
\begin{equation}
\mathcal A^{\rm N}\Gamma(\alpha {\rm N}+\beta)
\left[1+{\cal O}\left({1\over {\rm N}}\right)\right],
\nonumber
\end{equation}
with fixed $\beta$, which is more general than $(\alpha{\rm N})!\mathcal A^{\rm N}$. Such shifts and subleading powers do not alter the Gevrey order
$\alpha$ or the associated nonperturbative exponential scale, although they may modify
residues and fluctuation prefactors.
The other parameter ${\cal A}$ can be extracted from ${\bf A}_{00}$ and 
is defined in the following way:
\bg\label{ellaholly}
{\cal A} = {\cal A}(k_{\rm IR}, \mu) \equiv \prod_{i = 1}^4\int_{k_{\rm IR}}^\mu d^{11}k_i~{\sigma(k_i)\over a(k_i)}, 
\nd
where $k_{\rm IR} \le k \le \mu$ with $k_{\rm IR}$ being the IR cut-off and $\mu$ is the accessible energy scale.  
Of course we have ignored some details like the contribution from the denominator of the path integral et cetera. We will rectify this soon.
 Sparing us of all these details and performing the Borel-\'Ecalle resummation \cite{borelborel, ecalle}, after the dust settles we get \cite{borel}:
\bg\label{jeswest}
\langle\varphi\rangle_{\sigma} = \varphi_o({\bf x}) + {1\over g^{1/\alpha}}\left[\int_0^\infty d{\rm S} ~{\rm exp}\left(-{{\rm S}\over g^{1/\alpha}}\right) {1\over 1 - {\cal A} {\rm S}^{\alpha}}\right]_{\rm P.V} \int_{k_{\rm IR}}^\mu d^{11}k ~{\sigma(k)\over a(k)}~e^{-ik.x}, 
\nd
where P.V is the principal value of the integral over S and $a(k)$ is the massless propagator. Note that in writing \eqref{jeswest}, we have replaced $g$ by $-g$ as this would be more natural in the action from \eqref{chukamuk}. (It would also avoid introducing unnecessary factors of $\sqrt{-1}$ in the exponential and elsewhere.) We now make a few observations.
 \vskip.1in
\noindent $\bullet$ Within the assumed large-order Gevrey behavior, the asymptotic series may be given a
nonperturbative Borel--\'Ecalle completion.  This goes beyond any finite
order-by-order truncation and is the essential ingredient needed for the result obtained
here.

 \vskip.1in
 
\noindent $\bullet$ The final answer is a wave-function renormalization of the tree-level answer. This relies on the subtlety that we alluded to earlier underlying the no-go theorem from \cite{nogo} which prohibits the generation of four-dimensional de Sitter at the tree-level.

 \vskip.1in
 
\noindent $\bullet$ In our analysis we kept $\alpha$ undetermined. Interestingly this doesn't effect the result significantly because  whether ${\alpha}$ is even or odd, there is only one pole in the Borel axis. This may be inferred from figures 5 and 6 in \cite{borel}.

\vskip.1in

\noindent All what we said above is interesting, but the main goal of this section, and in particular by following the procedure of the Borel-\'Ecalle resummation, is two-fold. One, is to determine the back-reaction effects, and two, is to determine a closed form expression for the cosmological constant. Somewhat unexpectedly, both these are tied to each other as we shall see in the following.

We will start by analyzing the backreaction problem, but before that let us demonstrate that the closed form expression for the wave-function renormalization term takes the following expected instanton series format:
\begin{eqnarray}\label{audplaza}
&&\int_0^\infty d{\rm S}~ {{\rm exp}\left({-{\rm S}/g^{1/\alpha}}\right)\over 1 - {\cal A} {\rm S}^{\alpha}} \nonumber\\
&= & \sum_{{\rm N} = 1}^\infty 
~~\underset{\rm saddle}{\underbrace{{\rm exp}\left(-{{\rm N}\over g^{1/\alpha}}\right)}_{\rm instanton}} ~~\underbrace{\left[{1\over 1 - {\rm N}^{\alpha} {\cal A}} + \sum_{n = 1}^\infty {{\cal A}^n \Big(\sum\limits_{r = 1}^\infty {}^\alpha{\bf C}_r~{\rm N}^{\alpha - r} (s - {\rm N})^r\Big)^n\over \left(1- {\rm N}^{\alpha} {\cal A}\right)^{n+1}}\right]}_{\rm fluctuation~determinant},
\end{eqnarray}
where ${}^\alpha{\bf C}_r = \Big(\begin{matrix} \alpha \\ r\end{matrix}\Big)$. One may easily see that \eqref{audplaza} 
takes the expected form of an infinite number of instanton saddles along-with their corresponding fluctuation determinants. These instanton saddles appear from both real and the complex turning points around the supersymmetric Minkowski minimum in {\bf figure \ref{bogoliubov}}.  Around the zero instanton sector, the result is just 1 with no fluctuations, at least for the simple case we study here. In fact, since $s - {\rm N}$ is a small fluctuation over the instanton number ${\rm N}$, we can safely take $r = 1$ in \eqref{audplaza}. For such a case we have:
\bg\label{catkim}
\int_0^\infty d{\rm S}~ {{\rm exp}\left({-{\rm S}/g^{1/\alpha}}\right)\over 1 - {\cal A} {\rm S}^{\alpha}} = \sum_{{\rm N} = 0}^\infty
{\rm exp}\left(-{{\rm N}\over g^{1/\alpha}}\right)
\left[{1\over 1 - {\rm N}^{\alpha} {\cal A}} + \sum_{n = 1}^\infty {\alpha^n{\cal A}^n ~{\rm N}^{n(\alpha - 1)} \over \left(1- {\rm N}^{\alpha} {\cal A}\right)^{n+1}} \left(s - {\rm N}\right)^n\right], \nd
where we have taken $\alpha > 1$. 
It is easy to see that for $\alpha = 2$ we reproduce precisely the instanton series from \cite{joydeep}. The above analysis reveals that all the instantons conspire to give us a closed form expression for $\langle\varphi\rangle_\sigma$. 

This powerful technique of the non-perturbative summation now gives us a way to implement the backreactions quantitatively. We want our final answer to look like \eqref{wownerpas1} as we have identified the scalar field with the ${\bf g}_{00}$ metric component. But the path integral gives us the expectation value as in \eqref{jeswest} which contains two distinct pieces: one coming from the solitonic background which is basically the vacuum configuration, and the other coming from the full non-perturbative completion as we showed above. To reproduce the result from \eqref{wownerpas1}, we can now split $\sigma(k)$ into two pieces: 
\bg\label{westofjessi}
\sigma(k) \equiv \sigma(k_0, {\bf k}) = \sigma_1(k_0, {\bf k}) + \sigma_2(k_0, {\bf k}), \nd
where the first choice of $\sigma_1(k_0, {\bf k})$ creates a Glauber-Sudarshan state that does not vary with time by imposing a delta function constraint on $k_0$. This is an off-shell state ({\it i.e.} a state with zero frequency but non-zero ${\bf k}$), and such a state can be arranged to kill off the solitonic background $\varphi_o({\bf x})$. Of course this cancellation is only precise when we take the most probable amplitude of the Glauber-Sudarshan state\footnote{Not to be confused with the most probable amplitude for the wave-functional $\Psi(\langle\varphi\rangle_\sigma)$ or, in a more generic setting, for the wave-functional $\Psi(\langle{\bf\Xi}\rangle_\sigma)$. Recall that 
$\Psi(\langle\varphi\rangle_\sigma)$ (or $\Psi(\langle{\bf\Xi}\rangle_\sigma)$) is the envelope wave-functional over all the on- and off-shell Glauber-Sudarshan states \cite{wdwpaper}.}. The other piece, namely $\sigma_2(k_0, {\bf k})$, with a delta function constraint on ${\bf k}$, can now be arranged to reproduce \eqref{jeswest} using the Fourier model $e^{-ik_0 t}$. Additionally $\sigma(k)$ should also satisfy the following equation:
\bg\label{maristomie}
{\delta \check{\bf S}(\langle\varphi\rangle_\sigma)\over \delta \langle\varphi\rangle_\sigma} = 0 = \left\langle {\delta \check{\bf S}(\langle\varphi\rangle_\sigma)\over \delta\langle\varphi\rangle_\sigma}\right\rangle, \nd
which are precisely the {\it classical} and the {\it quantum} EOMs from \eqref{korina} (since we are only taking a scalar field $\varphi$). Putting everything together, we see that the backreaction issue can be quantified precisely once the Glauber-Sudarshan state $\vert\sigma\rangle$ satisfies the following two conditions.
\vskip.1in

\noindent $\bullet$ $\sigma(k)$ should be split into two pieces, one of which remains constant with respect to time. 

\vskip.1in

\noindent $\bullet$ The resulting $\sigma(k)$ should satisfy the remnant of the Schwinger-Dyson equations given in \eqref{korina} including the ghost EOM \eqref{seldena}. 

\vskip.1in

\noindent Both the aforementioned conditions are {\it necessary} otherwise no solution exists. A simpler quantitative discussion of how this is achieved has appeared in \cite{borel, joydeep} which the readers may want to look at.  
However an even more surprising result appears when we identify 
$\langle\varphi\rangle_\sigma$ with $\langle{\bf g}_{00}\rangle_\sigma$.  In other words:
\begin{eqnarray}\label{sorina}
\langle \varphi\rangle_\sigma & \equiv &  \langle {\bf g}_{00}\rangle_\sigma = \left({1\over \Lambda t^2}\right)^{4\over 3} = \Lambda^{-4/3} t^{-8/3} \nonumber\\
 & = & \underbrace{{1\over g^{1/\alpha}}\left[\int_0^\infty d{\rm S} ~{\rm exp}\left(-{{\rm S}\over g^{1/\alpha}}\right) {1\over 1 - {\cal A} {\rm S}^{\alpha}}\right]_{\rm P.V}}_{\Lambda^{-4/3}} \underbrace{\int_{k_{\rm IR}}^\mu d^{11}k ~{\sigma_2(k_0, {\bf k})\over a(k)}~e^{-ik.x}}_{t^{-8/3}},
\end{eqnarray}
where expectedly only $\sigma_2(k_0, {\bf k})$ from \eqref{westofjessi} appears, as $\sigma_1(k_0, {\bf k})$ is used up to remove the solitonic background $\varphi_o({\bf x})$. The identification of $\sigma_1(k_0, {\bf k})$ to the solitonic background $\varphi_o({\bf x}) = \eta_{00}$
of course ignores the warp-factor that is necessary to allow for a supersymmetric warped-Minkowski vacuum solution. It's insertion however doesn't change much so we will avoid making the analysis more generic. 
The important point is that the identifications in \eqref{sorina}
not only fixes the form for $\sigma(k)$ but also provides a closed form expression for the four-dimensional cosmological constant $\Lambda_{4d} \equiv {\rm M}_p^2 \Lambda$!
 This means the four-dimensional cosmological constant is given by the following expression:
 \begin{equation}\label{valentisab}
\Lambda_{4d} = {{\rm M}_p^2\over {1\over g^{3/4\alpha}}\left[\int_0^\infty d{\rm S} ~{\rm exp}\left(-{{\rm S}\over g^{1/\alpha}}\right) {1\over 1 - {\cal A} {\rm S}^{\alpha}}\right]^{3/4}_{\rm P.V}} 
\end{equation}
which is positive definite irrespective of the sign of ${\cal A}$ \cite{borel}. As was discussed in \cite{borel}, this is {\it not} a small number because the principal value integral in the denominator just provides an ${\cal O}(1)$ factor. However as discussed in \cite{joydeep}, 
\eqref{valentisab} can be made small if we include the contributions from 
$({\bf A}^{\rm N})_{1j}$ for $j > 0$ in \eqref{tagchinke}. This involves new construction with ``\textcolor{blue}{Borel Boxes}" described in section 5.2 and Tables 4 and 5 of \cite{joydeep}. If we also include higher order interactions beyond $\varphi^4$, one can show that the four-dimensional cosmological constant could be lowered significantly. As to how low it can get, or whether we can come even closer to a value like $10^{-120}{\rm M}_p^2$, will be discussed elsewhere.

Before ending this section let us clarify a few things that we kept under the rug so far. \textcolor{blue}{First}, note that the principal value integral appearing in \eqref{jeswest} can be rewritten in the following suggestive way:
\bg\label{jesuzuki}
{1\over g^{1/\alpha}}\left[\int_0^\infty d{\rm S} ~{\rm exp}\left(-{{\rm S}\over {g}^{1/\alpha}}\right) {1\over 1 - {\cal A} {\rm S}^{\alpha}}\right]_{\rm P.V} = {1\over \hat{g}^{1/\alpha}}\left[\int_0^\infty d{u} ~{\rm exp}\left(-{u\over \hat{g}^{1/\alpha}}\right) {1\over 1 - u^{\alpha}}\right]_{\rm P.V}, \nd
where $\hat{g} = g{\cal A}$, and therefore there is only one pole at $u = 1$ on the real axis in the Borel plane, irrespective of the choice of ${\cal A}$. This also means that any changes to ${\cal A}$ can be absorbed in the definition of the renormalized coupling $\hat{g}$. This is good because, although the definition of ${\cal A}$ in \eqref{ellaholly} appears to be sensitive to the choice of $\sigma(k)$ in \eqref{westofjessi}, in the limit the renormalized coupling $\hat{g} \equiv g{\cal A} \to 0$ the denominator in \eqref{valentisab} approaches identity.

\textcolor{blue}{Secondly}, it is instructive to specify the precise values for $\sigma_1(k_0, {\bf k})$ and $\sigma_2(k_0, {\bf k})$ in \eqref{westofjessi}. It is clear from \cite{borel, joydeep} that we can always express $\sigma_i(k_0, {\bf k})$ as powers of the eleven-momentum $k$ as $\sigma_i(k_0, {\bf k}) = {k^{{}^{\rho_i(k_0, {\bf k})}}\over 2{\rm V}}$, with ${\rm V}$ being the eleven-dimensional volume and $k^2 = k^2_0 + {\bf k}^2$. The form for $\rho_i(k_o, {\bf k})$ appears in eq. (5.4) and (5.2) respectively in \cite{joydeep} which allows us to make the split in the second line of \eqref{sorina}. However this raises the question whether one can justify the uniqueness of such a split. Additionally the appearance of $\Lambda$ in the definition of $\rho_1(k_o, {\bf k})$ might be a bit disconcerting. The latter is fortunately not a problem because of the insensitivity of ${\cal A}$ in the limit of $\hat{g} \to 0$ for the principal value integral. Thus the {\it smallness} of the four-dimensional cosmological constant only depends on distribution of the rows and columns in the \textcolor{blue}{Borel Boxes} as evident from eq. (5.19) of \cite{joydeep}. 

The uniqueness of split in the second line of \eqref{sorina} is still an issue. The ambiguity can appear from replacing the $\log~2$ factors in eq. (5.2) and (5.4) of \cite{joydeep} by arbitrary positive logarithmic numbers. They have to be  the same for both cases because we can match the two solutions at $t = -{1\over \sqrt{\Lambda}}$. This matching tells us that, if we replace the $\log~2$ factor in eq. (5.4) of \cite{joydeep}, $\Lambda$ needs to be changed by exactly the inverse factor to be equal (albeit with an opposite sign) to the solitonic solution. Thus eq. (5.4) cannot have any arbitrary logarithmic number replacing $\log~2$, and from the matching condition mentioned earlier, eq. (5.2) of \cite{joydeep} also cannot have any arbitrary positive logarithmic number replacing $\log~2$ there.

The aforementioned analysis then tells us that, at least for this simple toy model, one may quantify both the back-reaction effects as well as the value of the four-dimensional cosmological constant precisely. Note that both the results are interconnected and without using the powerful technique of Borel-\'Ecalle resummation procedure we would not have been able to infer them correctly here. In fact the resummation procedure is absolutely necessary to quantify the value of the four-dimensional cosmological constant and in turn argue for it's positivity. This suggests that maybe the cosmological constant is an {\it emergent} phenomena that is completely invisible at order-by-order perturbative level. Once we sum up all the non-perturbative effects, the positive cosmological constant emerges out of it. This fascinating idea clearly needs more elaboration, and in particular we would like to see how we can increase the denominator of \eqref{valentisab} so that the four-dimensional cosmological constant may become very small. This and other related topics will be discussed elsewhere.

\section{Discussion and conclusion \label{sec3}}

One of the misconception that has propagated for a long time is the belief that quantum gravity can be treated in the same footing as the non-gravitational quantum field theories. This becomes abundantly clear once we try to define localized states in quantum gravity using canonical methods.
In this work we have pointed out the difficulty in defining the Glauber-Sudarshan states using such canonical formalism. Due to various constraints, and especially the ones coming from the Wheeler-De Witt equation, it appears that only way to define them properly is to use the path integral approach. The result is that the Glauber-Sudarshan states take similar forms as the vertex operators in string theory, which is how string theory beautifully overcomes these issues\footnote{As discussed here, vertex operators in string theory are defined as integrated functions over both world-sheet (or world-volume) space and time directions in a path integral approach, thus taking care of all constraints emerging from the Wheeler-De Witt equation.}. 
The key difference, however, is that these states have their origin in M-theory---where
there is no known comparably simple world-sheet vertex-operator formalism---and their
allowed configurations are governed by the quantum Schwinger--Dyson/bootstrap
conditions of the effective theory.  Conventional perturbative string vertex operators,
by contrast, are characterized by the corresponding world-sheet physical-state or
marginality conditions, which reproduce the spacetime equations of motion in the
appropriate semiclassical limit.
 We can then make duality transformations to construct the corresponding Glauber-Sudarshan states in dimensionally reduced string theories \cite{desitter2, coherbeta, coherbeta2, gshet}. 

The vertex-operator like formalism, albeit from eleven-dimensional point of view, also dispels any notion that the M-theory action in the far IR is augmented by new interactions. In fact the picture that emerges from the canonical point of view is that the displacement operator is defined at every instant of time (within the allowed trans-Planckian temporal bound)
without invoking a bulk Hamiltonian (as it vanishes due to the diffeomorphism constraints) and only in the presence of a boundary Hamiltonian. Such a picture naturally extends to the path integral as sum over histories, but also reveals the necessity of adding the Faddeev-Popov ghosts. The latter is very important because, according to a theorem by Weinberg \cite{weinberg}, once we take the gravitational degrees of freedom there does not exist a gauge where ghosts can be completely decoupled. Such a consideration leads to the fascinating possibility of (a) the ghosts appearing in the Wheeler-De Witt equation itself, or (b) the Wheeler-De Witt equation being expressed in terms of a physical Hamiltonian with decoupled BRST ghosts. 

Regarding the Wheeler-De Witt equation, we show that it appears at two levels. One at the supersymmetric warped-Minkowski level, and the other at the emergent non-supersymmetric de Sitter level. The former does involve the Faddeev-Popov ghosts and interestingly they provide a way to define temporal evolutions at the supersymmetric Minkowski level once the boundary Hamiltonian is taken into account. The latter, {\it i.e.} the Wheeler-De Witt equation at the emergent de Sitter level, does not involve the Faddeev-Popov ghosts, although we expect the BRST ghosts to appear in the construction. This somewhat surprising result was first noticed in \cite{wdwpaper} and here we provide more detailed elaboration. The Hamiltonian constraints at {\it both} levels tell us that one needs to carefully study the dynamics at the supersymmetric warped-Minkowski and the transient de Sitter cases. In fact it appears that only the path integral approach can correctly capture the dynamics, and other popular attempts using canonical methods may lead to somewhat ambiguous results.  

The path integral method also leads to two rather interesting outcomes. When we try to compute the expectation value of an on-shell field component $-$ here we take a scalar field for simplicity $-$ the factorial growth of the resulting so-called nodal diagrams \cite{borel, joydeep} allows us to perform the Borel-\'Ecalle resummation \cite{borelborel, ecalle} to express the answer in a closed form. This non-perturbative resummation technique then gives us a way to quantify the four-dimensional cosmological constant as well as the back-reaction effects simultaneously.

The fact that the two outcomes $-$ the closed form expression for the cosmological constant and the backreaction effects $-$  are interconnected is not a surprise. The backreaction effects convert the vacuum supersymmetric Minkowski background to the transient de Sitter phase thus imbibing it with a net positive energy. One would therefore anticipate such a connection\footnote{The non-existence of a positive energy minima in the potential for the fields, as shown in {\bf figure \ref{bogoliubov}}, serves as a precursor to the aforementioned connection between the two outcomes. This in turn helps us avoid pitfalls like the ones coming from considerations related to the Bogoliubov transformations.}. The surprising thing however is that not only we could get a closed form expression of the cosmological constant, albeit for a toy model, but we could also argue for its {\it smallness} once we bring in the detailed construction of the Borel Boxes \cite{joydeep}. Such a consideration suggests that maybe the cosmological constant is truly an emergent phenomena and is completely invisible at the order-by-order perturbative level. 

Two things however stops us to declare this as a solution to the cosmological constant problem. The first one is of course the simple minded toy model approach to the problem. The actual consideration with 256 field components in the far IR of M-theory is a far cry from the one scalar field model with $\varphi^4$ interaction that we took here. Once we go to an actual set-up, the full Borel-\'Ecalle resummation \cite{borelborel, ecalle} in the presence of Batalin-Vilkovisky ghosts \cite{batalin} will be a tremendous technical challenge. And so would be to work out the Borel Boxes for all these cases, {\it i.e.} to sum over all the columns in each of these Borel Boxes following the strategy laid out here. 

The second one is the precise determination of the Glauber-Sudarshan states associated to the on-shell degrees of freedom. This is because it requires us to solve the set of equations in \eqref{korina}. What makes it hard is that the action appearing therein, {\it i.e.} $\check{\bf S}(\langle{\bf \Xi}\rangle_\sigma)$ is {\it not} Wilsonian. The Wilsonian, or the Exact Renormalization Group, approach can only give us ${\bf S}(\langle{\bf \Xi}\rangle_\sigma)$ as in the first three steps of \eqref{aspenb}. The final step that converts ${\bf S}(\langle{\bf \Xi}\rangle_\sigma)$ to $\check{\bf S}(\langle{\bf \Xi}\rangle_\sigma)$ relies on the renormalization technique that we pointed out in \cite{wdwpaper}, which in turn depends on solving the Wheeler-De Witt equation \eqref{ryanfurn}. This interlinked nature makes the system non-trivial, and we hope to report progress on this in near future.

\section*{Acknowledgements}
We would like to thank Suddhasattwa Brahma, Robert Brandenberger, Simon Caron-Huot, Joydeep Chakravarty, Renata Kallosh, Archana Maji, Adel Rahman, P. Ramadevi, Savdeep Sethi and Radu Tatar for many helpful discussion and correpondence; and Gary Shiu, Timm Wrase and Irene Valenzuela for organising a stimulating ``The Swampland and the Landscape'' conference at the ESI, Vienna this summer. The work of KD, FG and BK is supported in part by the NSERC grant.


\end{document}